\documentclass[11pt]{article}
\usepackage{graphicx}
\usepackage{dcolumn}
\usepackage{amsmath}
\usepackage{epsfig}
\usepackage{multirow}
\newcommand{\BABARConfNumber} {09/003}
\newcommand{\SLACPubNumber}{13758}

\input {pubboard/babarsym}
\newcommand{\bei}{\begin{itemize}}
\newcommand{\eei}{\end{itemize}}
\newcommand{\beq}{\begin{equation}}
\newcommand{\eeq}{\end{equation}}
\newcommand{\beqn}{\begin{eqnarray}}
\newcommand{\eeqn}{\end{eqnarray}}
\newcommand{\beqns}{\begin{eqnarray*}}
\newcommand{\eeqns}{\end{eqnarray*}}

\def\epem{e^+e^-}
\def\lplm{l^+l^-}
\def\lplmP{l^{\prime +}l^{\prime -}}
\def\mupmum{\mu^+\mu^-}
\def\Wp{W^\prime}
\def\dm{\Delta m}
\def\mbar{\overline{m}}
\def\foure{\epem\epem}
\def\fourmu{\mupmum\mupmum}
\def\twoetwomu{\epem\mupmum}

\def\ea{{\em et al.}}
\def\min{{\rm min}}

\def\rPTbarkappa{\kern 0.18em\overline{\kern -0.18em r}{}^{\kappa}{}}
\def\rPTbarsigma{\kern 0.18em\overline{\kern -0.18em r}{}^{\sigma}{}}

\def\deltabarkappa{\kern 0.18em\overline{\kern -0.18em \delta}{}_r^{\kappa}}
\def\deltabarsigma{\kern 0.18em\overline{\kern -0.18em \delta}{}_r^{\sigma}}
\def\deltaTbarkappa{\kern 0.18em\overline{\kern -0.18em \delta}{}_T^{\kappa}}
\def\deltaTbarsigma{\kern 0.18em\overline{\kern -0.18em \delta}{}_T^{\sigma}}

\def\OC{X}
\def\OCbar{{\kern 0.18em\overline{\kern -0.18em \OC}}}

\def\a{\kappa}

\def\ab{\a\b}

\def\Amptpbar{\kern 0.18em\overline{\kern -0.18em {\cal A}}_{{\overline B^0} \rightarrow K^-\pi^+\pi^0}}

\def\Amptpbarkappa{\kern 0.18em\overline{\kern -0.18em A}{}^{\kappa}{}}
\def\Amptpbarsigma{\kern 0.18em\overline{\kern -0.18em A}{}^{\sigma}{}}

\def\Tbarkappa{\kern 0.18em\overline{\kern -0.18em T}{}^{\kappa}{}}
\def\Tbarsigma{\kern 0.18em\overline{\kern -0.18em T}{}^{\sigma}{}}
\def\Pbarkappa{\kern 0.18em\overline{\kern -0.18em P}{}^{\kappa}{}}
\def\Pbarsigma{\kern 0.18em\overline{\kern -0.18em P}{}^{\sigma}{}}

\def\Nbpm{{\kern 0.18em\overline{\kern -0.18em N}}^{+-}}
\def\Nbmp{{\kern 0.18em\overline{\kern -0.18em N}}^{-+}}

\def\Mu{\mu}
\def\Chi2MinaMu{\chi^2_{\min ;\a,\Mu}}
\def\Chi2MinMu{\chi^2_{\min ;\Mu}(a)}

\def\fscfave{\kern 0.18em\overline{\kern -0.18em f}_{\rm SCF}}

\def\abar{\bar{a}}

\def\Bbar{\kern 0.18em\overline{\kern -0.18em B}{}\xspace}

\def\BRpmb{{\cal \kern 0.18em\overline{\kern -0.18em  B}}{}_{\rho\pi}^{+-}}
\def\BRmpb{{\cal \kern 0.18em\overline{\kern -0.18em  B}}{}_{\rho\pi}^{-+}}

\def\BRipmb{{\cal \kern 0.18em\overline{\kern -0.18em  B}}{}_{\rho^+\pi^-}}
\def\BRimpb{{\cal \kern 0.18em\overline{\kern -0.18em  B}}{}_{\rho^-\pi^+}}

\def\Abar{\kern 0.18em\overline{\kern -0.18em A}{}}
\def\abar{\kern 0.18em\overline{\kern -0.18em a}{}}

\def\ab   {\ensuremath{\mbox{\,ab}}\xspace}
\long\def\inst#1{\par\nobreak\kern 4pt\nobreak
    {\it #1}\par\vskip 10pt plus 3pt minus 3pt}

\begin{document}
{\pagestyle{empty}

\begin{flushright}
%\babar-CONF-\BABARPubYear/\BABARConfNumber \\
\babar-CONF-\BABARConfNumber \\
%\babar-PUB-\BABARPubYear/\BABARPubNumber \\
SLAC-PUB-\SLACPubNumber \\
%hep-ex/\LANLNumber \\
%BAD \#2231 Version 5
August 2009 \\
\end{flushright}

\begin{center}
\Large \bf Search for a Narrow Resonance in $e^+e^-$ to Four Lepton Final States
\end{center}

\begin{center}
\large The \babar\ Collaboration\\
\mbox{ }\\
\today
\end{center}
\bigskip \bigskip

% Abstract
\begin{center}
\large \bf Abstract
\end{center}
Motivated by recent models proposing a hidden sector with $\sim\gev$
scale force carriers, we present a search for a narrow dilepton
resonance in 4 lepton final states using $536\invfb$ collected by the
\babar\ detector.  We search for the reaction,
$\epem\to\Wp\Wp\to(\lplm)(\lplmP)$, where the leptons carry the full
4-momentum and the two dilepton pair invariant masses are equal.  We
do not observe a significant signal and we set 90\% upper limits of
$\sigma(\epem\to\Wp\Wp\to\foure)<(15-70)\ab$,
$\sigma(\epem\to\Wp\Wp\to\twoetwomu)<(15-40)\ab$, and
$\sigma(\epem\to\Wp\Wp\to\fourmu)<(11-17)\ab$ in the  $\Wp$ mass range between $0.24$ and $5.3\gevcc$.
Under the assumption that the $\Wp$ coupling to electrons and muons is the same, we obtain a combined upper limit of
$\sigma(\epem\to\Wp\Wp\to\lplm\lplmP)<(25-60)\ab$. Using these limits,
we constrain the product of the SM-dark sector mixing and the dark
coupling constant in the case of a non-Abelian Higgsed dark sector.
\vfill
\begin{center}

Submitted to the XXIV International Symposium on Lepton Photon \\ Interactions at High Energies, \\
17 August---22 August 2009, Hamburg, Germany.

\end{center}

\vspace{1.0cm}
\begin{center}
{\em SLAC National Accelerator Laboratory, Stanford University, 
Stanford, CA 94309} \\ \vspace{0.1cm}\hrule\vspace{0.1cm}
Work supported in part by Department of Energy contract DE-AC02-76SF00515.
\end{center}

\newpage
} % end of pagestyle{empty}

% Input author list file
%
%author list removed temporarily to save trees 7/9/04 RNC
%
\begin{center}
\small

The \babar\ Collaboration,
\bigskip

%% Special author list in plain latex for LP 2009
%
{B.~Aubert,}
{Y.~Karyotakis,}
{J.~P.~Lees,}
{V.~Poireau,}
{E.~Prencipe,}
{X.~Prudent,}
{V.~Tisserand}
\inst{Laboratoire d'Annecy-le-Vieux de Physique des Particules (LAPP), Universit\'e de Savoie, CNRS/IN2P3,  F-74941 Annecy-Le-Vieux, France}
{J.~Garra~Tico,}
{E.~Grauges}
\inst{Universitat de Barcelona, Facultat de Fisica, Departament ECM, E-08028 Barcelona, Spain }
{M.~Martinelli$^{ab}$,}
{A.~Palano$^{ab}$,}
{M.~Pappagallo$^{ab}$ }
\inst{INFN Sezione di Bari$^{a}$; Dipartimento di Fisica, Universit\`a di Bari$^{b}$, I-70126 Bari, Italy }
% ...

\end{center}\newpage

 \section{Introduction}
\label{sec:introduction}

Recent cosmic ray measurements of the electron and positron flux from
ATIC\cite{atic}, FERMI\cite{fermi}, and PAMELA\cite{pamela} have
spectra which are not well described by galactic cosmic ray models
such as GALPROP\cite{galprop}.  For instance,  PAMELA shows an
increase in the positron/electron fraction with increasing energy.  No
corresponding increase in the antiproton spectrum is observed.  There
have been two main approaches attempting to explain these features:  astrophysical sources (particularly from undetected, nearby  pulsars)\cite{pulsars} and annihilating or decaying dark matter.  

Arkani-Hamed \ea\cite{arkani-hamed} have introduced a class of
theories containing a new ``dark force'' and a light, hidden sector.
In this model, the ATIC and PAMELA signals are due to dark matter
particles with mass $\sim400-800\gevcc$ annihilating into the gauge
boson force carrier with mass $\sim 1\gevcc$, which they dub the $\phi$, which subsequently decays to Standard Model particles.  If the $\phi$ mass is below twice the proton mass, decays to $p\overline{p}$ are kinematically forbidden allowing only decays to states like $\epem$, $\mupmum$, and $\pi\pi$.  If the dark force is non-Abelian, this theory can also accommodate the 511 keV signal found by the INTEGRAL satellite \cite{integral} and the DAMA modulation data \cite{dama}.

The dark sector couples to the Standard Model through kinetic mixing with the photon.  
Thus low-energy/high luminosity $\epem$ experiments like \babar~are in excellent position to 
probe these theories. Recent papers by Batell \ea~\cite{batell} and Essig \ea~\cite{essig} have
discussed the prospects for finding evidence for the dark sector at the B-Factories 
in the Abelian and non-Abelian cases, respectively.  In the Abelian case, the signatures would be 
$\epem \to \gamma\phi\to\gamma l^+ l^-$  or $\epem \to \phi h^\prime \to  3(l^+ l^-)$ (where 
$h^\prime$ is a ``dark Higgs'').  There are actually two non-Abelian scenarios: the Higgsed case and the confined case (``dark QCD'').  In the Higgsed case there are at least three dark particles in play: $A^\prime$ which mixes with the photon, another gauge boson $W^\prime$, and the dark Higgs $h^\prime$.  In this regime, signatures 
are $\epem\to\Wp\Wp\to\lplm\lplm$ (via a virtual $A^\prime$) and $\epem\to\gamma A^\prime(\to\Wp\Wp)\to\gamma\lplm\lplm$, plus ``Higgs$^\prime$-strahlung'' processes which may lead to missing energy.  Finally, the confined case could lead to a proliferation of ``dark mesons'', whose lowest mass states decay to leptons.  
Depending on the scenario and the coupling between the Standard Model and dark sectors, cross sections could
be as large as a few femtobarns at \babar\, which would translate to hundreds of events observed in the detector.  

In this note we describe a  search  for the $\Wp$ in the reaction
$\epem\to\Wp\Wp\to\lplm\lplm$ in exclusive 4-lepton  final states,
where we require that the four leptons carry the full center of mass energy and that the two dilepton pairs have the same invariant mass.

 \section{The \babar\ Detector and Dataset}
\label{DataSets}

The data used in this analysis were collected with the \babar\ detector
at the \pep2\ asymmetric energy $\epem$\ storage rings between 1999 and 2008 
and correspond to an integrated luminosity of 536 \invfb.
This data was mostly at the 
$\Upsilon (4S)$ peak but it also includes collisions at the $\Upsilon (2S)$ and $\Upsilon (3S)$
as well as off-resonant data. 
%The  data set is summarized in Table~\ref{tab:moderato}.

%\begin{table}[htb]
%\begin{center}
%\begin{tabular}{lccccc}
%\hline\hline
%Sample &  $\mathcal{L}_{on} (\invfb)$ &   $\mathcal{L}_{off} (\invfb)$ & Resonance\\
%\hline
%Run 1  &  20.80          &      2.65       &    $\Upsilon (4S)$  \\
%Run 2  &  61.64          &      6.99       &   $\Upsilon (4S)$  \\
%Run 3  &  32.52          &      2.49       &   $\Upsilon (4S)$  \\
%Run 4  &  100.64         &      10.30      &   $\Upsilon (4S)$  \\
%Run 5  &  133.90         &      14.68      &   $\Upsilon (4S)$  \\
%Run 6  &  78.79          &      6.94       &   $\Upsilon (4S)$  \\
%Run 7  &  64.08          &       ---       &    $\Upsilon (2S)$, $\Upsilon (3S)$,$>\Upsilon (4S)$ \\
%\hline
%Total  & 492.37         &      44.05      &     \\
%\hline\hline
%\end{tabular}
%\end{center}
%\caption{Summary of the integrated on-resonance and off-resonance data. }\label{tab:moderato}
%\end{table}

To study signal efficiency and resolution,  $\epem\to\Wp\Wp\to\lplm\lplm$ Monte Carlo (MC) samples were generated  (where $l$=$e$ or $\mu$) for different values of $\Wp$ mass using the MadGraph event generator\cite{Alwall:2007st}. 
There were $10^4$ events generated at each mass value of:  0.3, 0.4, 0.5,
0.7, 1.0, 1.5, 2.0, 3.0, 4.0, and 5.0 $\gevcc$.   
To study backgrounds, we have inspected $B\overline{B}$ ($\sim$3x luminosity), $uds$, $c\overline{c}$, 
and $\tau\tau$ MC samples (each $\sim$1x luminosity).  In addition we created 4-lepton QED samples using the {\texttt diag36} event generator\cite{diag36}.  

A detailed description of the \babar\ detector is given in~\cite{babarNim}. 
Charged-particle trajectories are measured by a five-layer, double-sided silicon vertex tracker (SVT) and a 40-layer
drift chamber (DCH) coaxial with a 1.5~T magnetic field. Charged-particle identification is achieved by combining
the information from a ring-imaging Cherenkov device (DIRC)  with the ionization energy loss (\dedx ) measurements 
from the DCH and SVT. Photons are detected in a CsI(Tl) electromagnetic calorimeter (EMC) inside the coil. 
Muon candidates are identified in the instrumented flux return (IFR)
of the superconducting solenoid.  We use \textsc{GEANT}4-based~\cite{geant4} software to simulate the detector response and account for
 the varying beam and environmental conditions.

 \section{Event Selection}
\label{sec:eventSelection}

We search for the exclusive pair production of a narrow resonance, consistent with the detector resolution, decaying to leptons and with a mass in the range between  $240\mevcc$ to $\sqrt{s}/2$.  
The signature is 4 leptons with zero total charge carrying the full beam momentum where the two dilepton invariant masses are equal.  This topology, particularly the equal invariant masses, is quite unique and the only backgrounds are from 4-lepton QED processes.  The full selection criteria are described below.  
We used 10\% of the data as a test ({\it blind}) sample to choose our selection and signal extraction procedures before looking at the full dataset.

We begin by selecting events with:

\bei
	\item 4 charged tracks 
	\item two leptons  with $p_{\rm CM}>1.5\gevc$
	\item sum of the absolute value of  momentum of all tracks$>6\gevcc$ or the total visible energy (lab)$>8\gevcc$
\eei

We reconstruct 4-lepton candidates from combinations of two $\Wp\to\lplm$ candidates.  The lepton candidates are chosen by their signitures in the  EMC and IFR.   The $\Wp$ candidates are formed from $\epem$ or $\mupmum$ pairs.  We then select events which satisfy the following criteria:
\bei
\item $[N_e,N_\mu]=[4,0],[2,2],~or~[0,4]$
\item $M_{4lepton}>10\gevcc$
\item the helicity angle of a lepton pair, defined as the angle between the positive lepton and the lepton-pair flight direction, is required to be $|cos(\theta_H)|<0.95$ for each pair
\item  to reduce background from photon converstions, we require the flight significance, defined as the $\Wp$ candidates decay length from the interaction point divided by the error, is $<4\sigma$ for each pair
\item to reduce background from radiative Bhabha events, we require the angle between the decay planes of the lepton pairs, $\phi_{DPN}>0.2$ 
\eei
The 4-lepton candidate is then fit constraining the four-momentum to the total beam momentum   and the vertex to the interaction point. 

\begin{figure}[tb]
  \centerline{\epsfxsize6cm\epsffile{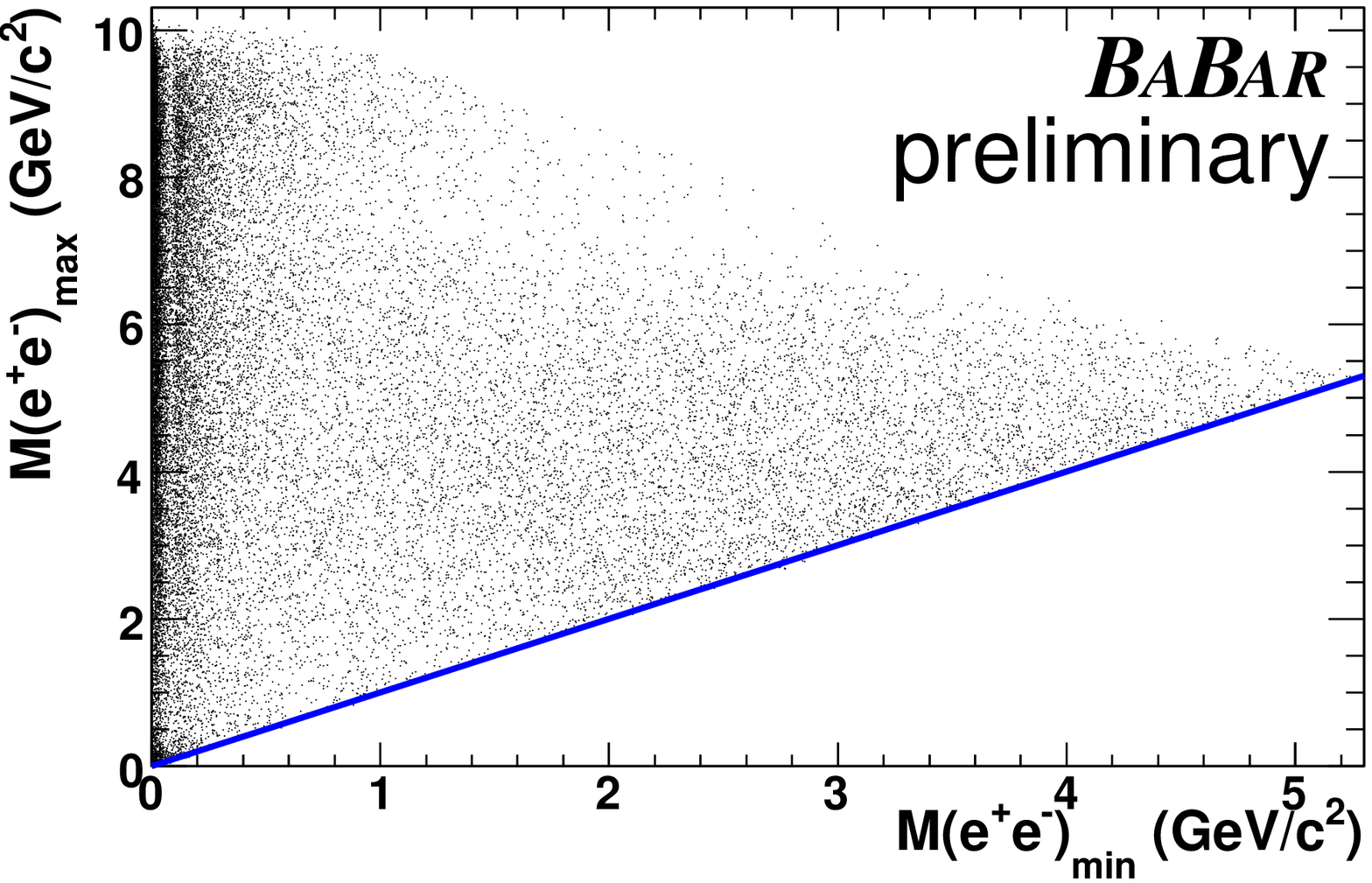}
              \epsfxsize6cm\epsffile{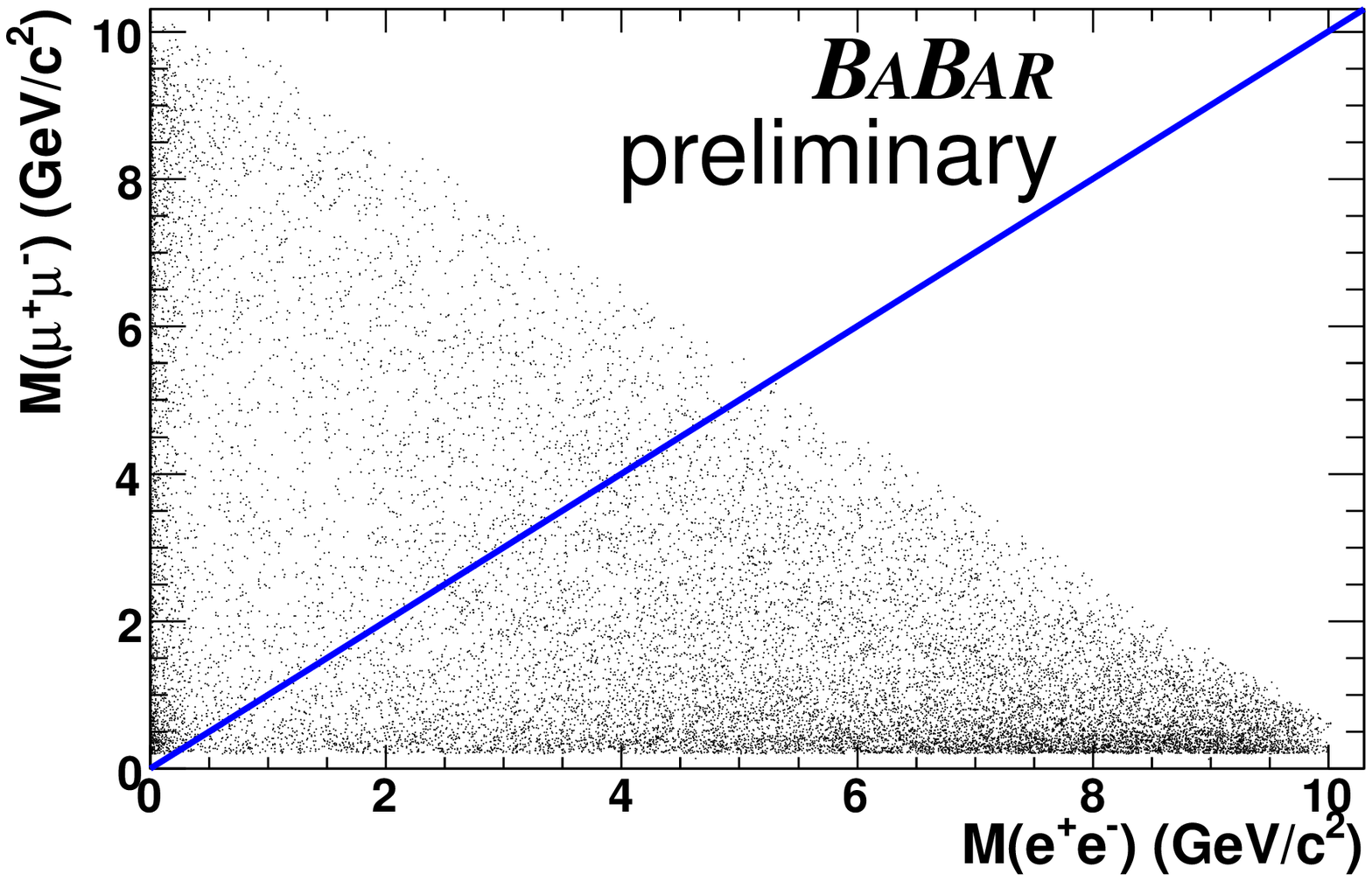}
              \epsfxsize6cm\epsffile{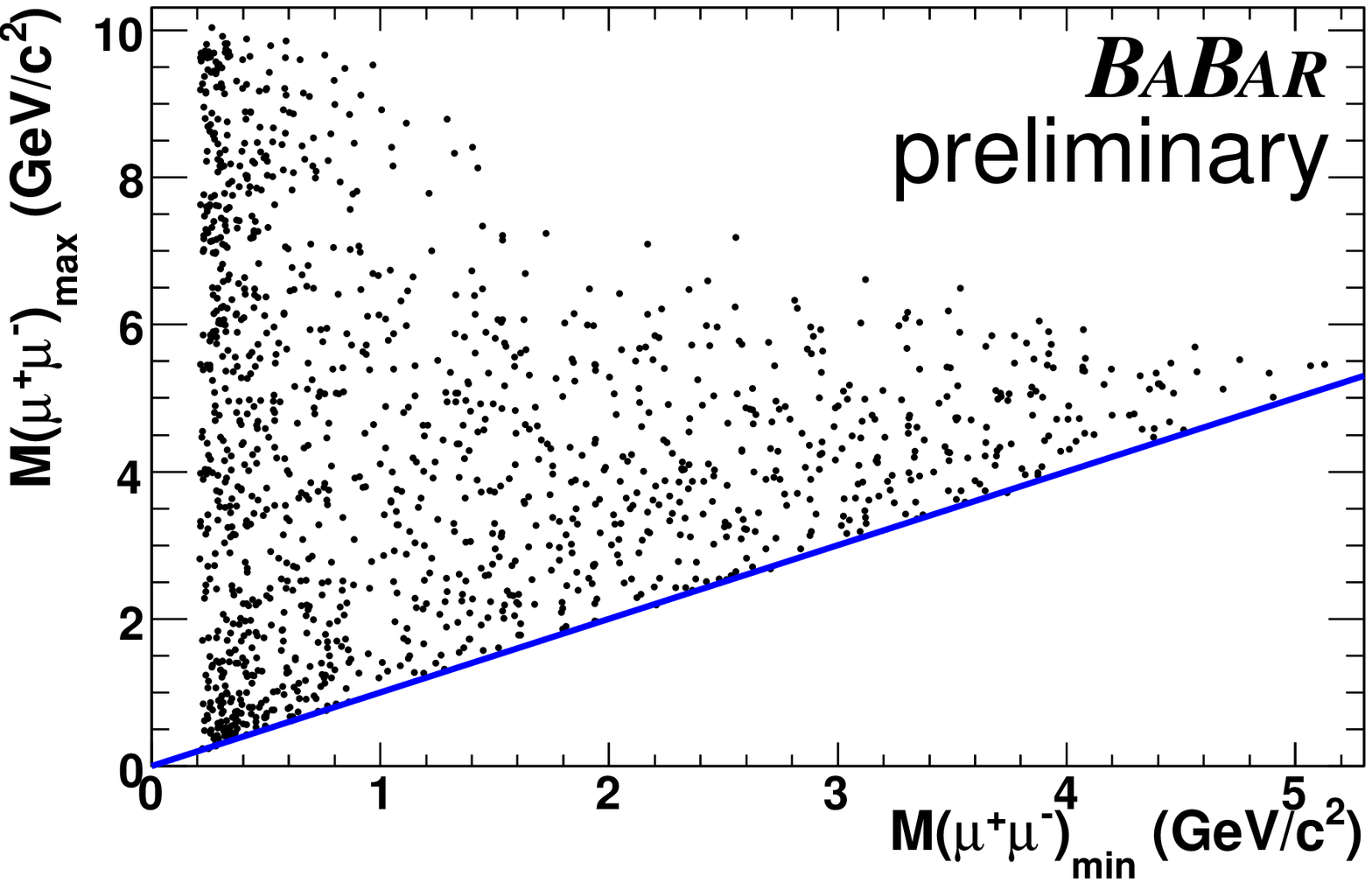}}
  \vspace{-0.1cm}
\caption{\label{fig:m1vsm2}
The dilepton invariant mass distributions from data for (left to right) $\epem\epem$, $\epem\mupmum$, and $\mupmum\mupmum$ after all other cuts.  The solid lines denotes $m_1=m_2$.
}
\end{figure}

\begin{figure}[tb]
  \centerline{\epsfxsize8cm\epsffile{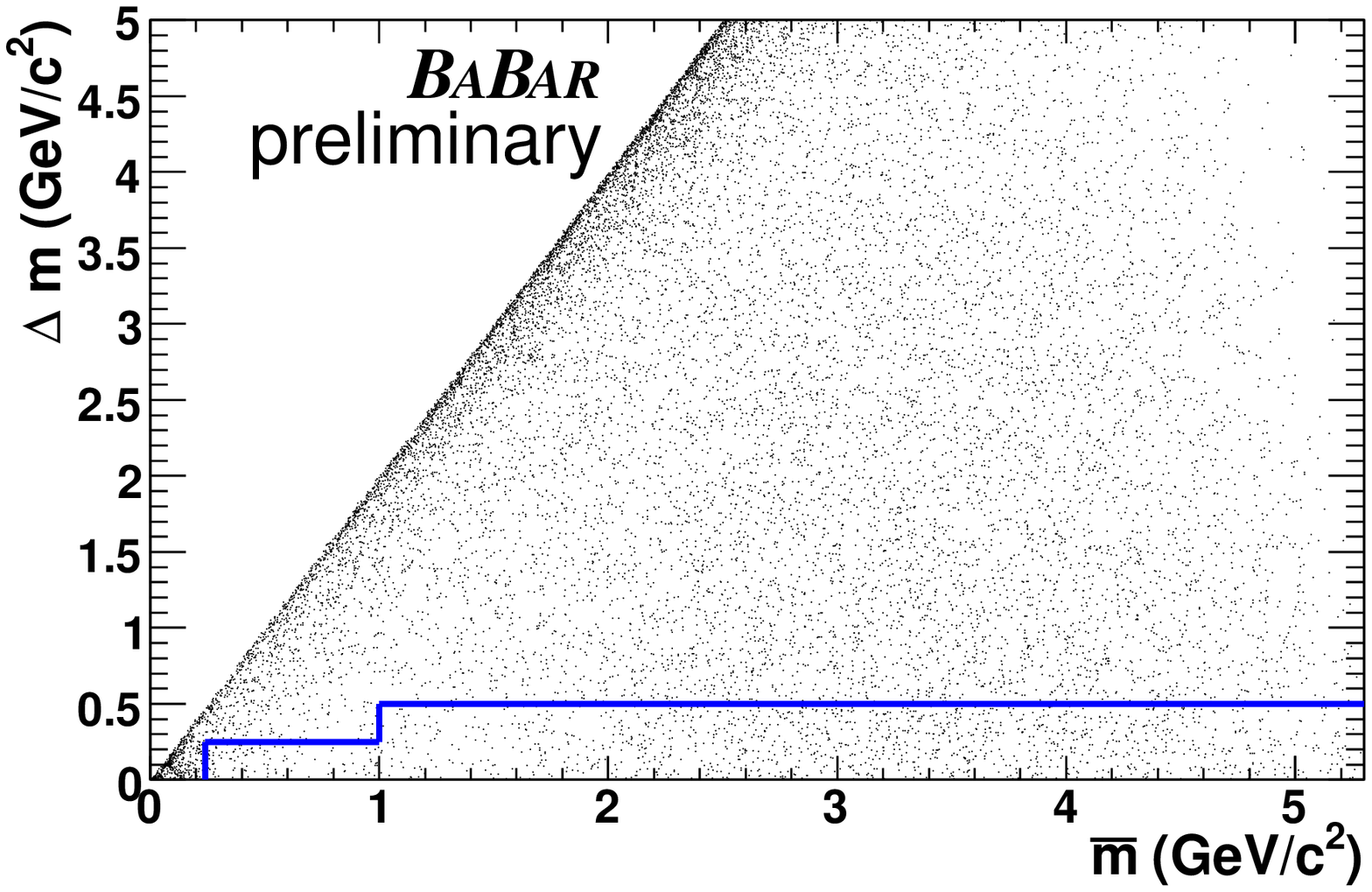}
              \epsfxsize8cm\epsffile{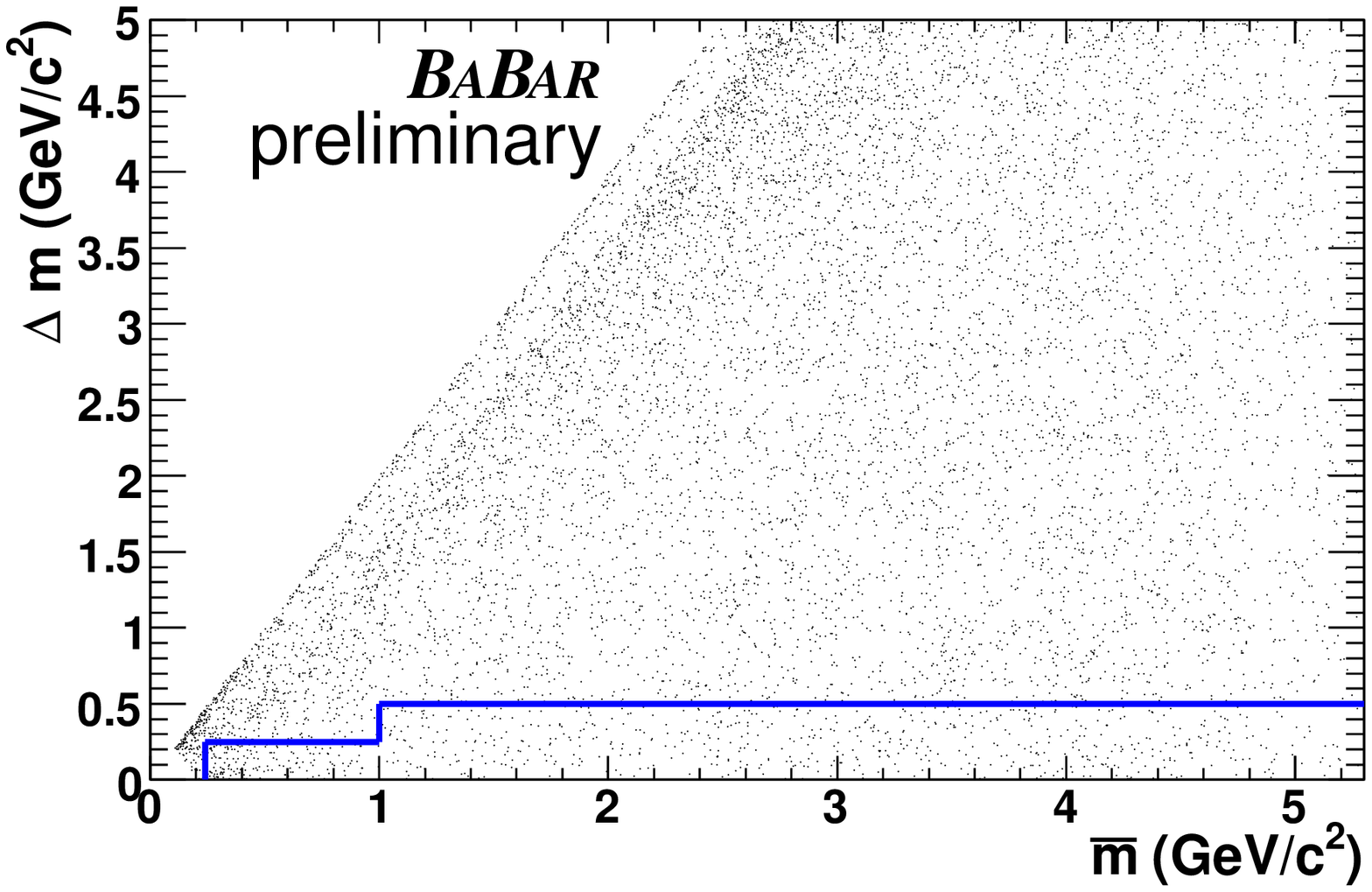}}
    \centerline{          \epsfxsize8cm\epsffile{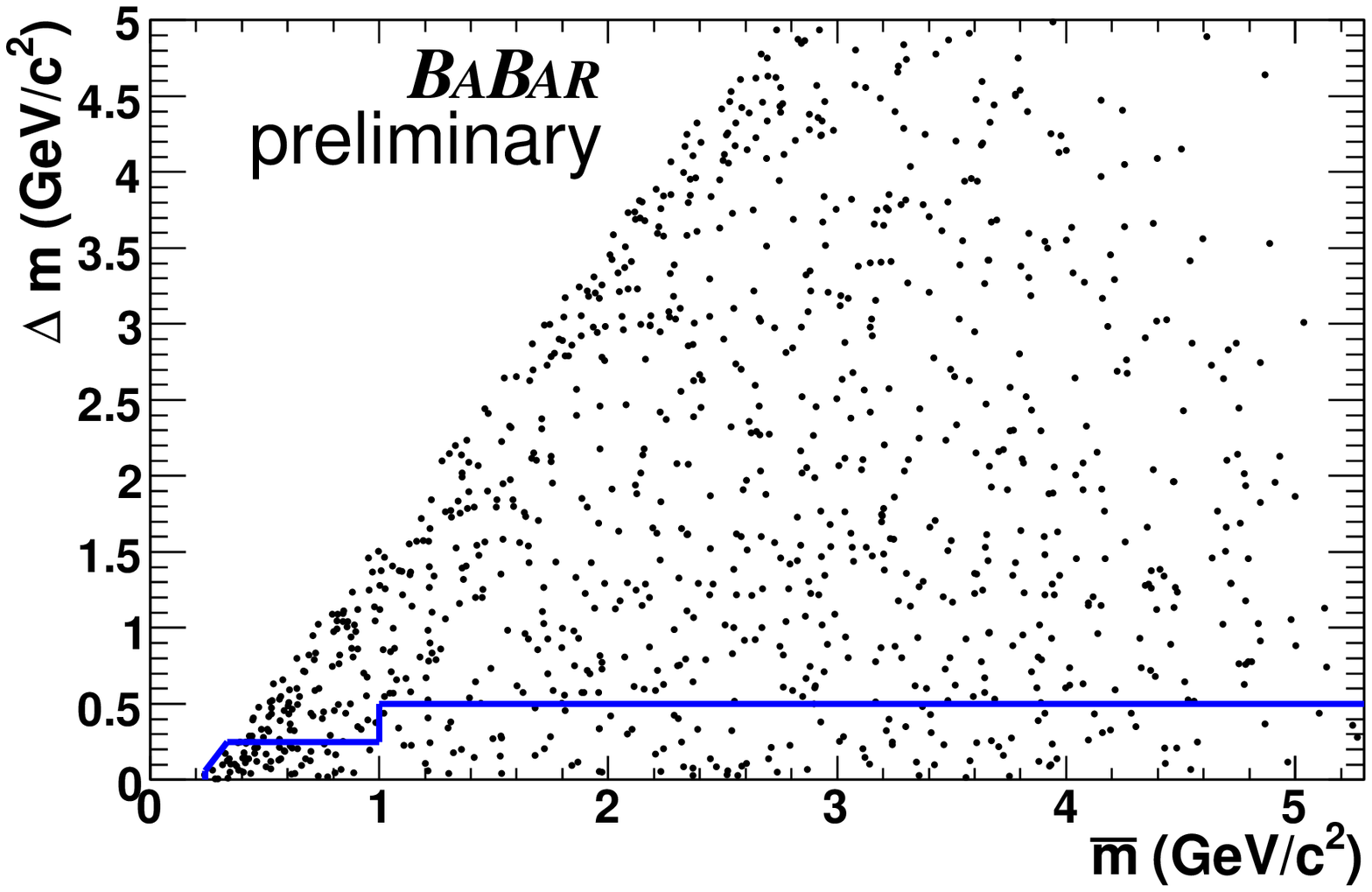}}
  \vspace{-0.1cm}
\caption{\label{fig:delMvsmBar}
The transformed mass distributions,$\dm$ vs $\mbar$, from data for (left to right) $\epem\epem$, $\epem\mupmum$, and $\mupmum\mupmum$ after all other cuts.  The solid lines denotes the $\dm$ cut value.
}
\end{figure}

At this point, we can exploit the fact that both dilepton pairs for our signal events  have the same invariant mass.  The 2-dimensional distributions of dilepton masses for 
each final state after all of the above cuts for the blinding sample is shown in Figure \ref{fig:m1vsm2}.  We define the transformed masses:
\beqn
\mbar=\left(m_1+m_2\right)/2\\
\dm=\left|m_1-m_2\right|
\eeqn
where $m_1$ and $m_2$ are the dilepton invariant masses.  The distribution of events  for these variables is shown in Figure \ref{fig:delMvsmBar}.  We impose a cut on $\dm$ (shown as the solid line in Figure \ref{fig:delMvsmBar}) of $\dm<0.25\gevcc$ for $\mbar<1.0\gevcc$ and  $\dm<0.50\gevcc$ for  $\mbar>1.0\gevcc$.  Because of the threshold effects in   $\mupmum\mupmum$, we tighten the $\dm$ cut in a linear fashion below $\mbar<4\times M(\mu)$.  

In the case of the $\epem\epem$ and $\mupmum\mupmum$ final state, there are two possible $\lplm$ pair combinations.   If both pairings pass all cuts,  the pair with the smallest value of $\dm$ is used.  For data, we see two pairings passing all cuts except the $\dm$ cut for 25\% of  $\epem\epem$ events and for 44\% of the $\mupmum\mupmum$ events. 
Table \ref{tab:eff} shows the progressive and total efficiencies for the three different final states of $\Wp\Wp\to\lplm\lplm$ (assuming the mass of the $\Wp$ is $1\gevcc$) as well as the progressive efficiency for the data. As shown in the table, the loose cut on $\dm$ is extremely powerful at reducing the background while not affecting the signal efficiency.  
After all selection, there are 28303 events remaining in our data sample; of these  16531 are $\epem\epem$ events, 9592 are $\epem\mupmum$ events, and 2180 are $\mupmum\mupmum$ events.

\begin{table}[thb]
  \begin{center}
    \caption[Relative Cut efficiencies]{\label{tab:eff}
      Selection efficiencies relative to the previous cut with binomial errors 
      for  the three signal decay modes assuming $M(\Wp)=1\gevcc$ and for onpeak  data.}
    \begin{tabular}{lcccc}
\hline
\hline
&&&&\\[-0.2cm]
Cuts & \multicolumn{4}{c}{Relative Efficiencies (\%)}   \\
& $\varepsilon_{\Wp\Wp\to 4e}$
& $\varepsilon_{\Wp\Wp\to 2e2\mu}$
& $\varepsilon_{\Wp\Wp\to 4\mu}$
& $\varepsilon_{data}$\\[0.2cm]
\hline
&&&&\\[-0.2cm]
\rule[-1.7mm]{0mm}{5mm}  Preselection                       & $61.4\pm0.5$  &  $68.2\pm0.5$ &  $73.4\pm0.4$  &  $-----$  \\
\rule[-1.7mm]{0mm}{5mm}  N(tracks)=4                        & $93.3\pm0.3$  &  $95.5\pm0.3$ &  $97.1\pm0.2$  &  $83.7\pm0.0$  \\
\rule[-1.7mm]{0mm}{5mm}  N(leptons)                         & $99.9\pm0.1$  &  $100.0\pm0.0$ &  $100.0\pm0.0$  &  $98.7\pm0.0$  \\
\rule[-1.7mm]{0mm}{5mm}  $M(4l)>10\gev$                     & $87.2\pm0.5$  &  $93.2\pm0.3$ &  $98.1\pm0.2$  &  $66.2\pm0.0$  \\
\rule[-1.7mm]{0mm}{5mm}  $|cos(\theta_H)|<0.95$             & $99.9\pm0.1$  &  $99.6\pm0.1$ &  $98.8\pm0.1$  &  $18.2\pm0.0$  \\
\rule[-1.7mm]{0mm}{5mm}  flt. sig.$<4\sigma$                & $97.2\pm0.2$  &  $96.5\pm0.4$ &  $99.1\pm0.1$  &  $54.6\pm0.1$  \\
\rule[-1.7mm]{0mm}{5mm}  $\mu$ PID		           & $100.0\pm0.0$ &  $81.9\pm0.6$ &  $70.3\pm0.6$  &  $74.2\pm0.1$  \\
\rule[-1.7mm]{0mm}{5mm}  $\phi_{DPN}>0.2$		           & $92.4\pm0.4$  &  $93.2\pm0.4$ &  $93.6\pm0.4$  &  $37.0\pm0.1$  \\
\rule[-1.7mm]{0mm}{5mm}  $\Delta(m)$ 		           & $98.3\pm0.2$  &  $99.8\pm0.1$ &  $99.6\pm0.1$  &  $~3.8\pm0.1$  \\
\hline
\rule[-1.7mm]{0mm}{5mm}   Total Efficiency                  & $43.7\pm0.5$  &  $44.5\pm0.5$ &  $44.8\pm0.5$  &  $-----$  \\
\hline
\hline
\end{tabular}

    \vspace{-0.8cm}
  \end{center}
\end{table}

 \section{Signal Extraction}
\label{sec:signal}

Our aim is to perform a search for a narrow peak in the $\mbar$ range 
from $240\mevcc$ up to $\sqrt{s}/2$.  After the selection described in the previous section, the expected backgrounds are quite low and we have decided to perform a cut-and-count analysis in bins of $\mbar$, using the the $\dm$ variable to define the signal and background regions.  The number of observed signal events in a $\mbar$ bin is then:
\beqn
N_{sig}=N_{signal~region}-N_{bkg~region}\times \frac{A_{signal}}{A_{background}}
\eeqn
where $A_{signal}$ ($A_{background}$) is the area of the signal (background) $\dm$ region. 

In this section, we will discuss the signal efficiency, $\dm$ shapes (including the definition of signal and background regions) and background rates as a function of  $\mbar$ and the method we plan to use in extracting the signal yields and setting limits. 

\subsection{Efficiency and $\dm$ resolution dependence on the $\Wp$ mass}
\label{sec:dmdependence}

\begin{figure}[tb]
  \centerline{\epsfxsize6cm\epsffile{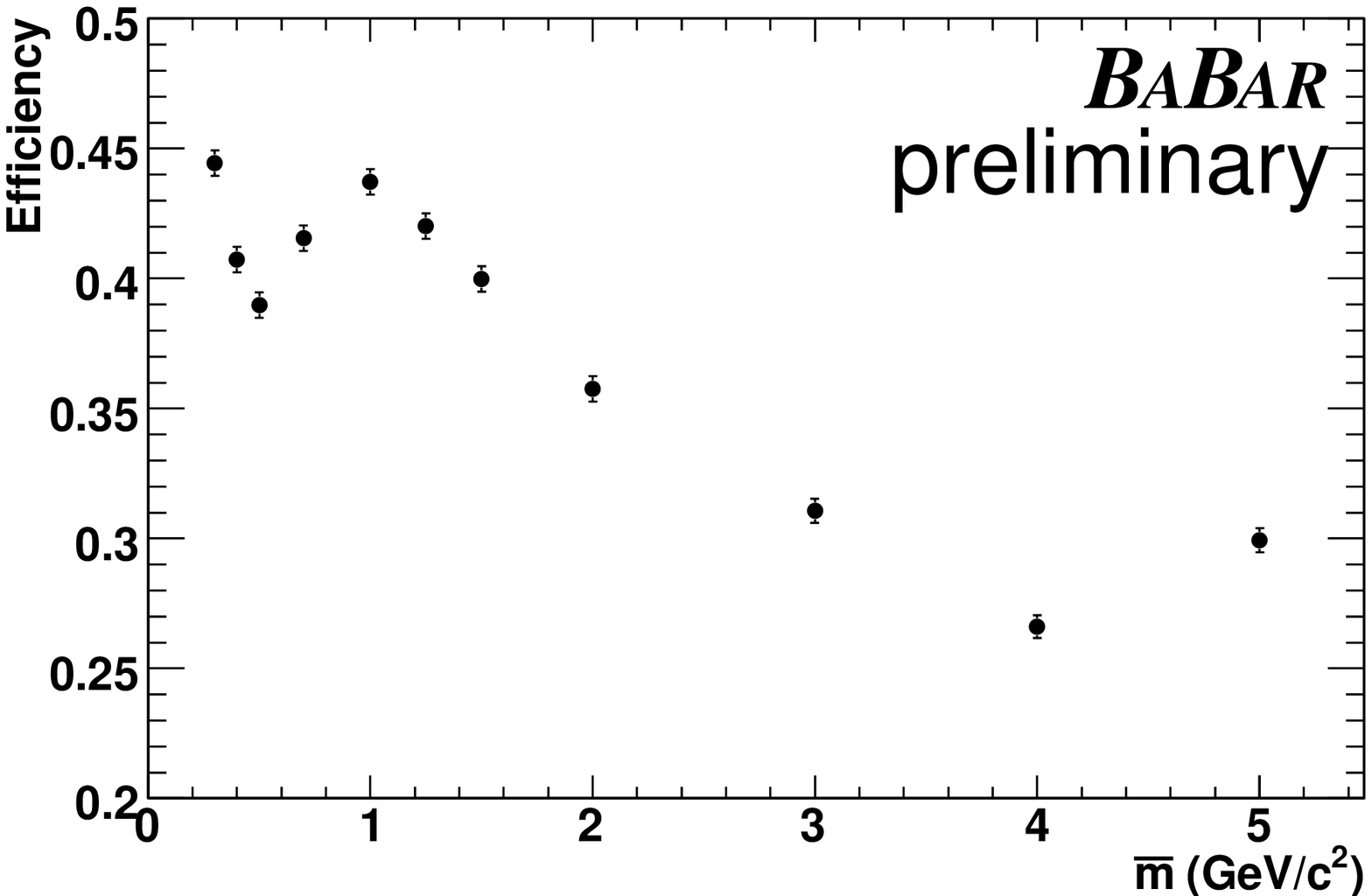}
              \epsfxsize6cm\epsffile{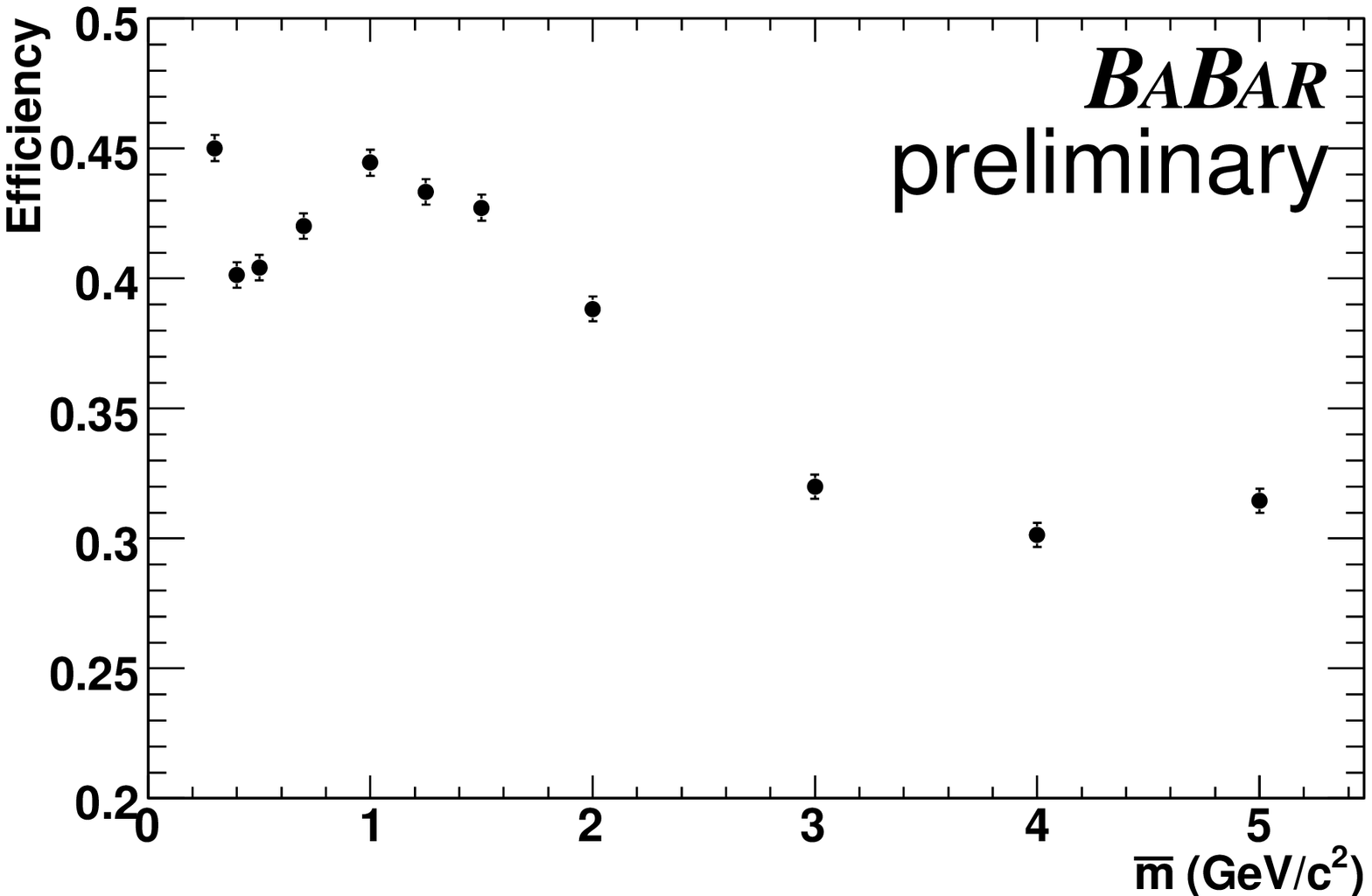}
              \epsfxsize6cm\epsffile{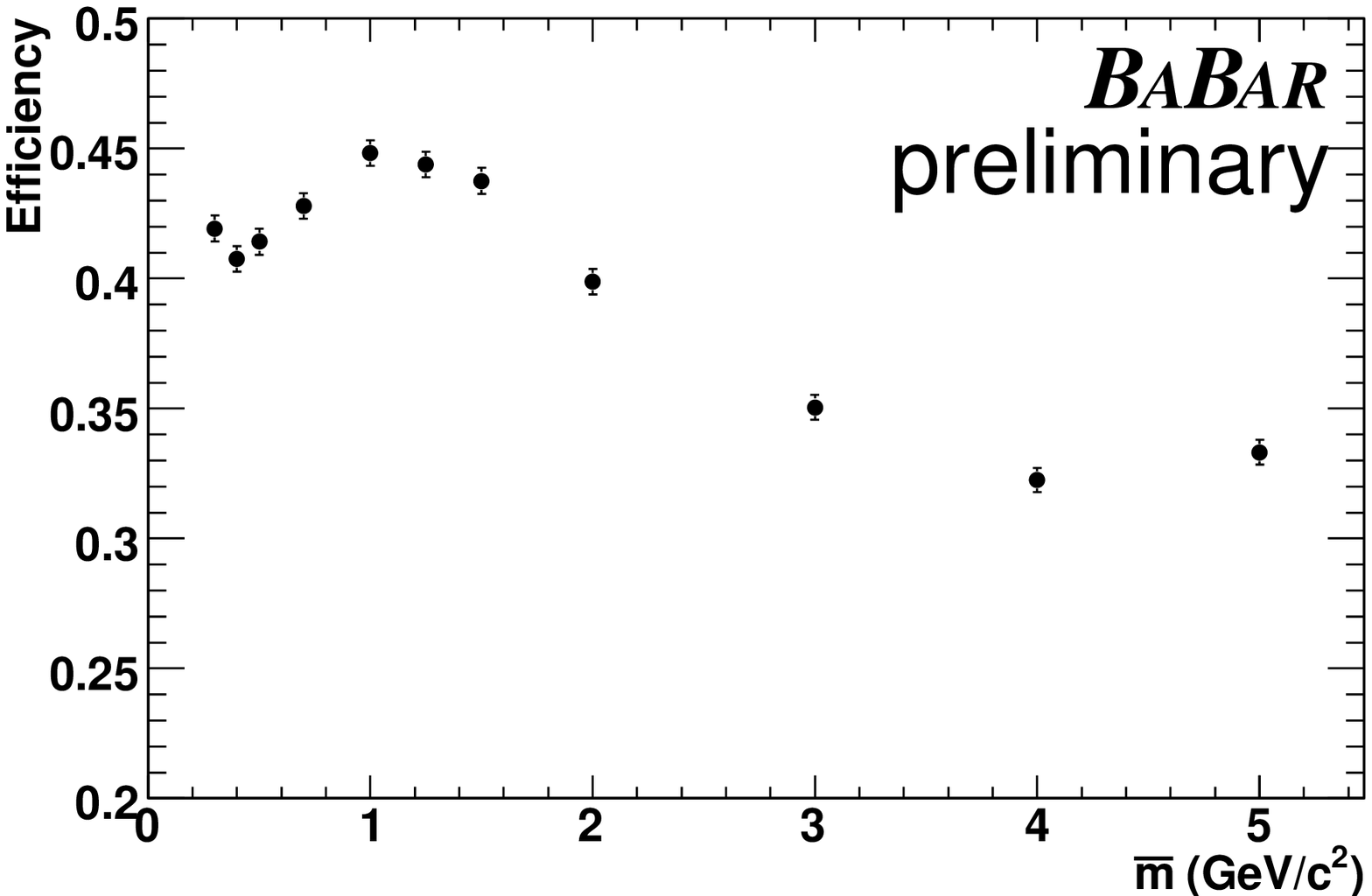}}
  \vspace{-0.1cm}
\caption{\label{fig:effvsmass}
The signal efficiency versus $\Wp$ mass for (left to right) $\Wp\Wp\to\epem\epem$, $\Wp\Wp\to\epem\mupmum$, and $\Wp\Wp\to\mupmum\mupmum$ after all cuts. 
}
\end{figure}

\begin{figure}[tb]
  \centerline{\epsfxsize6cm\epsffile{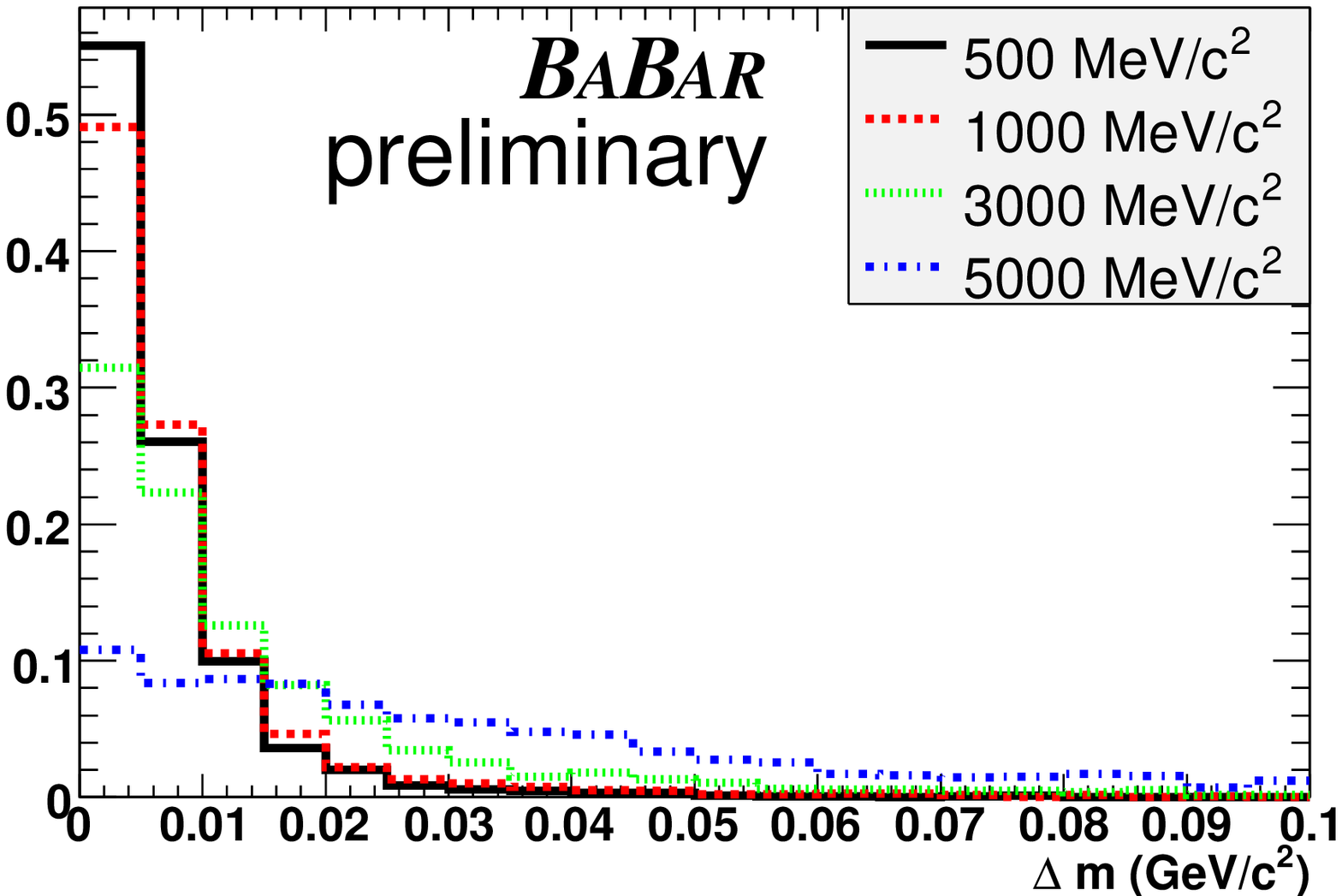}
              \epsfxsize6cm\epsffile{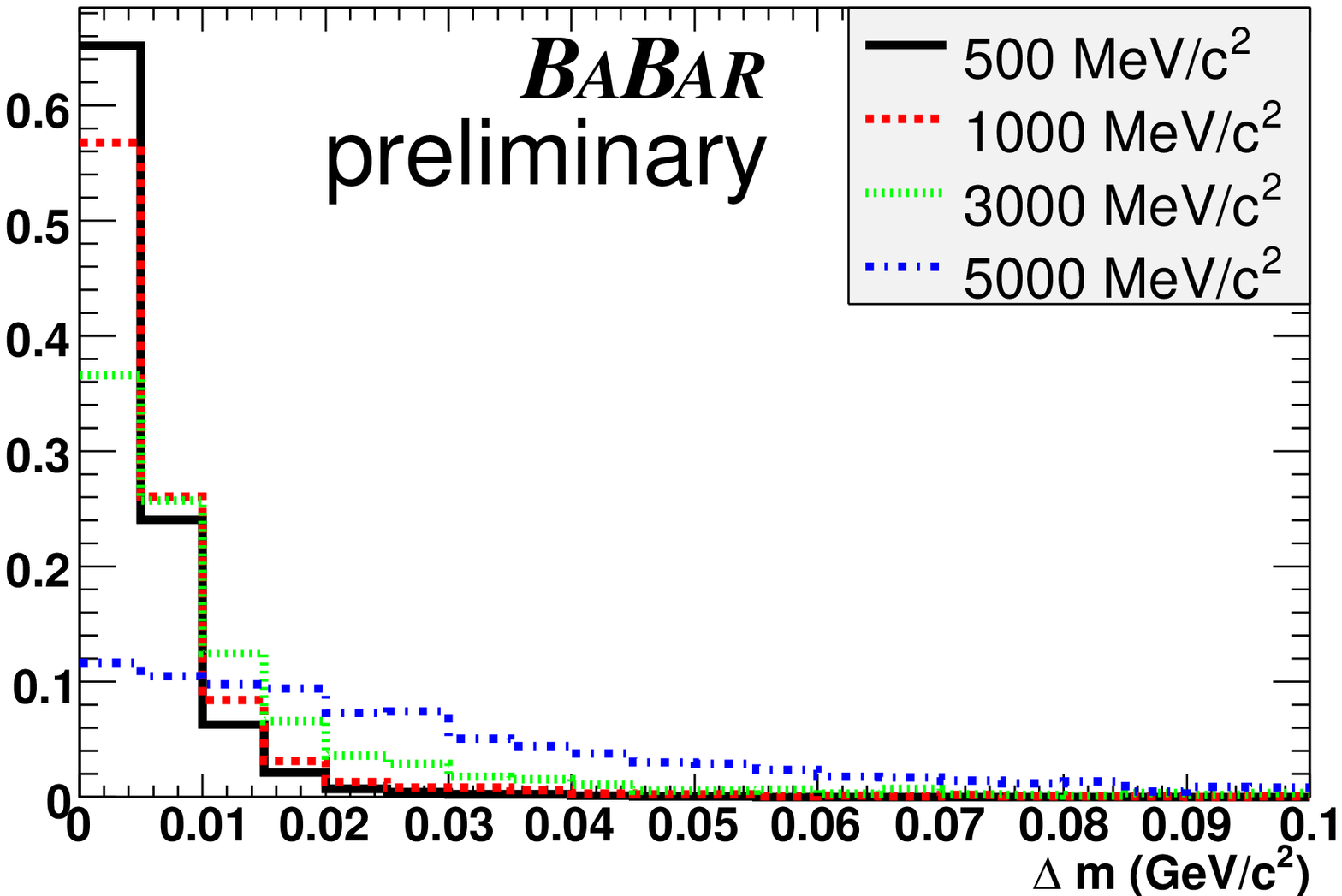}
              \epsfxsize6cm\epsffile{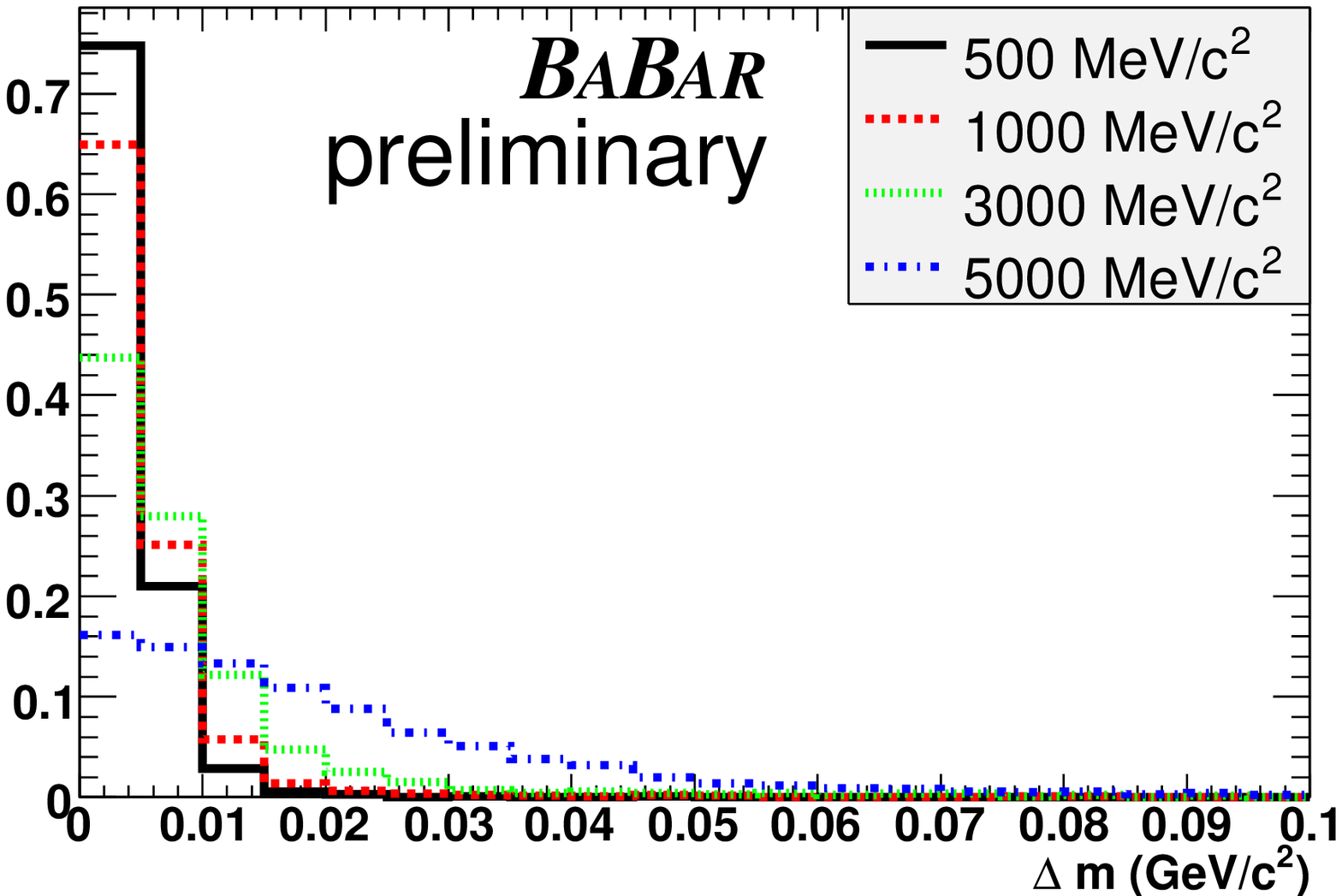}}
  \vspace{-0.1cm}
\caption{\label{fig:dmfourmass}
The $\dm$ distributions for four different $\Wp$ mass values (left to right) $\Wp\Wp\to\epem\epem$, $\Wp\Wp\to\epem\mupmum$, and $\Wp\Wp\to\mupmum\mupmum$ after all cuts. 
}
\end{figure}

The efficiency for different generated values of the $\Wp$ mass is shown in Figure \ref{fig:effvsmass}.  The efficiency decreases from $\sim45\%$ at $1\gev$ to $25-30\%$ at high masses depending on the decay mode. There is a dip in efficiency for  mass pairs around 500$\mevcc$ which is due to the opening angle of the lepton pair at this mass coinciding with the bending angle at the EMC, precluding us from identifying the two particles for a fraction of the events.  The $\dm$ resolution also varies significantly as a function of $\Wp$ mass.  Figure \ref{fig:dmfourmass} shows the distributions of $\dm$ for four different mass values. The resolution of $\dm$ increases with increasing $\Wp$ mass.  Since the background $\dm$ distribution is basically flat and roughly constant in $\mbar$ (see Section \ref{sec:background}), the effect is to reduce the sensitivity at higher masses.

\begin{figure}[tb]
  \centerline{\epsfxsize6cm\epsffile{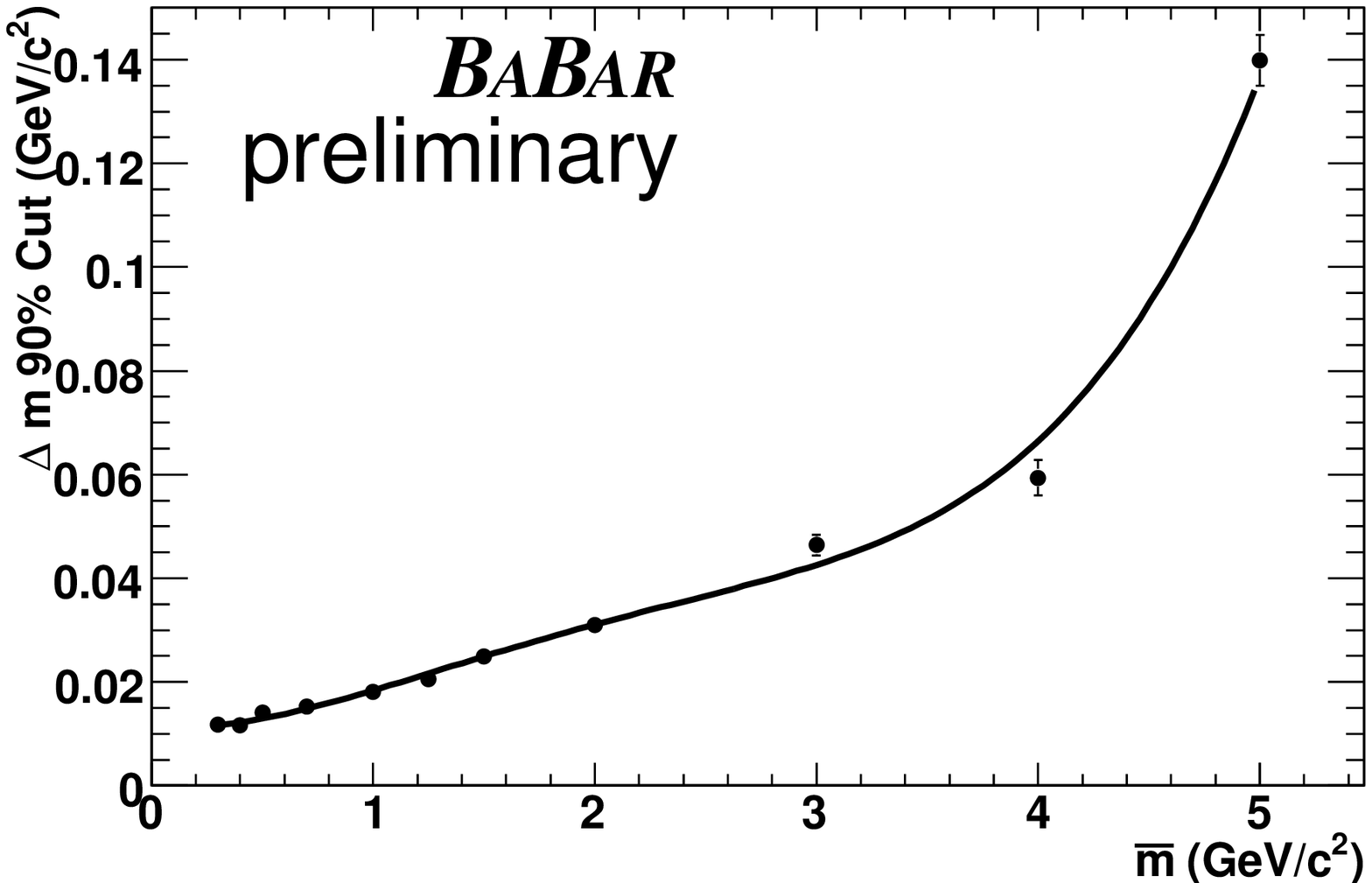}
              \epsfxsize6cm\epsffile{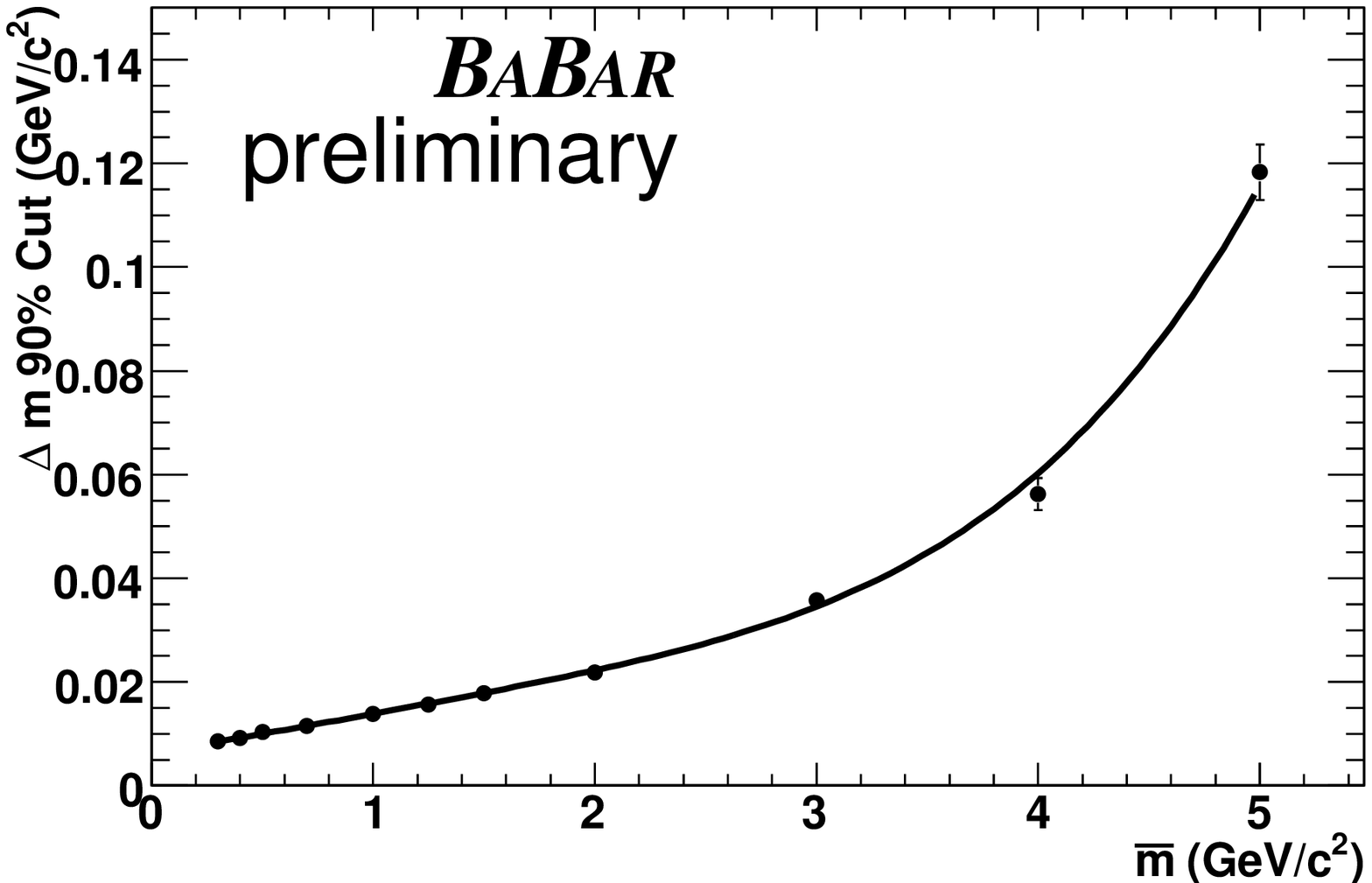}
              \epsfxsize6cm\epsffile{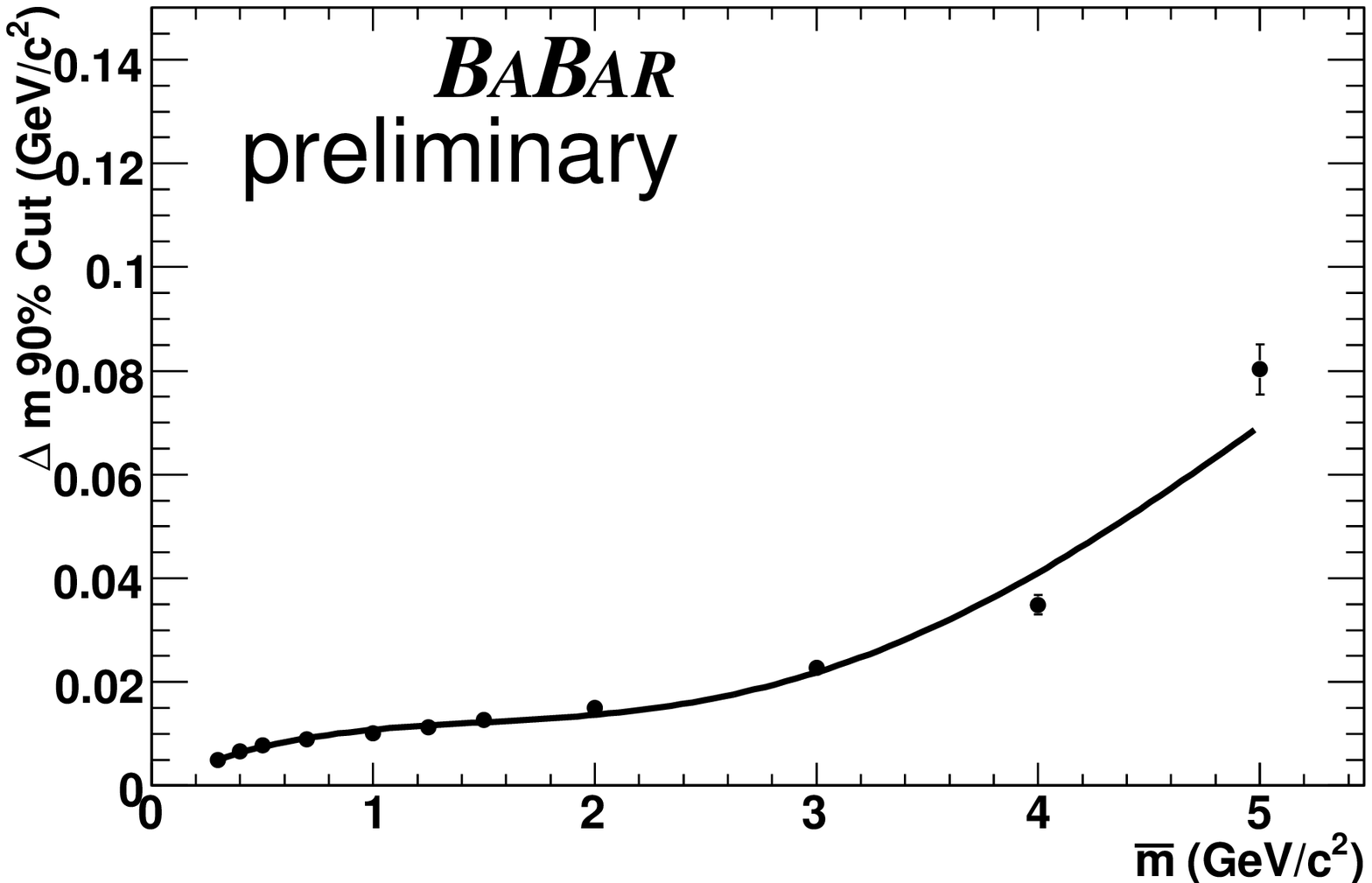}}
  \vspace{-0.1cm}
\caption{\label{fig:90perCut}
The values of the cut on $\dm$ keeping 90\% of signal events as a function of $\Wp$ mass for (left to right) $\Wp\Wp\to\epem\epem$, $\Wp\Wp\to\epem\mupmum$, and $\Wp\Wp\to\mupmum\mupmum$.  The line is a fit to a fourth order polynomial.  This cut defines our signal and background region. 
}
\end{figure}

Figure \ref{fig:90perCut} shows the values of the $\dm$ cut which retains 90\% of the signal 
as a function of $\mbar$.  We use this cut value to define the signal ($\dm<cutVal$) and background 
($dm>cutVal$) regions for the cut-and-count signal extraction.  Recall that the maximum 
value of $\dm$ is $0.25$$(0.5)\gevcc$ for $\mbar<(>)1.0\gevcc$.
The solid line is the result of a 4th-order polynomial fit which we use to extrapolate between
$\mbar$ points.

\subsection{Background composition}
\label{sec:background} 

While we ultimately use the $\dm$ sidebands to determine our background level, we have also used MC to study the composition of the background. In generic $q\overline{q}$, $B^0\overline{B}^0$, $B^+B^-$, and $\tau^+\tau^-$ samples we find only a single event passing the cuts (a $q\overline{q}$ event in the 4-electron final state).  From this we conclude that our background is dominated by QED processes.  

We have generated $\epem\mupmum$ and $\mupmum\mupmum$ samples\footnote{Due to the enormous $\epem\epem$ QED cross-section, this mode is difficult to generate efficiently.} using the {\tt diag36} generator and compared the MC to our selected dataset.  
We find  good agreement both in the scale and shape between data and the four-lepton QED MC.  
From the MC, we expect to observe $16241\pm 250$ $\mupmum\mupmum$ events in the full dataset  while we observe $15666\pm 125$ (statistical errors only).  For $\epem\mupmum$  we expect $219927\pm 3450$ and observe $185499\pm 431$ events.

\begin{figure}[tb]
\centerline{\epsfxsize6cm\epsffile{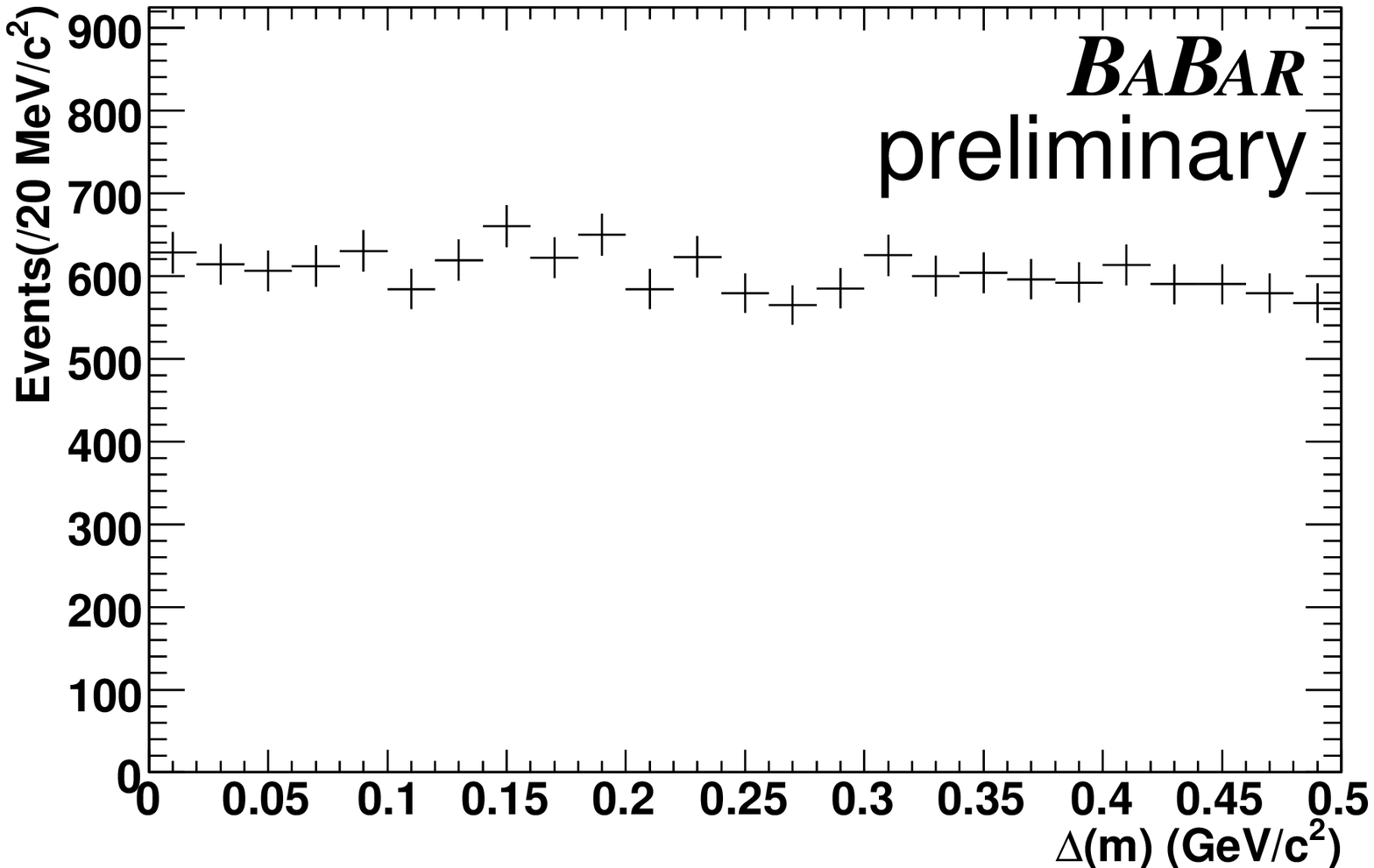}
              \epsfxsize6cm\epsffile{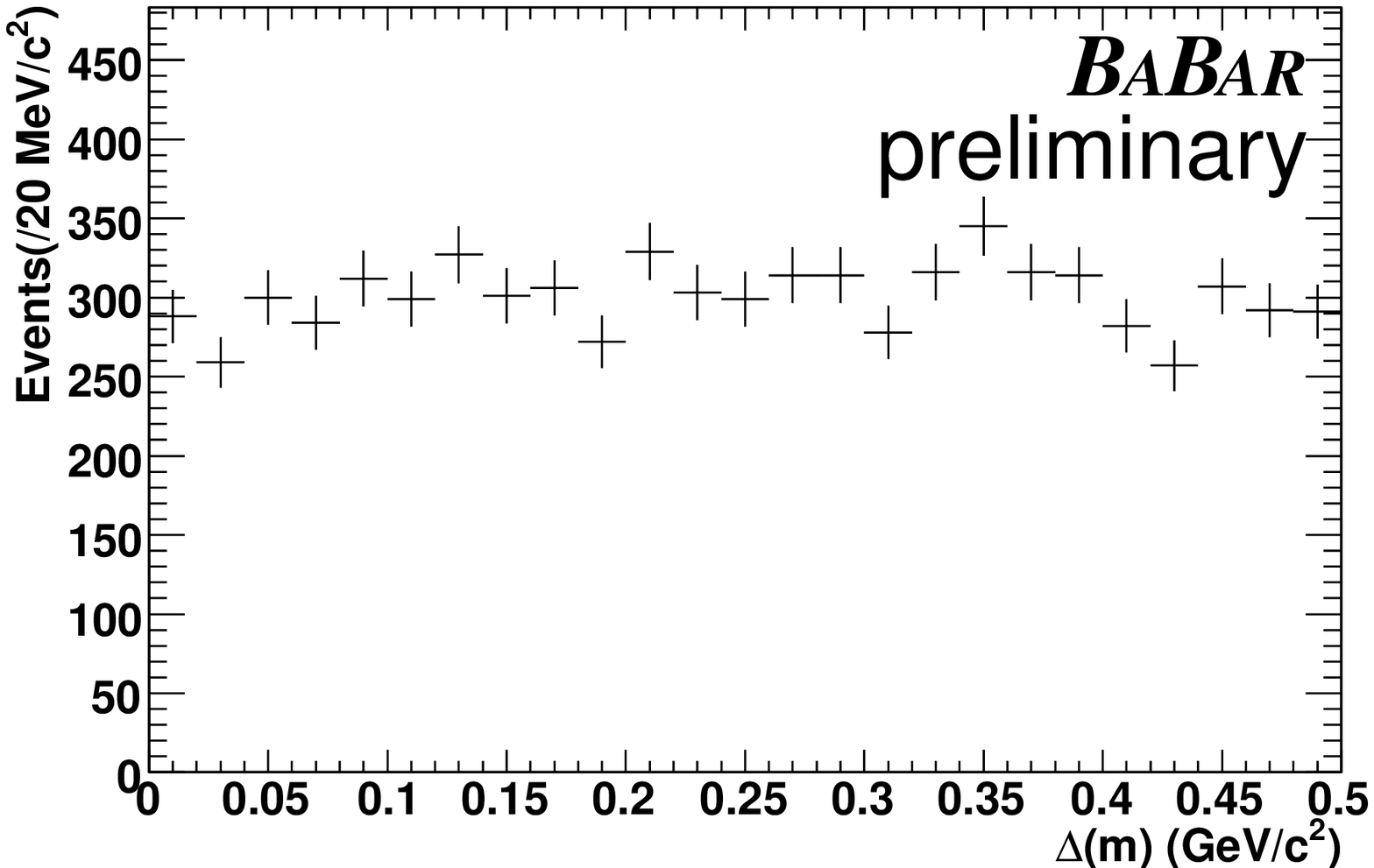}
              \epsfxsize6cm\epsffile{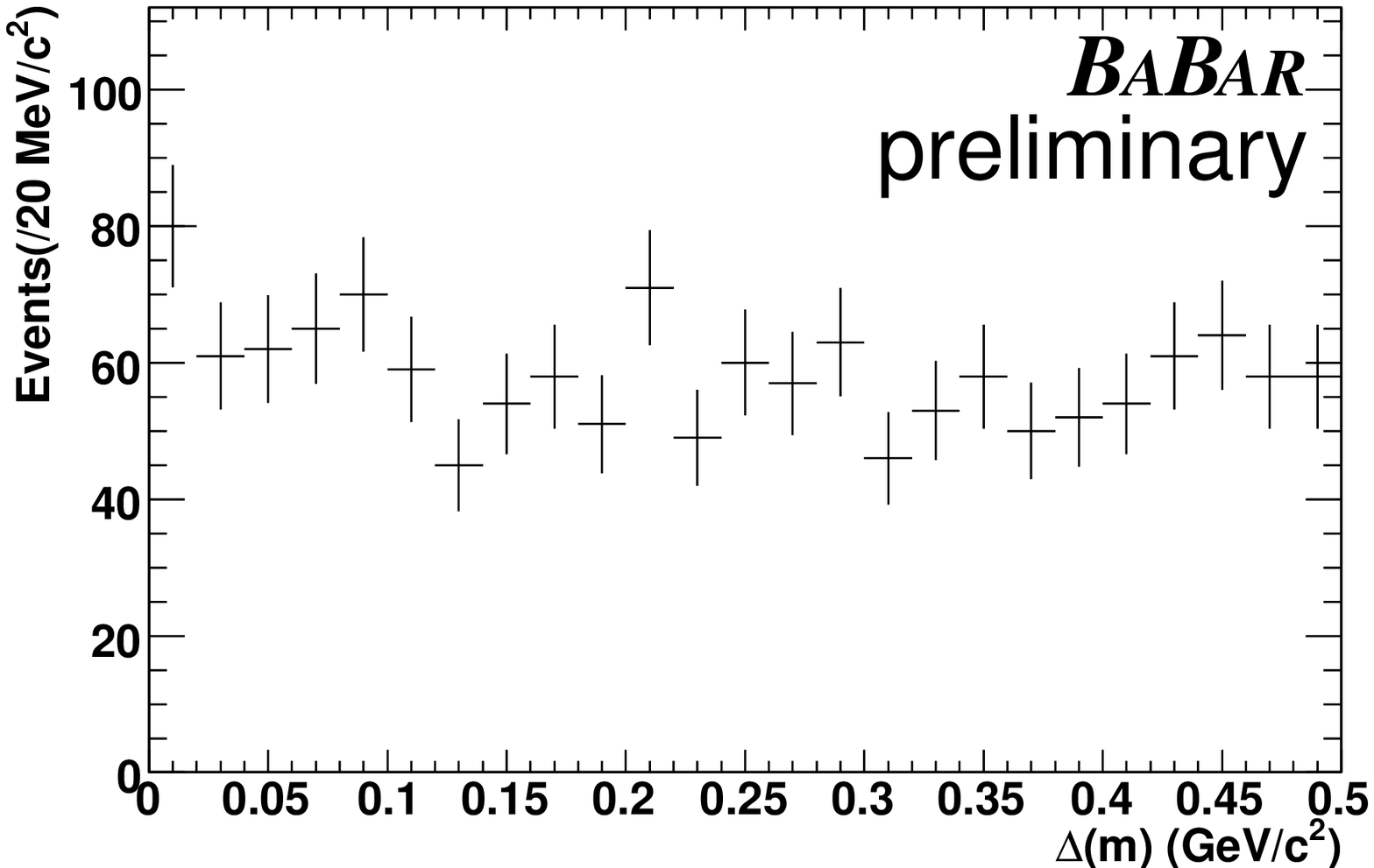}}
 \centerline{\epsfxsize6cm\epsffile{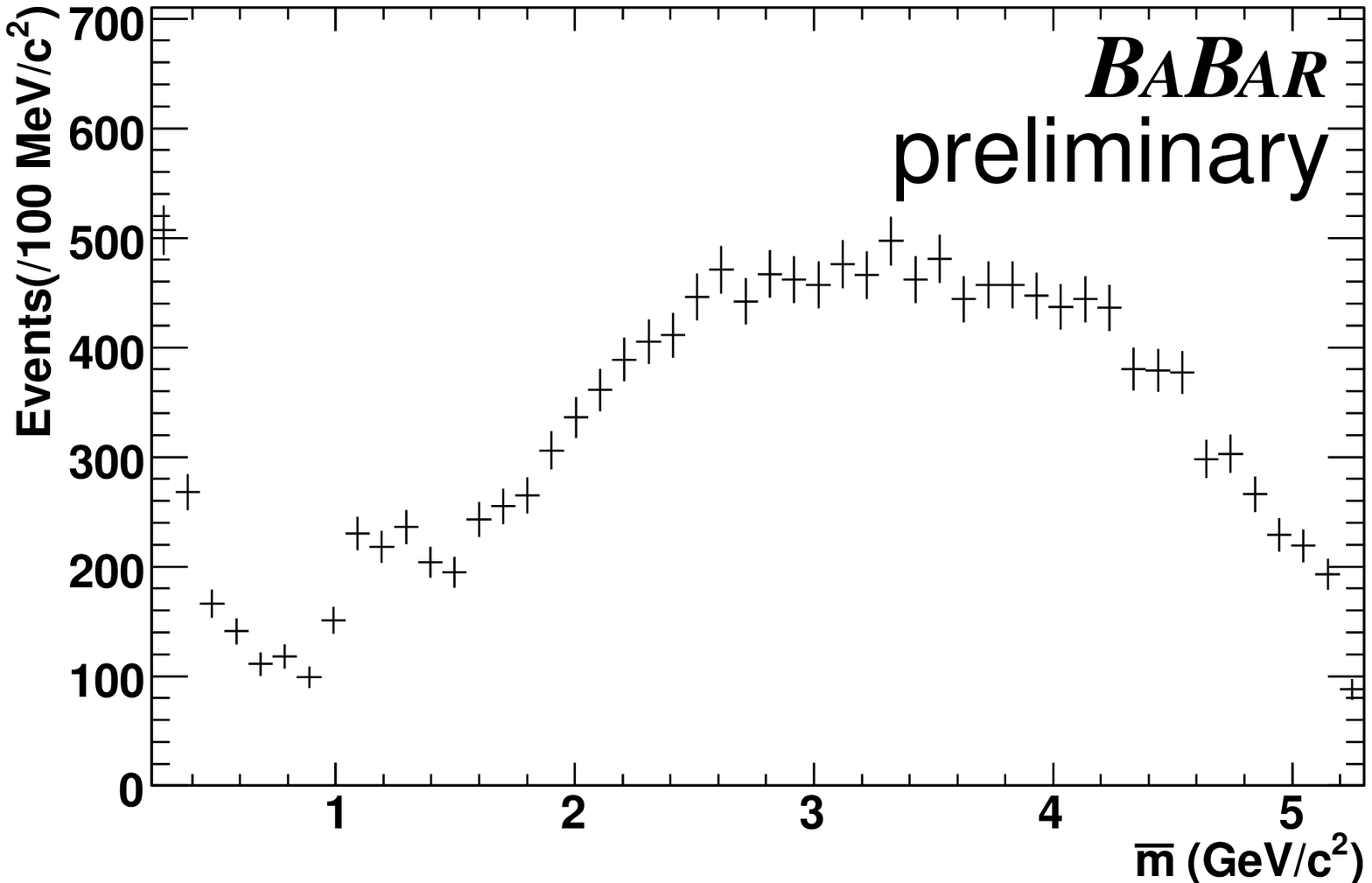}
              \epsfxsize6cm\epsffile{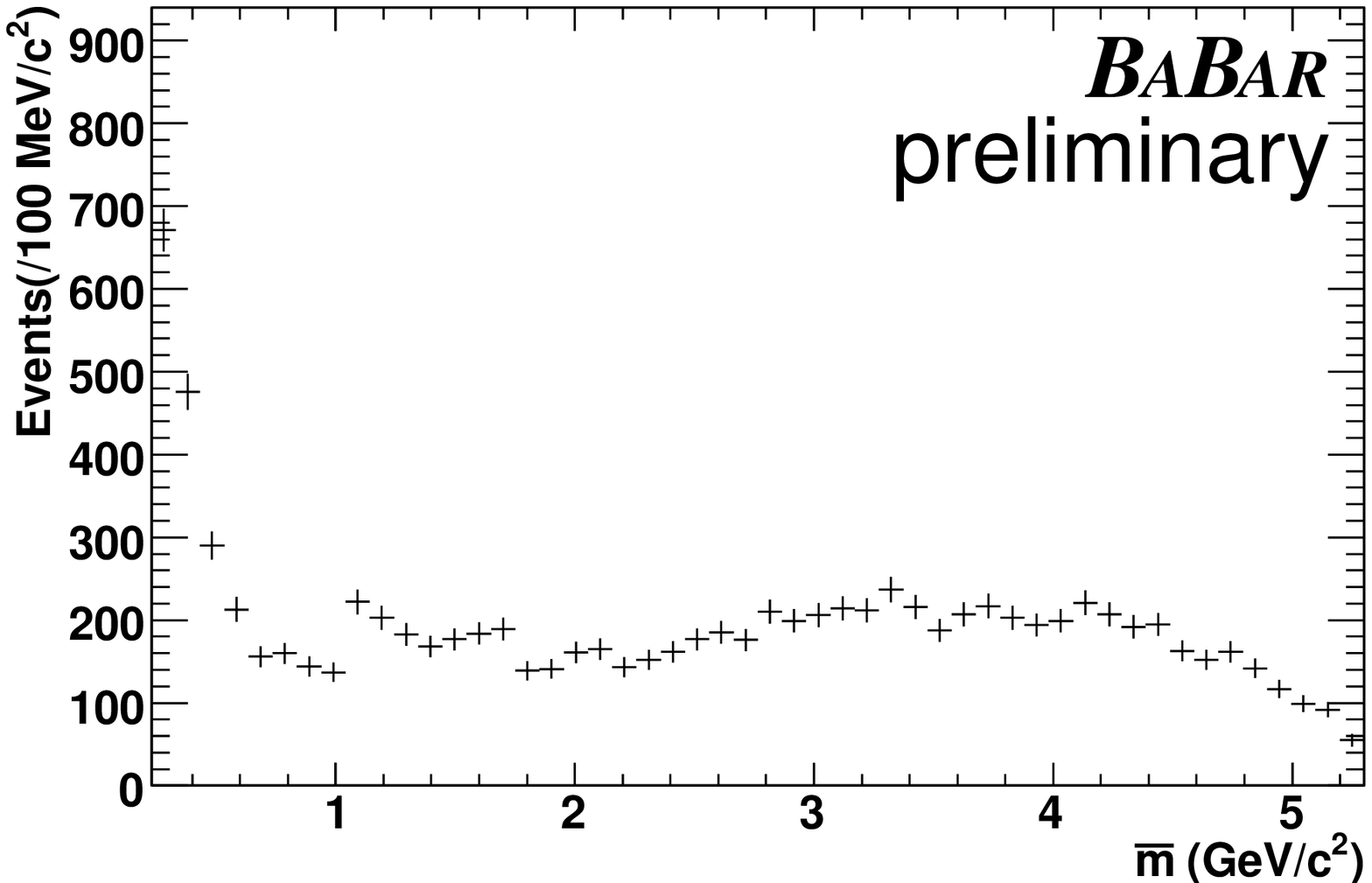}
              \epsfxsize6cm\epsffile{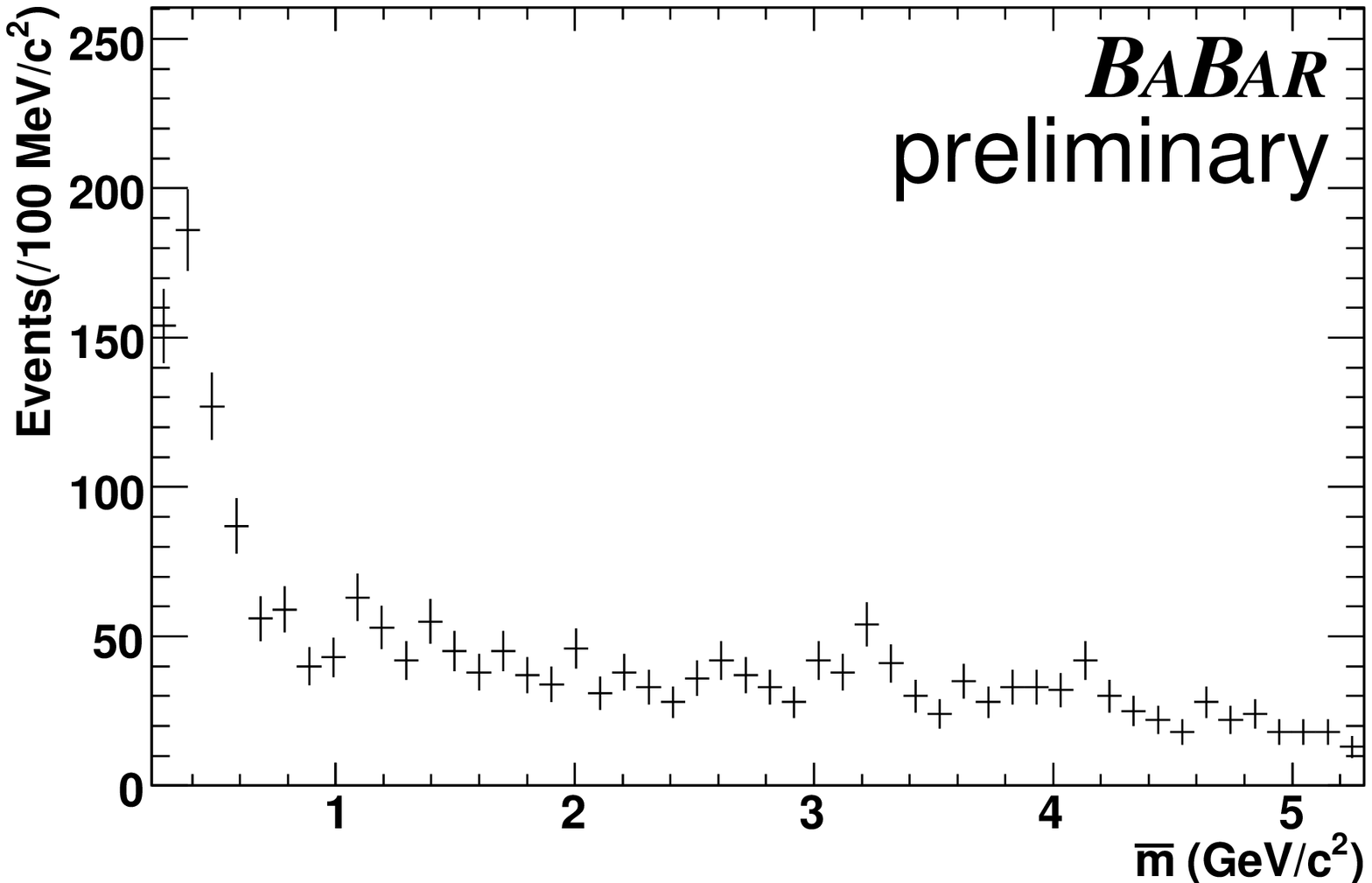}}
  \vspace{-0.1cm}
\caption{\label{fig:bkgDelM}
The background (top) $\dm$ and (bottom) $\mbar$ distributions for  (left-to-right) $\epem\epem$, $\epem\mupmum$, and $\mupmum\mupmum$ from the full dataset.  For the $\dm$ plots, we have required $\mbar>1GeV$ to that all events have the same upper $\dm$ value.  The effect of the $\dm$ cut increasing from $0.25\gevcc$ to $0.5\gevcc$ at $\mbar=1.0\gevcc$ can be seen in the $\mbar$ plots.  
}
\end{figure}

\begin{figure}[tb]
  \centerline{\epsfxsize6cm\epsffile{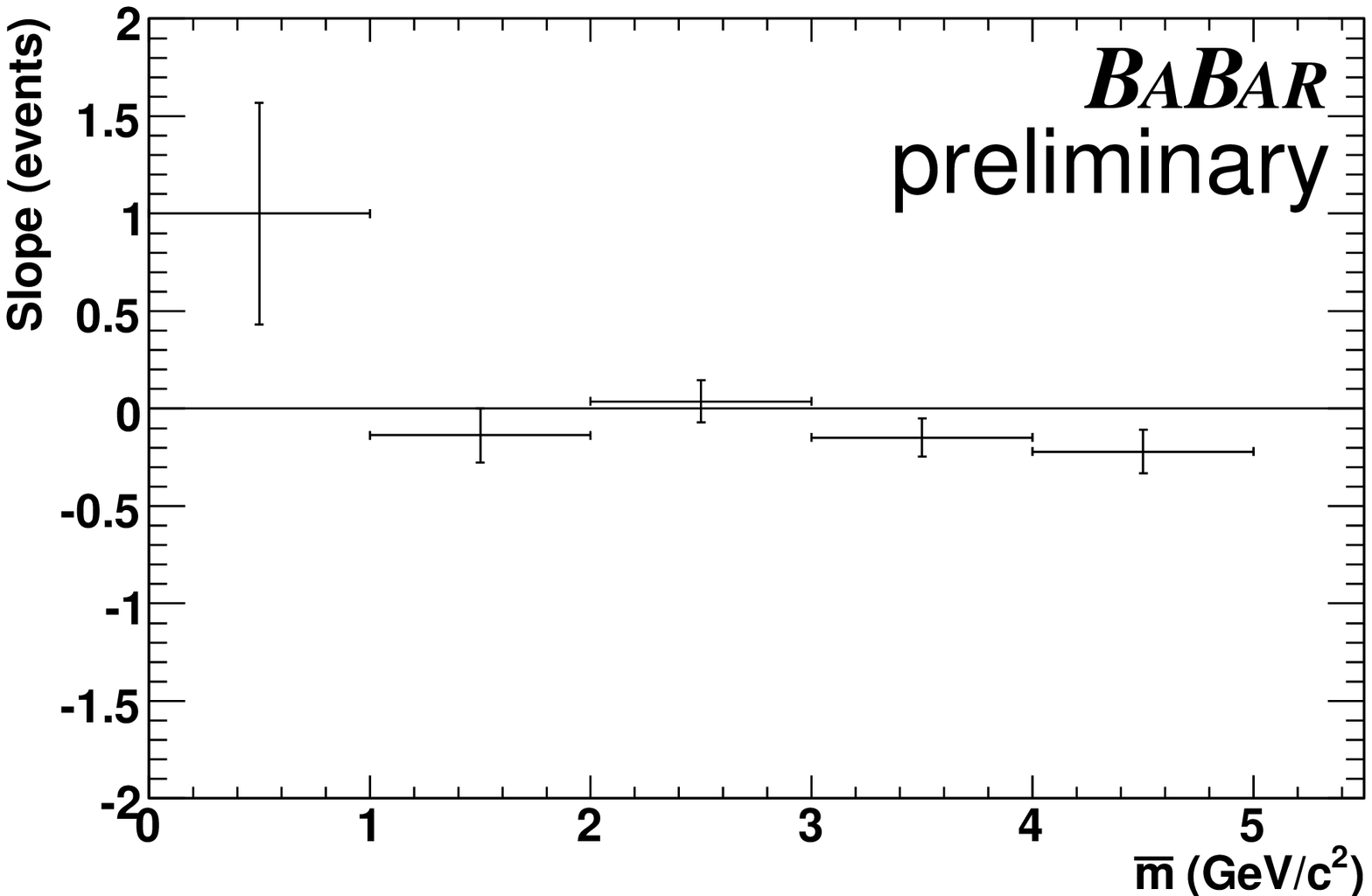}
              \epsfxsize6cm\epsffile{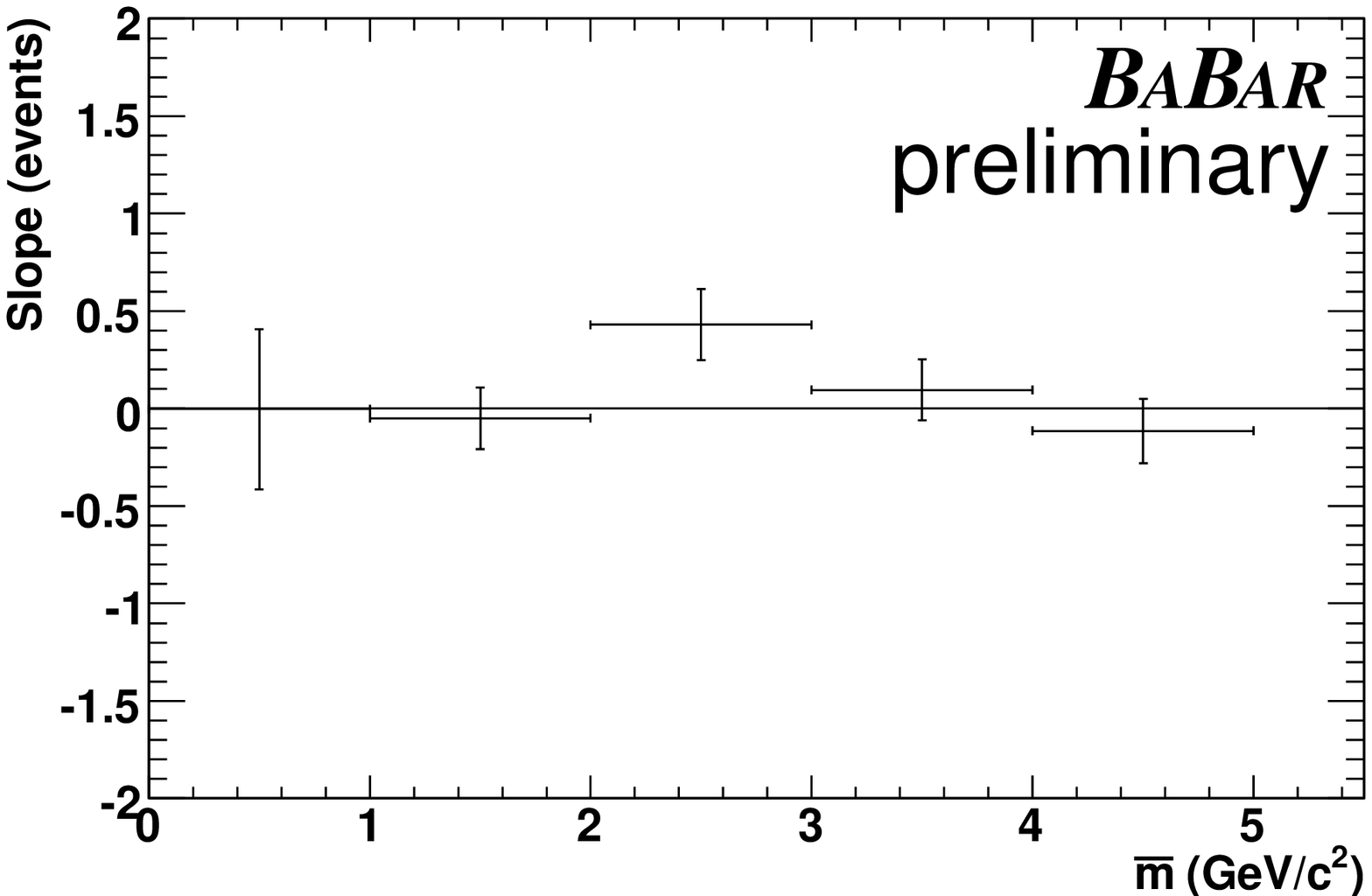}
              \epsfxsize6cm\epsffile{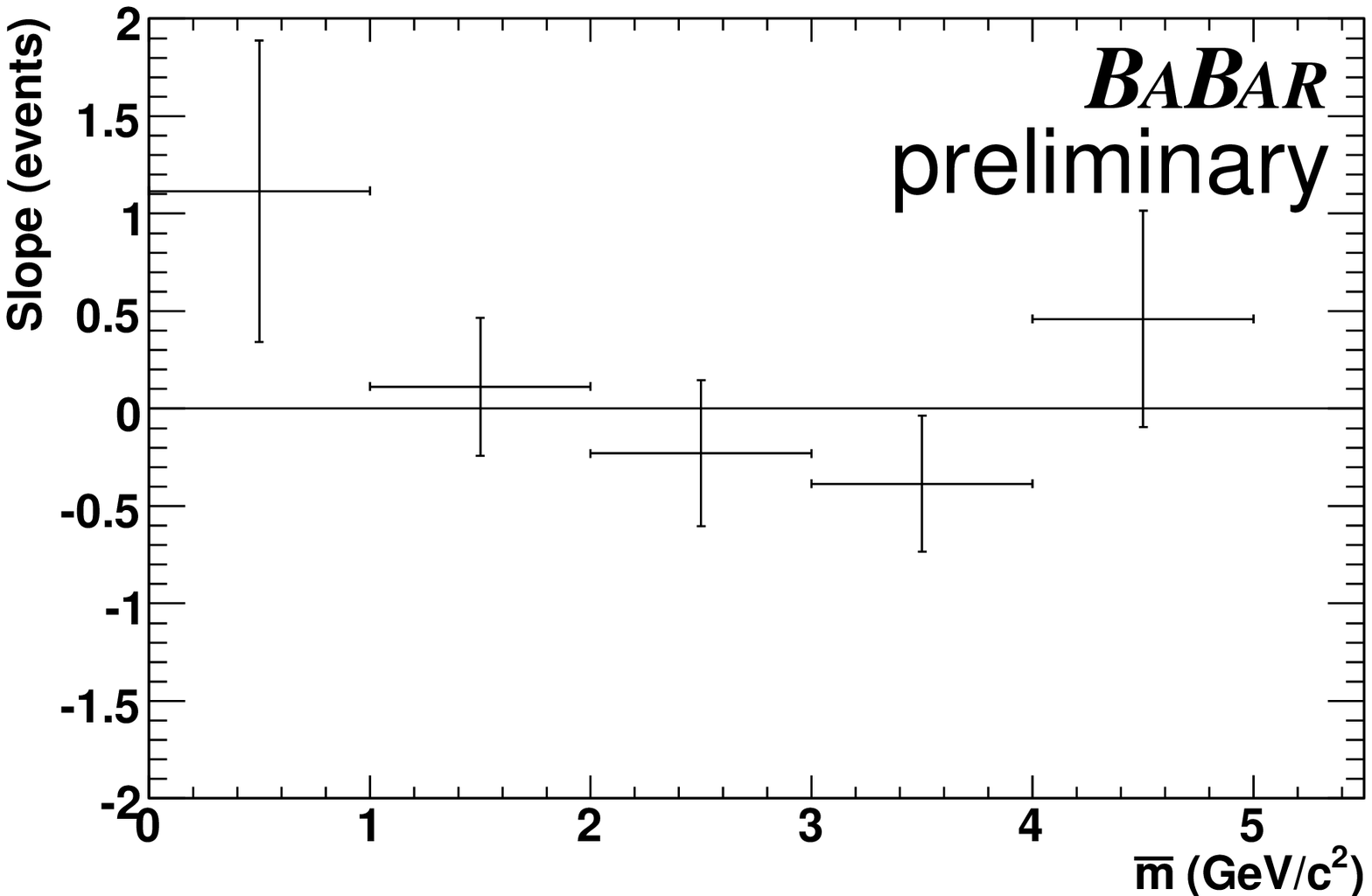}}
  \vspace{-0.1cm}
\caption{\label{fig:bkgSlope}
The slope of the background $\dm$ distributions for  (left-to-right) $\epem\epem$, $\epem\mupmum$, and $\mupmum\mupmum$ as a function of $\mbar$. The mean values of the slopes are:  $-0.11\pm0.06$, $0.07\pm 0.08$, and $-0.02\pm0.19$. 
}
\end{figure}

The background distributions in $\dm$ and $\mbar$, after all selection, are shown in Figure \ref{fig:bkgDelM}.  The background $\dm$ distributions were fit with a line in different slices of $\mbar$, the slopes of which are plotted for the three modes in Figure \ref{fig:bkgSlope}, and the slopes are consistent with 0.  When extracting the signal yields, we  assume a uniform background distribution, and take into account the uncertainties in the slope as a systematic error.    We use the full dataset for the above plots; any signal present would be completely washed out when projected onto the $\dm$ or $\mbar$ axis.   

\subsection{Signal extraction and limit setting}
\label{sec:limits}

\begin{figure}[tb]
  \centerline{\epsfxsize6cm\epsffile{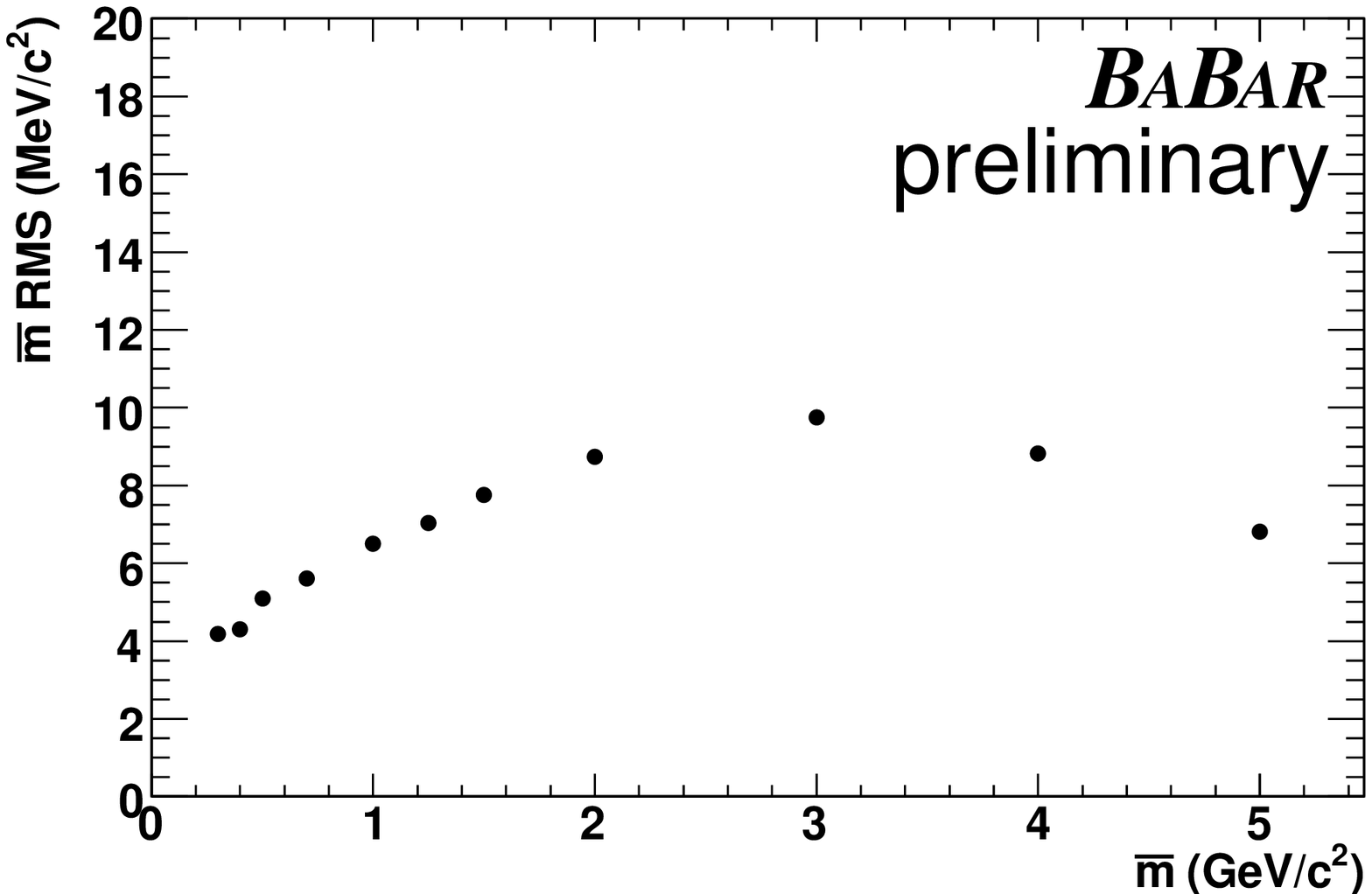}
              \epsfxsize6cm\epsffile{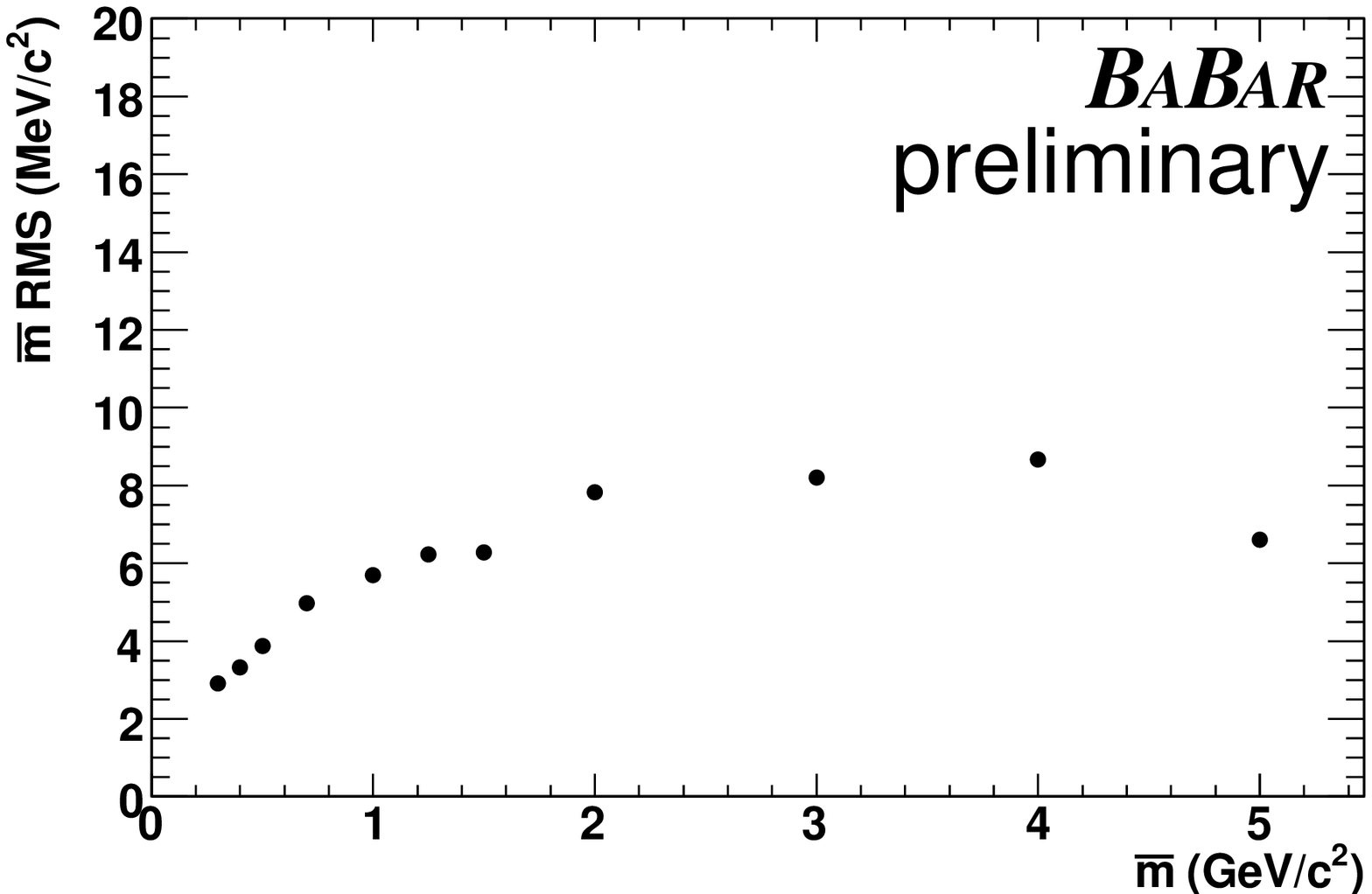}
              \epsfxsize6cm\epsffile{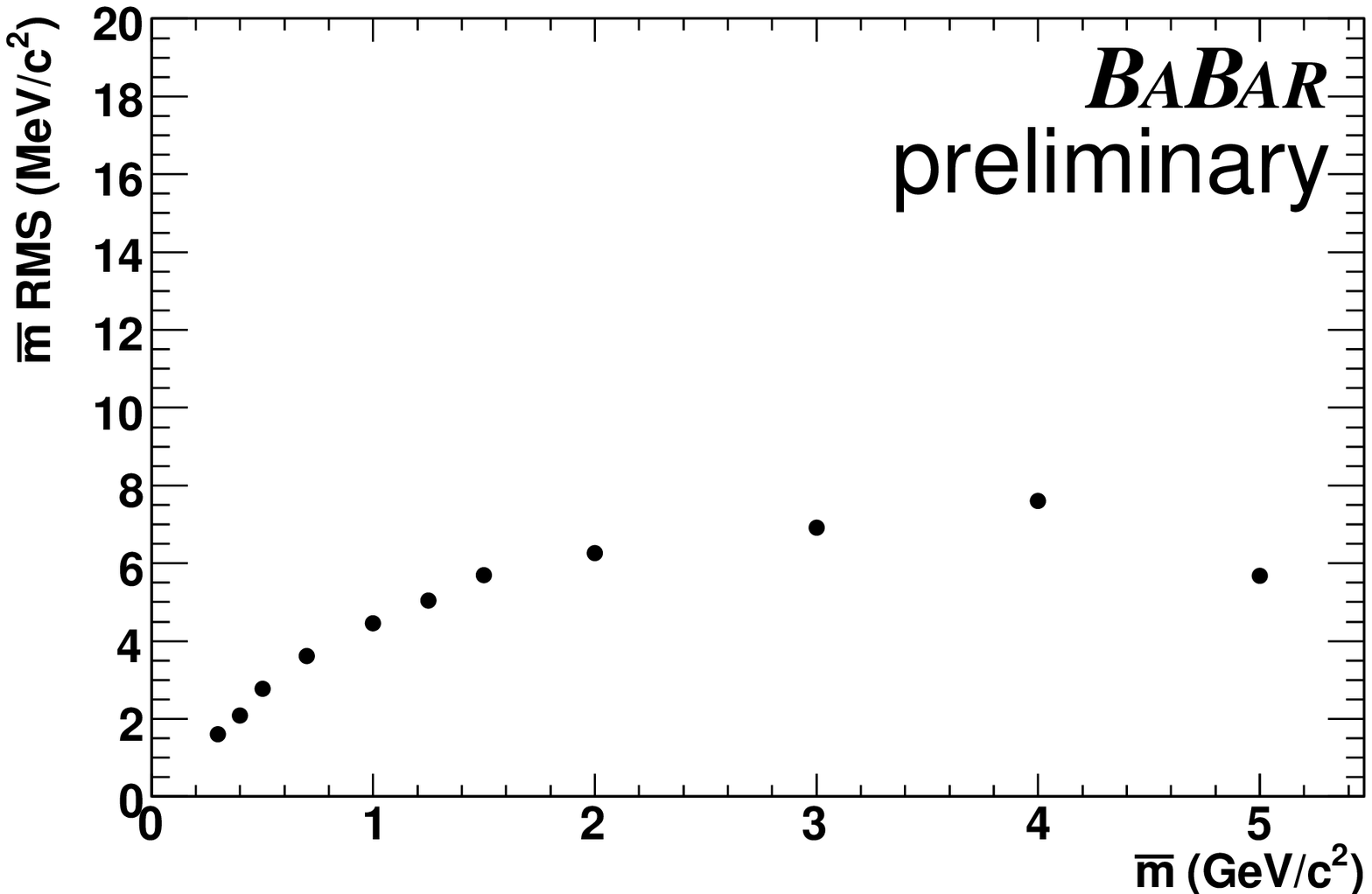}}
  \vspace{-0.1cm}
\caption{\label{fig:mbarRMS}
The $\mbar$ RMS versus $\Wp$ mass for (left to right) $\Wp\Wp\to\epem\epem$, $\Wp\Wp\to\epem\mupmum$, and $\Wp\Wp\to\mupmum\mupmum$.  
}
\end{figure}

In this analysis, our aim is to obtain a limit (or observe a signal) for $\epem\to\Wp\Wp$ as 
function of the presumed $\Wp$ mass. To this end, search for a  signal  in steps of the  
average dilepton mass $\mbar$.  We have chosen the $\mbar$ bin size to be $20\mevcc$, which is a large enough range 
to fully contain any signal.  Figure \ref{fig:mbarRMS} shows the RMS of $\mbar$ at the different mass points.   We scan $\mbar$ in steps of $10\mevcc$, half the bin size, so that at least one bin will fully contain the signal.  Thus, in the $\mbar$ range from $0.24-5.3\gevcc$, there are 507 total bins.   We define the signal and background 
regions in $\dm$ by cutting at a value of $\dm$ so that the signal is $90\%$ efficient, as discussed above.   

With this framework, the number of background events in a given $\mbar$ bin is quite small.  
  Except at low $\mbar$, the expected number of 
background events in the entire $\dm$ range is typically below 100 events in a $\mbar$ bin, particularly for the $\mupmum\mupmum$ mode where it is below 5 events.
Thus there will be relatively large fluctuations in the background due to Poisson statistics and 
the limit setting procedure must take this into account.  We  use a profile 
likelihood technique\cite{trolke} to set limits in the presence of nuisance parameters, such as the expected background 
yield.  Using this technique, we obtain a confidence level ($CL$) for the presence of signal defined as:
\beqn
CL = Prob\left(-2\log({\mathcal L_{s=0}})-2\log({\mathcal L_{max}})\right)
\eeqn
where ${\mathcal L_{s=0}}$ is the value of the likelihood at 0 signal events and ${\mathcal L_{max}}$ is the maximum value of the likelihood.  

\begin{figure}[tb] 
  \centerline{\epsfxsize8cm\epsffile{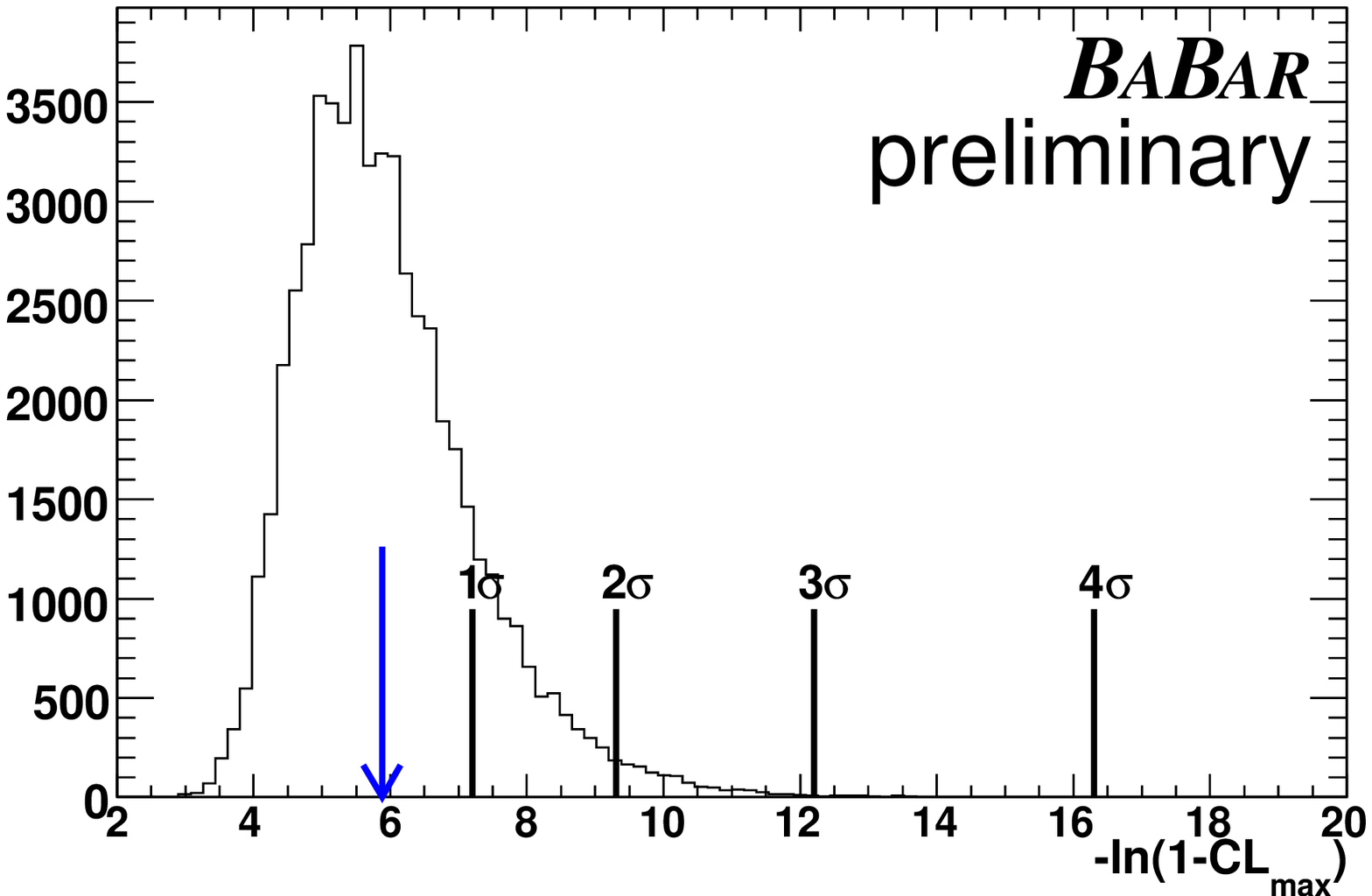}
              \epsfxsize8cm\epsffile{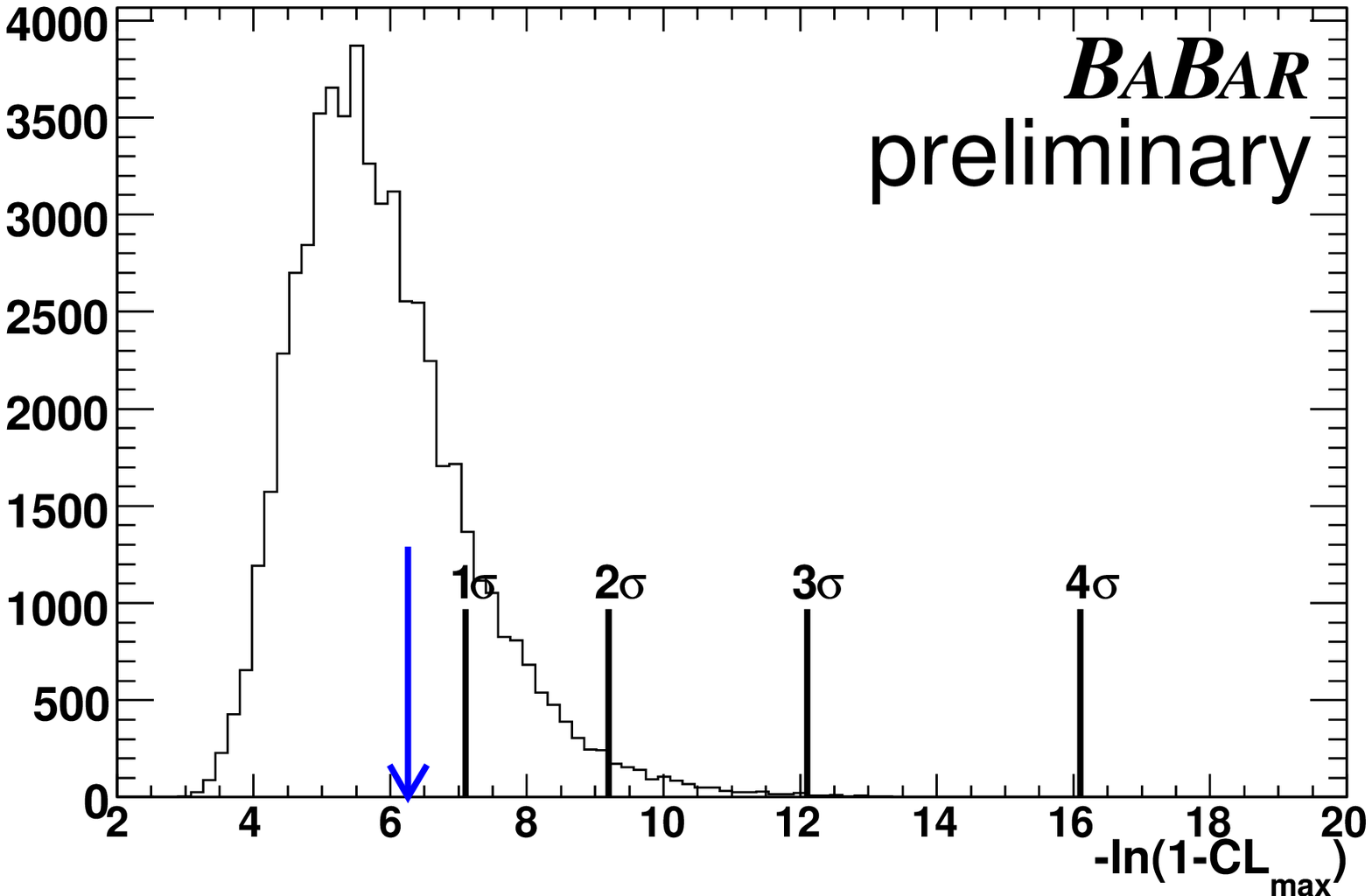}}
  \centerline{\epsfxsize8cm\epsffile{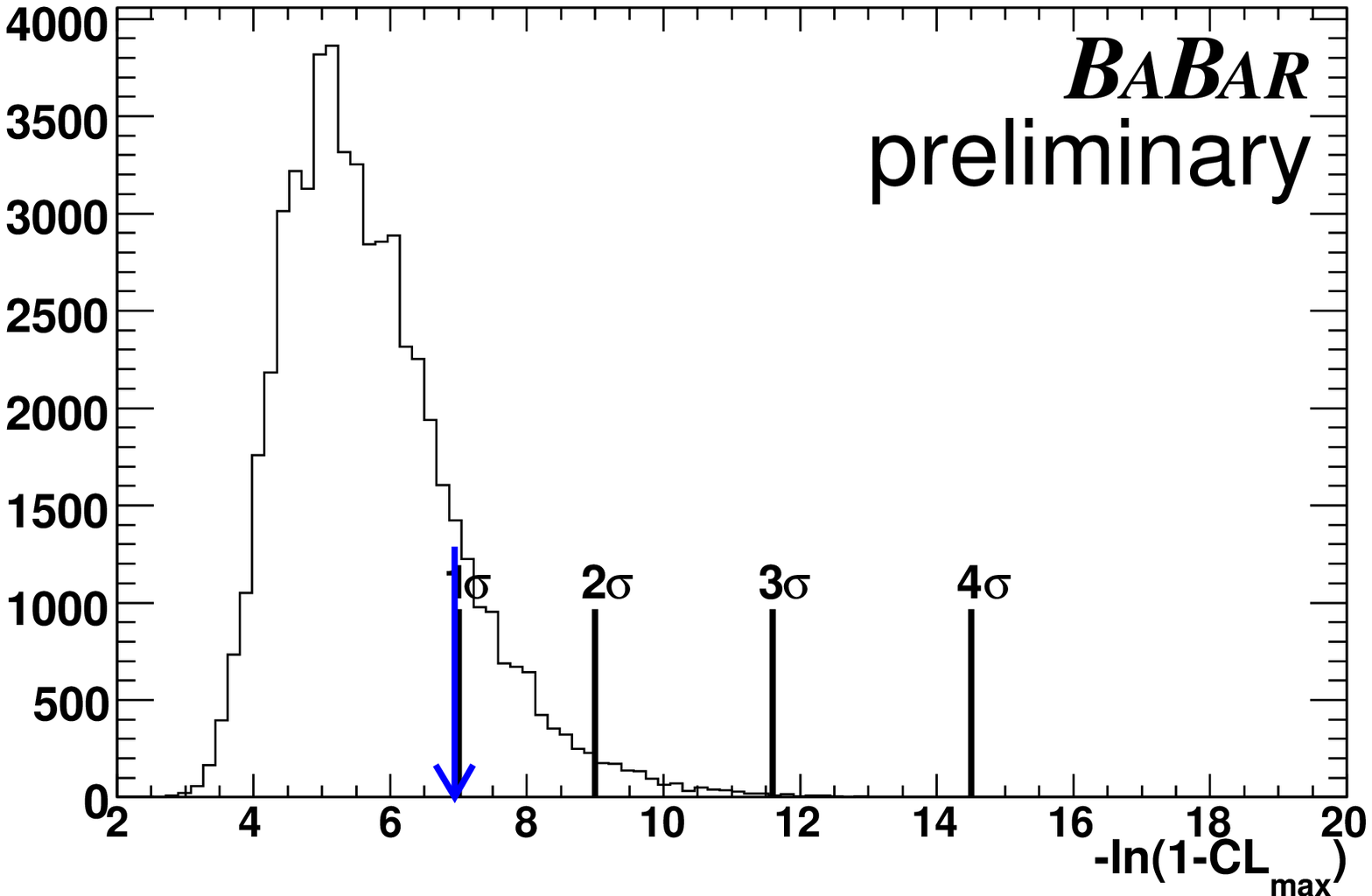}
              \epsfxsize8cm\epsffile{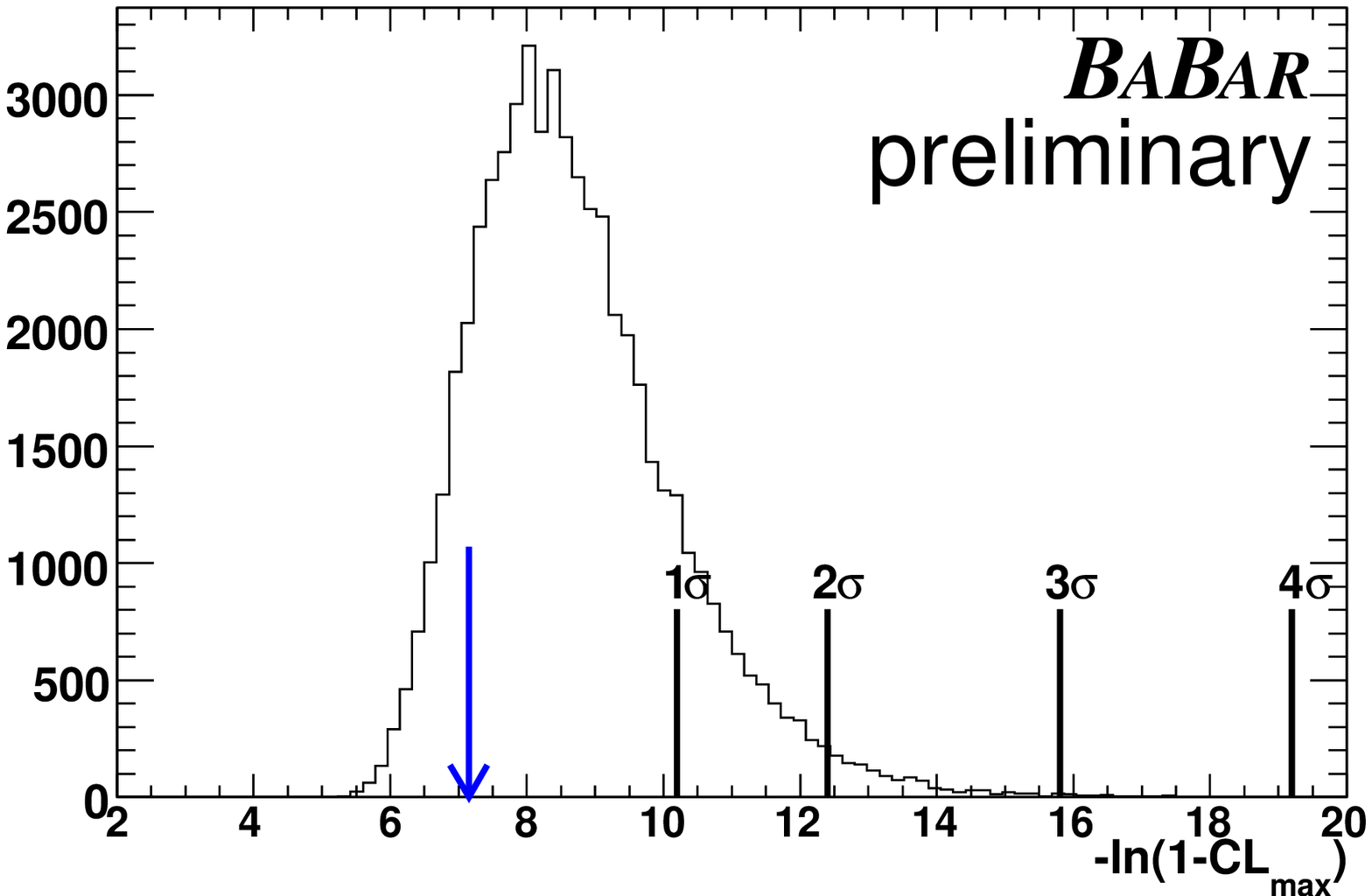}}
  \vspace{-0.1cm}
\caption{\label{fig:UnblindMaxSig}
The distribution of $-ln(1-CL_{max})$ from toy with the arrow showing the value of $-ln(1-CL_{max})$ observed in data.  The plots are (left to right, top to bottom)  $\epem\epem$, $\epem\mupmum$, $\mupmum\mupmum$, and the three modes combined. 
}
\end{figure}

\begin{figure}[tb]
  \centerline{\epsfxsize6cm\epsffile{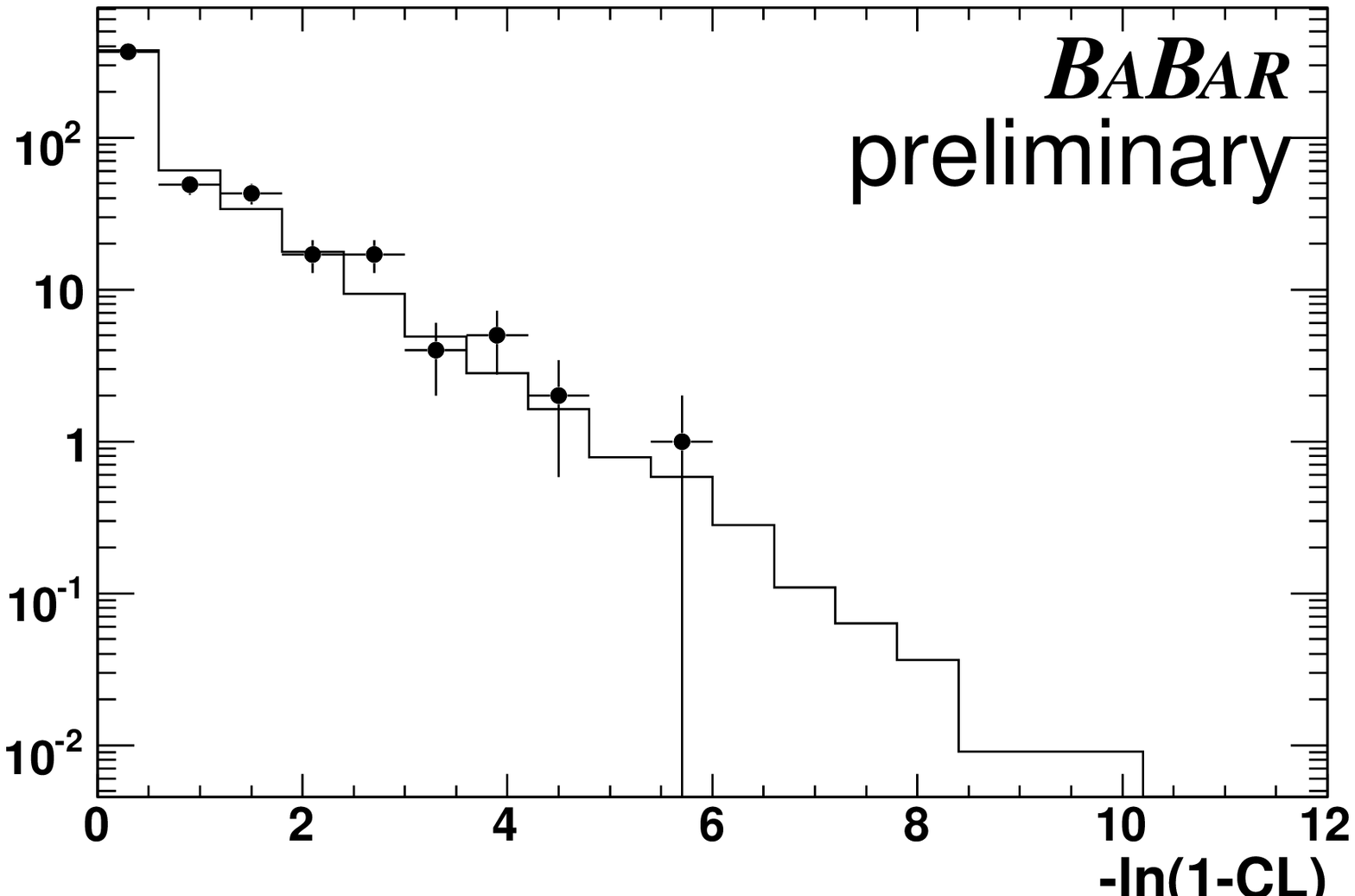}
              \epsfxsize6cm\epsffile{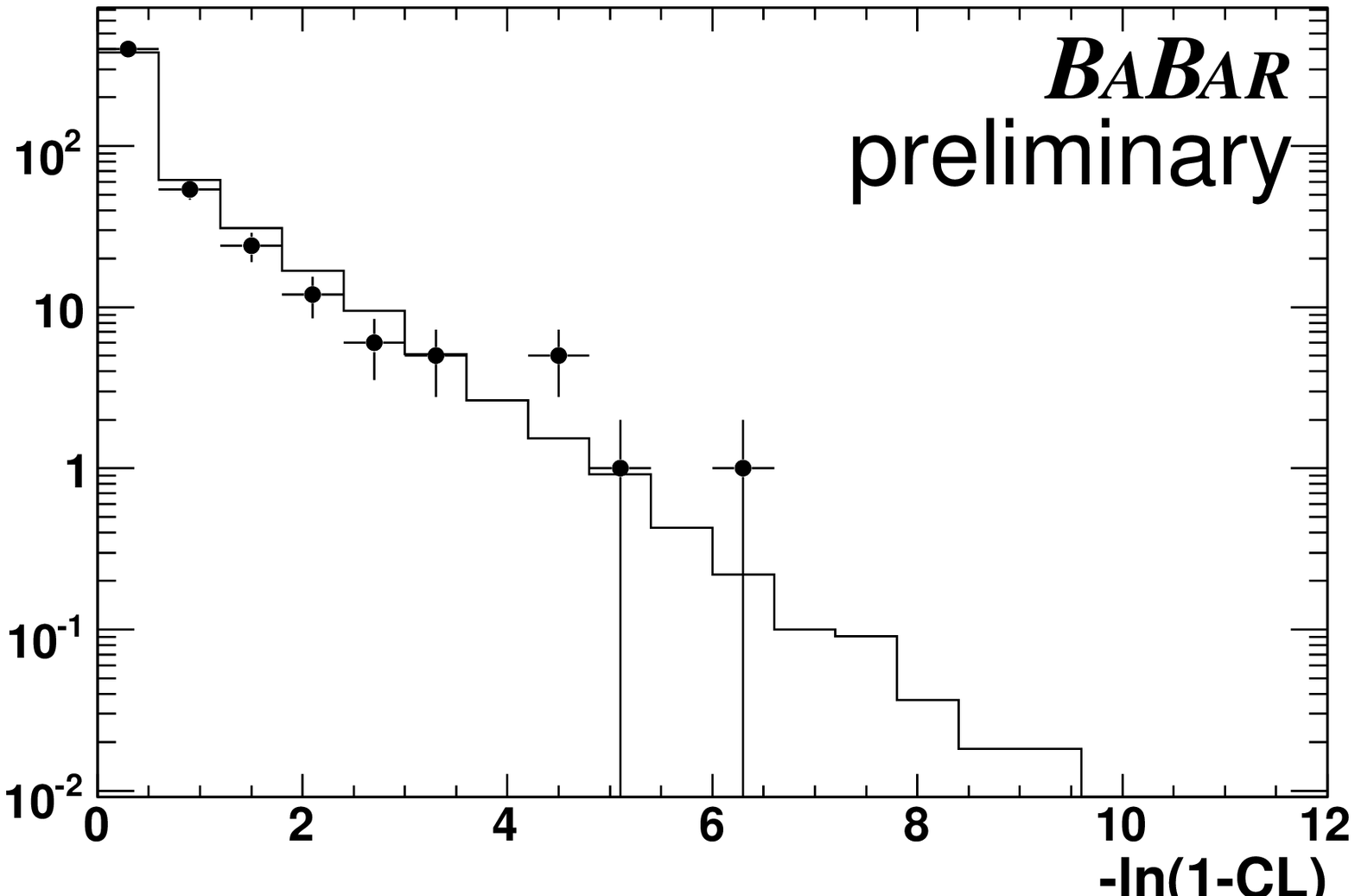}
              \epsfxsize6cm\epsffile{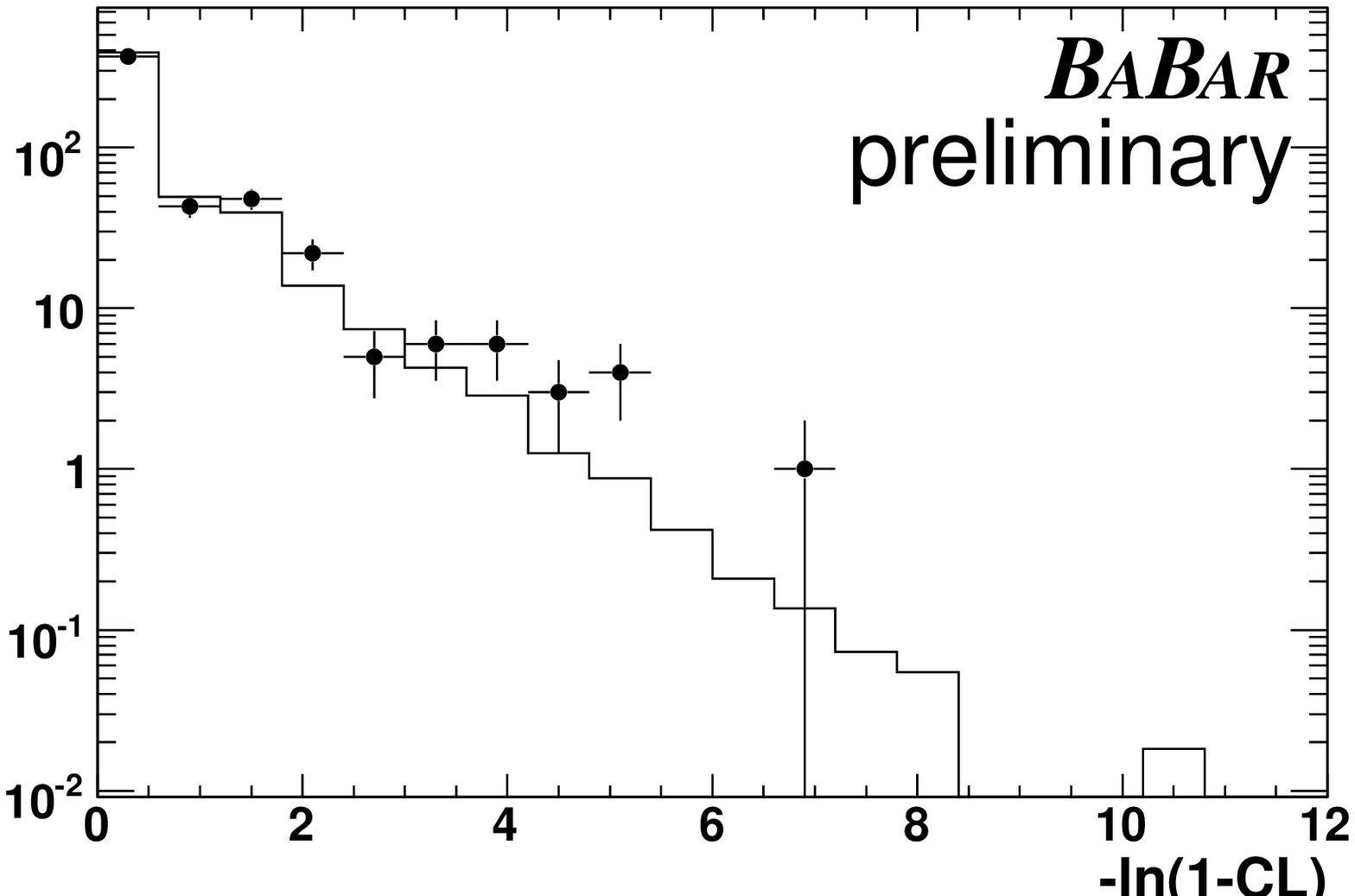}}
  \vspace{-0.1cm}
\caption{\label{fig:UnblindRawSig}
The distribution of values of $-ln(1-CL)$ from (error bars) data and (solid histogram) toy  for  (left to right) $\epem\epem$, $\epem\mupmum$, and $\mupmum\mupmum$. 
}
\end{figure}

Since in our dataset we will have $507$ correlated measurements (204 independent measurements), each at a different $\mbar$, 
we need to determine a criteria 
for a signal observation.  Simply asking whether an individual bin has an observed yield in it $>3\sigma$ above 0 is not enough  since the probability to observe at least 1 $>3\sigma$ fluctuation in one of the $\mbar$ bins is
$0.3$ (as determined from the simulation described below).  We need to redefine the $X\sigma$ levels for the new question ``What is the 
chance that I see a background fluctuation above $X\sigma$ in our 507 correlated trials?''.  We have done this by generating 
many simulated datasets (toys) with the expected $\mbar$ and $\dm$ background distributions
with 0 signal and plotting the highest value of the signal confidence level  observed over that dataset, which we 
call $CL_{max}$.  The
results of these simulations are shown in Figure \ref{fig:UnblindMaxSig}, plotting the more convenient variable $-ln(1-CL_{max})$. 
As a reference, the distribution of values $-ln(1-CL)$ from a single bin (i.e. not the largest value 
in an $\mbar$ scan) is shown in Figure \ref{fig:UnblindRawSig}. 
Table \ref{tab:observation} shows the values of $-ln(1-CL_{max})$ that correspond to 1-4$\sigma$ fluctuations of the
background (also displayed on the plot). Although the background levels are different, the values are consistent between the three modes.  

Additionally, we calculate the combined max confidence level, defined as:
\beqn
(1-CL_{max,C})=(1-CL_{max,4e})(1-CL_{max,2e2\mu})(1-CL_{max,4\mu})
\eeqn
whose distribution for background-only toys is shown in the bottom right plot of Figure \ref{fig:UnblindMaxSig}.
If lepton universality holds, this limit is potentially more sensitive than the individual confidence levels and allows us to catch a signal that is not significant in any single final state.  
Our criteria to claim evidence of a signal is 
to observe the largest value of $-ln(1-CL_{max})$ in any of $\epem\epem$, $\epem\mupmum$, $\mupmum\mupmum$ 
or in the combined confidence level that is greater than the $3\sigma$ values given in Table \ref{tab:observation}.  

\begin{table}[thb]
  \begin{center}
    \caption{ \label{tab:observation}
	Values of the 1-,2-,3-,4-$\sigma$ limits for $-ln(1-CL_{max})$ in the three final states.  
}
\begin{tabular}{lccccc}
\hline
\hline	
&&&&\\[-0.2cm]
Signif. & $P(CL_{max})<X$ & \multicolumn{3}{c}{$-ln(1-CL_{max})$}   \\
& &  $\epem\epem$
& $\epem\mupmum$
& $\mupmum\mupmum$
& Combined \\
\hline
&&&&\\[-0.2cm]
\rule[-1.7mm]{0mm}{5mm}  $1\sigma$  &   0.84135     & 	7.2	& 7.1	&  7.0	 & 10.2 \\
\rule[-1.7mm]{0mm}{5mm}  $2\sigma$  &    0.97725    & 	9.3	& 9.2	&  9.0   & 12.4\\
\rule[-1.7mm]{0mm}{5mm}  $3\sigma$  &    0.99865    & 	12.2	& 12.1	&  11.6  & 15.8\\
\rule[-1.7mm]{0mm}{5mm}  $4\sigma$  &    0.99997    & 	16.3	& 16.1	&  14.5  & 19.2 \\
\hline
\hline
\end{tabular}
    \vspace{-0.8cm}
  \end{center}
\end{table}

 \section{Systematic Errors}
\label{sec:syst}

There are two types of systematic errors in this analysis:  systematics that effect both the yield and cross-section
upper limits (e.g. errors due to uncertainties in the background shape) and systematics that just affect the 
cross-section (e.g. tracking efficiency errors).  The second type of error does not effect the signal significance.  
Table \ref{tab:syst} summarizes the values the systematic errors for the different sources described below. 

\bei

\item {\bf $\dm$ background shape}:  We assume that the background is uniform in $\dm$ and with our limited MC statistics
                                but we have no 
                                {\it a priori} reason to expect this.  While the background $\dm$ does look
                                quite flat and does not appear to depend on $\mbar$, see Figure~\ref{fig:bkgSlope}), we still need to account for uncertainties.  Consequently, we 
                                estimate the $\dm$ background shape from the data itself.
                                
\begin{figure}[tb]
  \centerline{\epsfxsize6cm\epsffile{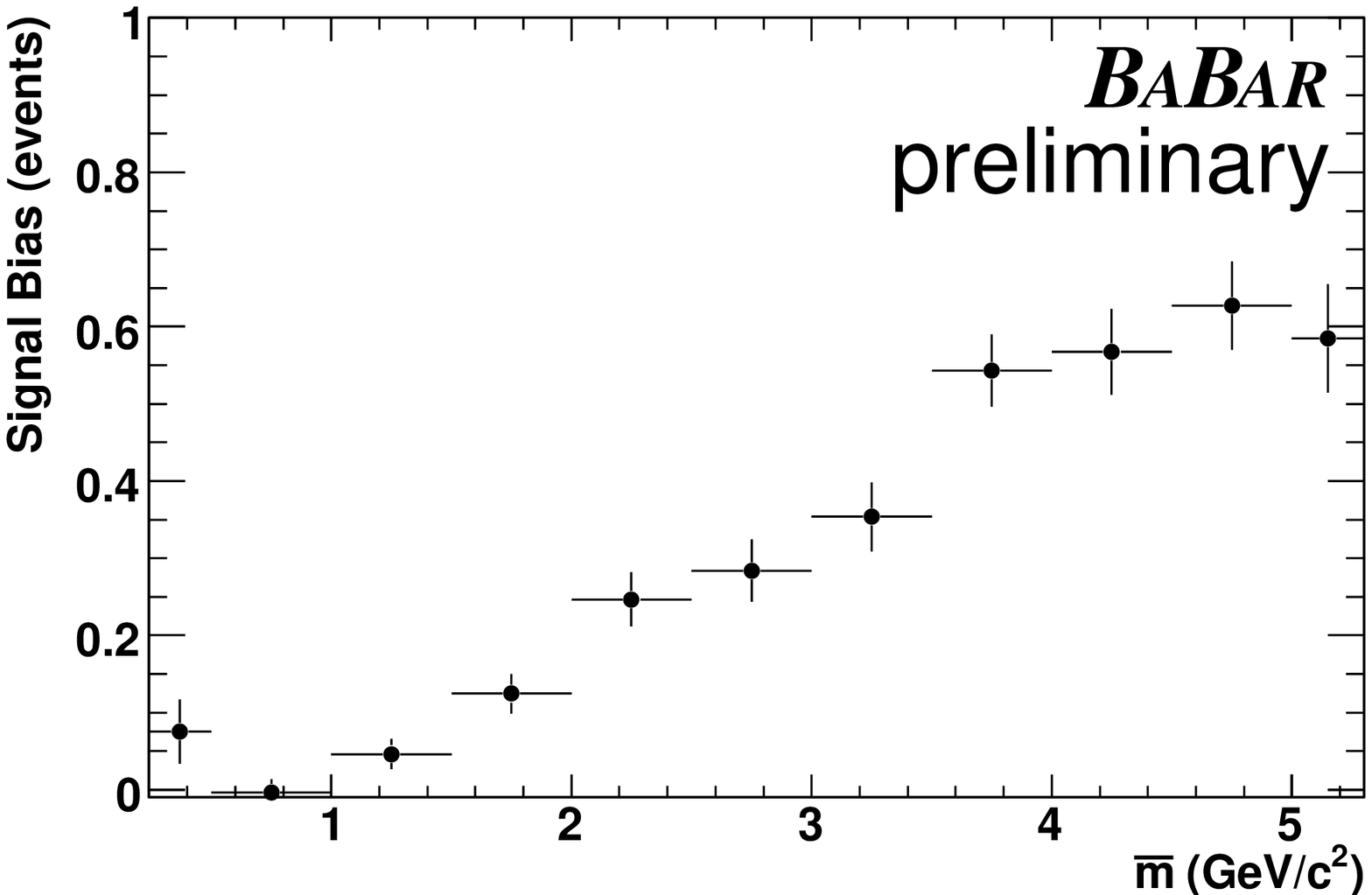}
              \epsfxsize6cm\epsffile{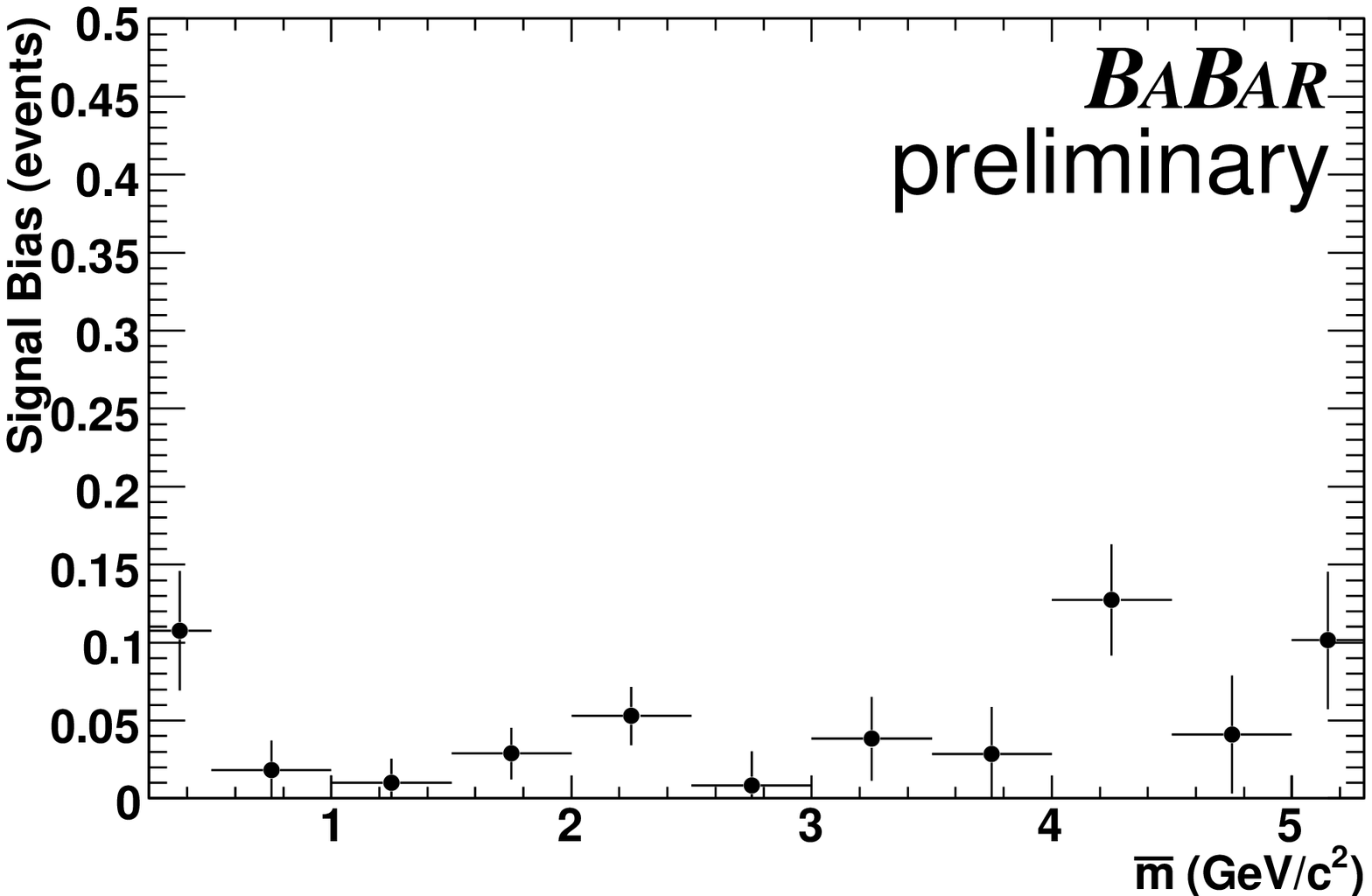}
              \epsfxsize6cm\epsffile{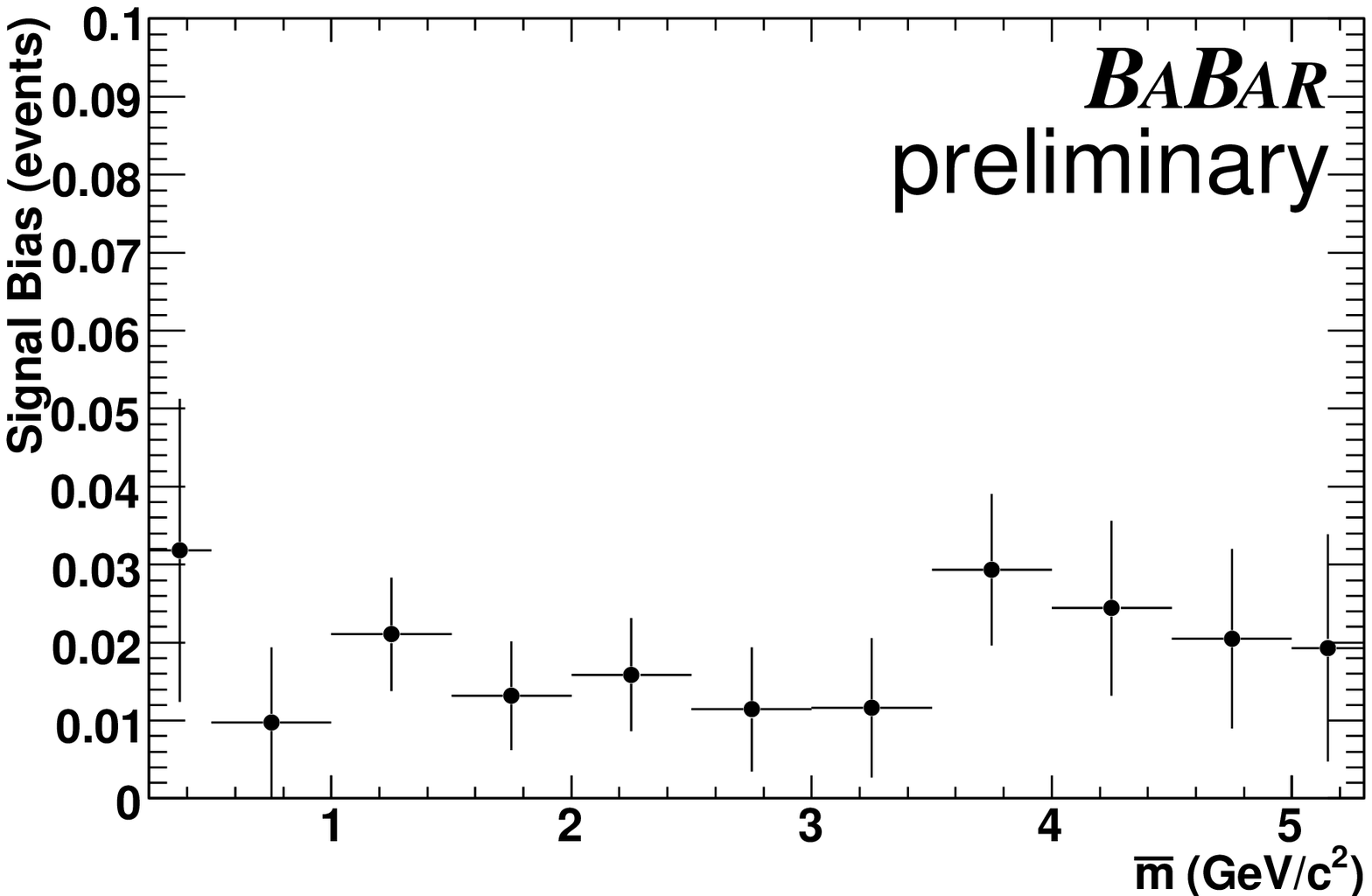}}
  \vspace{-0.1cm}
\caption{\label{fig:bkgshapesyst}
The positive signal yield bias due to the uncertainty in the background $\dm$ slope as a function of $\mbar$ for 
 (left-to-right) $\epem\epem$, $\epem\mupmum$, and $\mupmum\mupmum$. 
}
\end{figure}

                                In order to estimate the size of this uncertainty, we have generated toy $\mbar$ 
                                scans (background only) with a slope and calculated the signal yield assuming  
                                a slope of 0. 
                                The mean $\dm$ slopes are given in the caption of Figure \ref{fig:bkgSlope}. 
                                For this study, we shift the mean value  of the slope ($B_m$) by:
\bei
    \item for $B_m<0$, we  assign the slope to be $B_m-\sigma$
    \item for $B_m>0$, we  assign the slope to be $-\sigma$
\eei
                                where  $\sigma_m$ is the error on the mean.  
                                We only use the negative slope values because 
                                we are primarily interested in how this biases us  toward more signal.  
                                The results of this study are shown in Figure \ref{fig:bkgshapesyst} as the
                                observed signal yield bias vs $\mbar$ for the three modes.  The bias depends on 
                                $\mbar$ because both the number of background events in the full $\dm$ region  and 
                                because the $\dm$ signal/background region definitions depend on $\mbar$.       

                                We incorporate this bias into a systematic error on the cross section
                                by converting the bias in the number of events into a cross section in $\mbar$ bins. 
                                The error is largest for the $\foure$ mode where at high $\mbar$ is as large as $\sim
                                5~\ab$; for the other two modes this error is generally $<1 \ab$.

\item {\bf $\dm$ signal shape}:  We use MC at select mass values to interpolate the 90\% efficiency $\dm$ cut value to cover
                          all masses. The interpolation is done with a polynomial and we 
                                        vary the parameters of the polynomial within their errors 
                                        to get the error in the $\dm$ cut value.  This is then 
                                        translated into an efficiency error. The magnitude of this error is $\sim 1\%$ 
                                        and depends slightly on $\mbar$.

\item {\bf interpolation of total efficiency}:  We use MC at select mass values to interpolate the total efficiency 
                                        to all masses. The interpolation is done by interpolating the 
                                        efficiency linearly between the MC mass points.  We propagate the
                                        errors in the efficiency points due to MC statistics through the
                                        interpolation.  
                                        In addition, we take the difference between a linear 
                                        and quadratic interpolation and assign the difference, added
                                        in quadrature with the  statistical error, as the systematic.  
                                         The magnitude of this error is $\sim 3\%$ 
                                        and depends slightly on $\mbar$.

\item {\bf particle ID}: we assign a 1\% error per electron and 2\% error per muon on the cross sections to account for the systematic error in the PID efficiency. This is the dominant systematic error.  

\item {\bf tracking efficiency}:  we assign 0.21\% error per track on the cross sections to account for the systematic error in the charged track reconstruction efficiency.  

\item {\bf luminosity}:  we assign a 1.1\% error on the cross sections due to the uncertaintly in the total luminosity.  

\eei

We add these sources of systematic error in quadrature and scale the statistical 90\% upper limit by the fractional systematic error to obtain the final upper limit.  This error depends slightly on $\mbar$ but is around 5\% for $\foure$ (up to ~10\% for high $\mbar$), 6.5\% for $\twoetwomu$, and 8.2\% for $\fourmu$. 

\begin{table}[thb]
  \begin{center}
    \caption{ \label{tab:syst}
        Sources of systematic uncertainties and their contributions.  
}
%  \vspace{-0.8cm}
\begin{tabular}{lccc}
\hline
\hline  
&&&\\[-0.2cm]
Source &   $\epem\epem$ & $\epem\mupmum$ & $\mupmum\mupmum$ \\
\hline
&&&\\[-0.2cm]
$\dm$ bkg shape         & 0.4-5.5 \ab   &0.1-0.7 \ab    &0.1-0.3 \ab    \\
$\dm$ signal efficiency & 1\%           & 1\%           & 1\%\\
total signal efficiency & 3\%           & 3\%           & 3\%   \\
particle ID             & 4\%           & 6\%           &  8\%          \\
tracking efficiency     & 0.8\%         & 0.8\%         & 0.8\%         \\
luminosity              & 1.1\%         & 1.1\%         & 1.1\%         \\
\hline
\hline
\end{tabular}
  \end{center}
\end{table}

 \section{Results and Conclusions}
\label{sec:res}

\begin{figure}[tb]
  \centerline{\epsfxsize6cm\epsffile{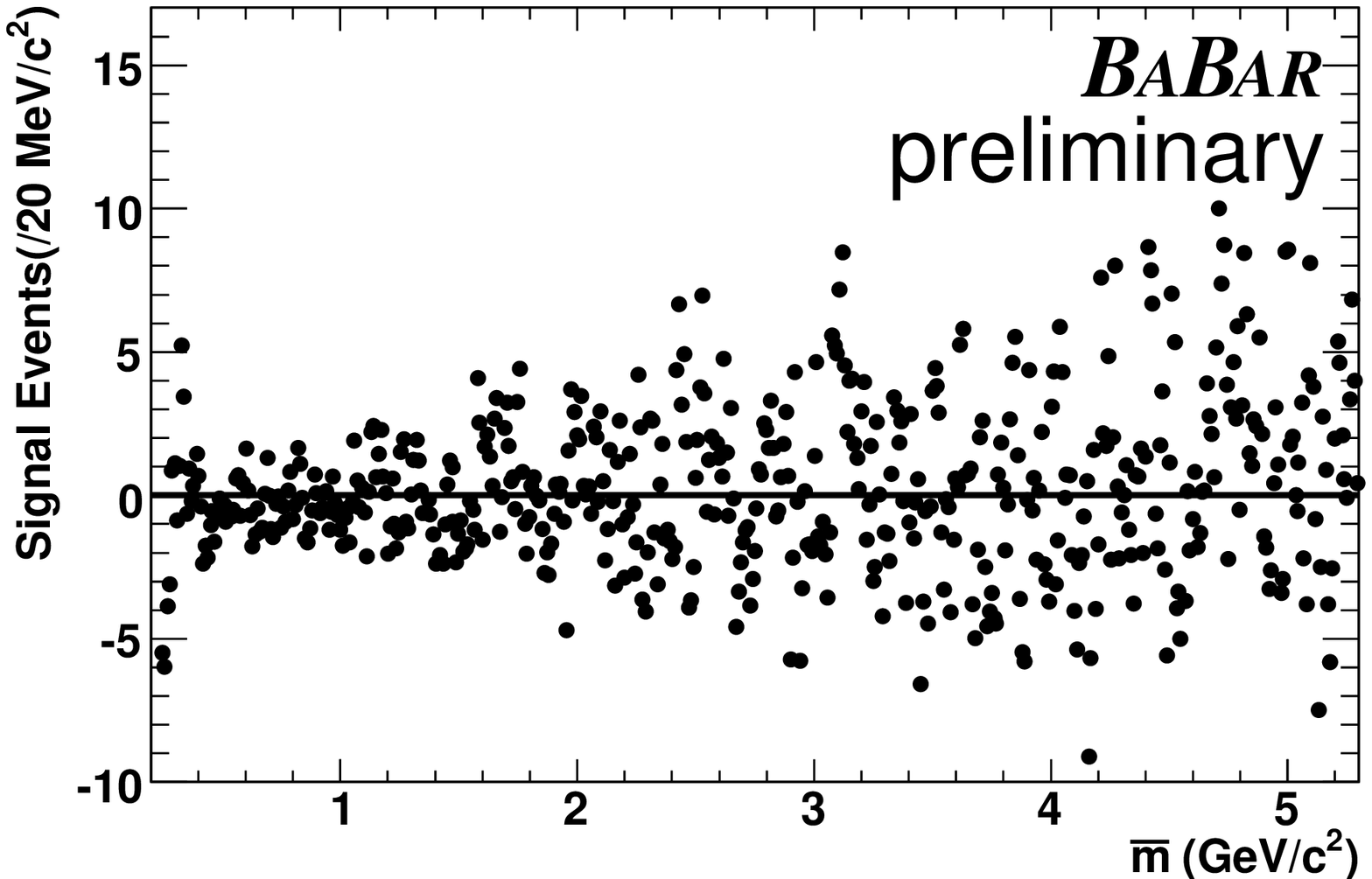}
              \epsfxsize6cm\epsffile{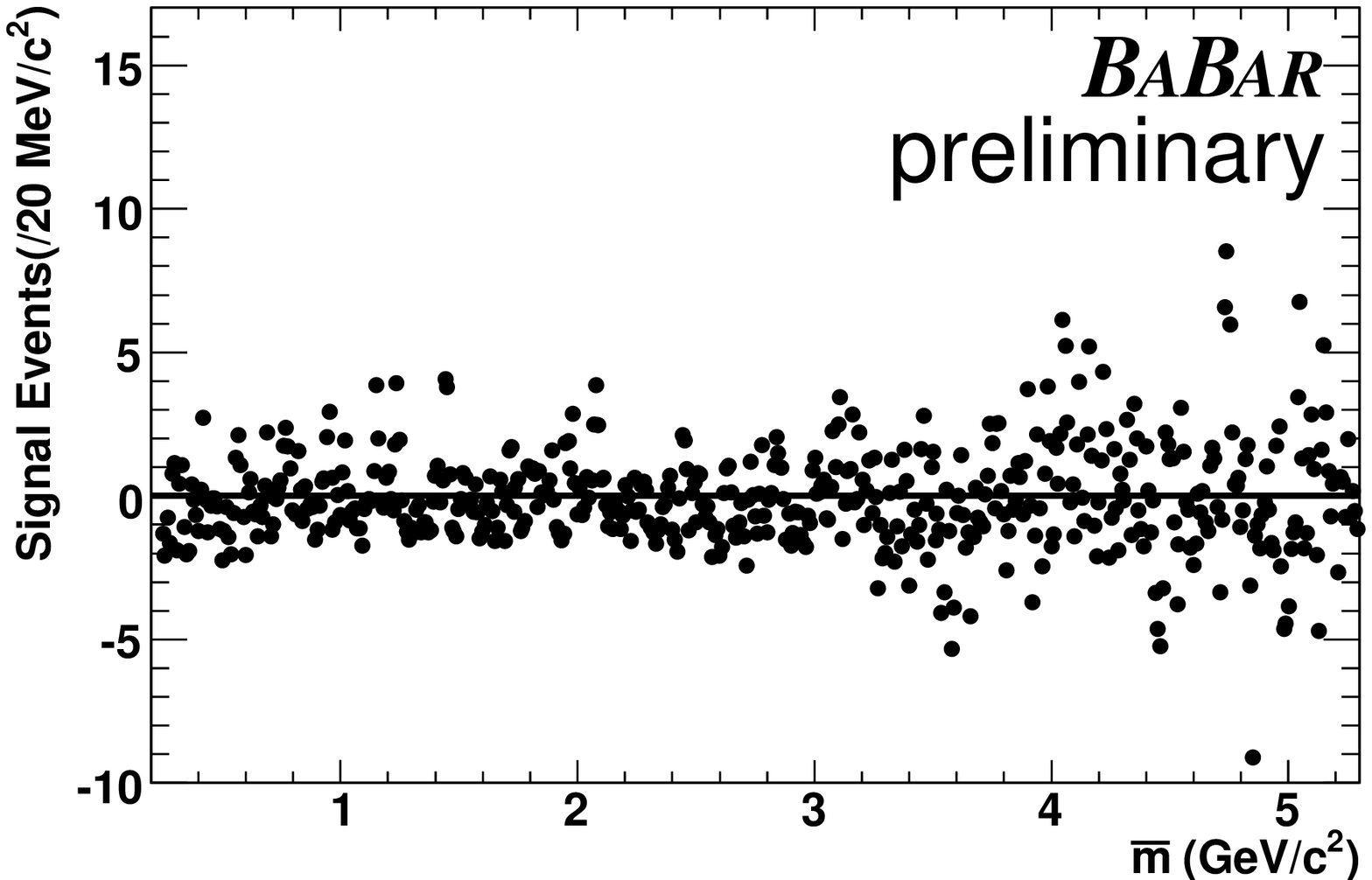}
              \epsfxsize6cm\epsffile{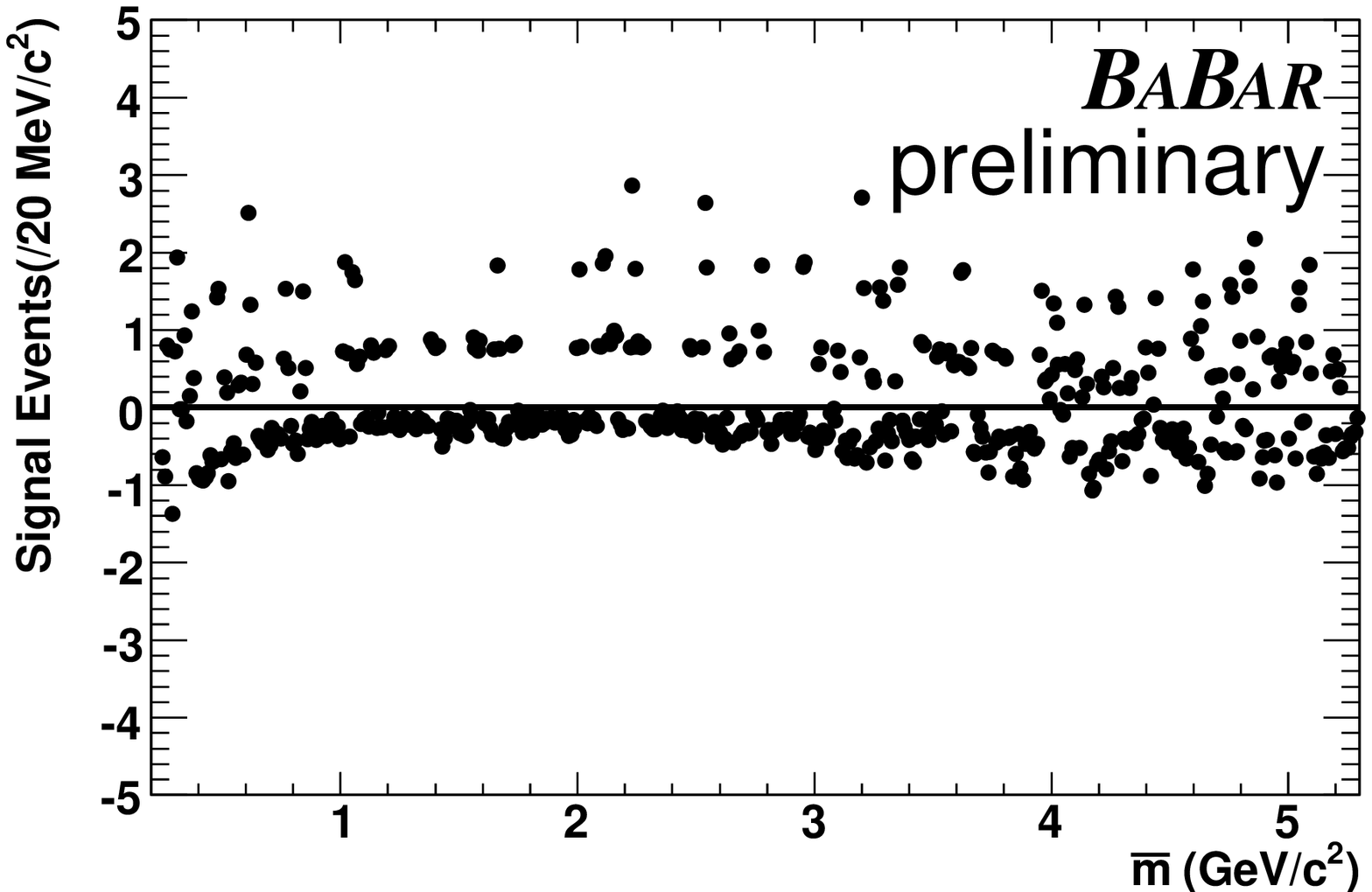}}
  \vspace{-0.1cm}
\caption{\label{fig:UnblindObs}
The number of signal events after background subtraction versus $\mbar$ for  (left to right) $\epem\epem$, $\epem\mupmum$, and $\mupmum\mupmum$.   The band structure evident in the $\fourmu$ plot is due to the very low number of events in this mode. 
}
\end{figure}

\begin{figure}[tb]
  \centerline{\epsfxsize6cm\epsffile{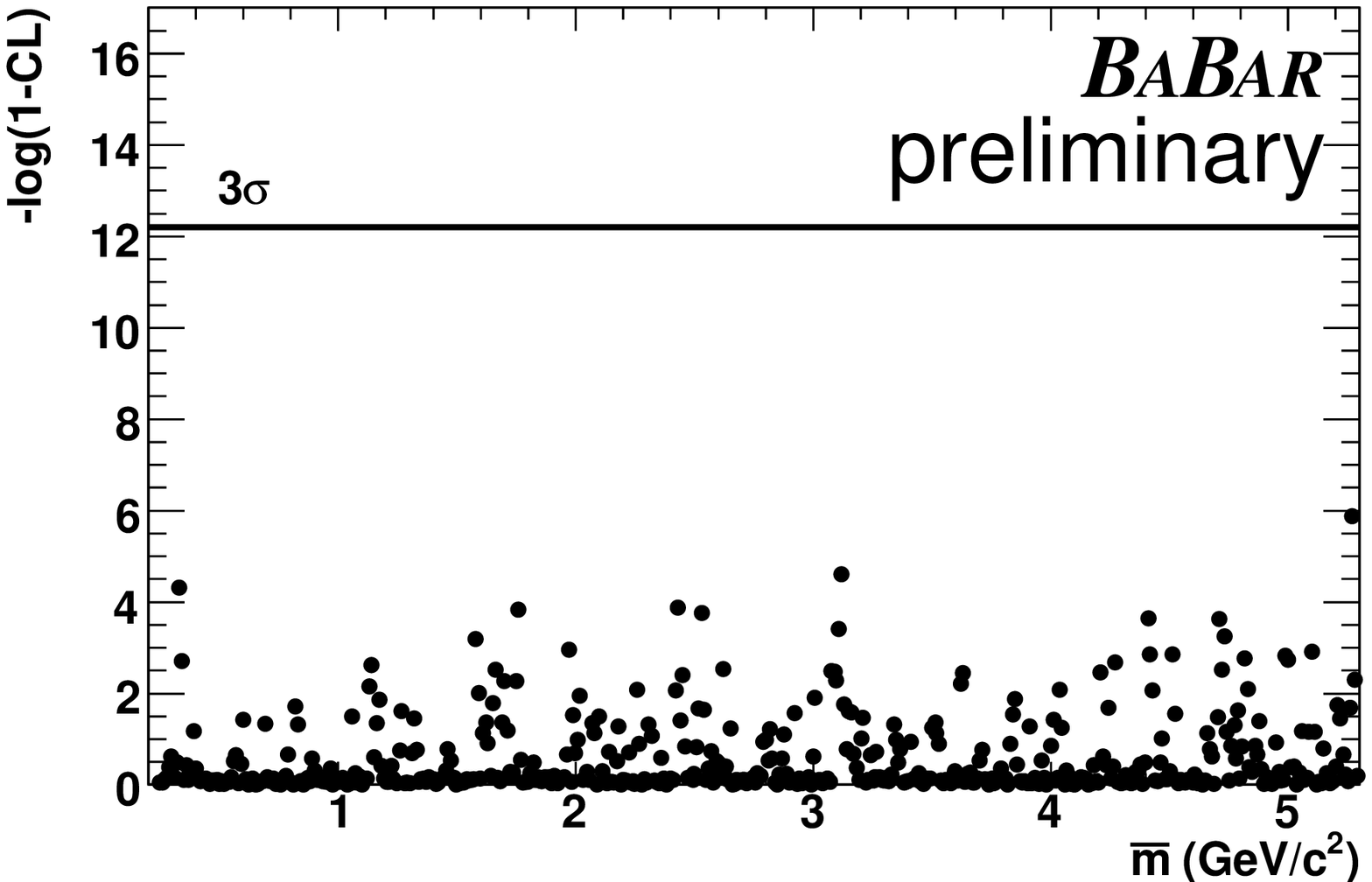}
              \epsfxsize6cm\epsffile{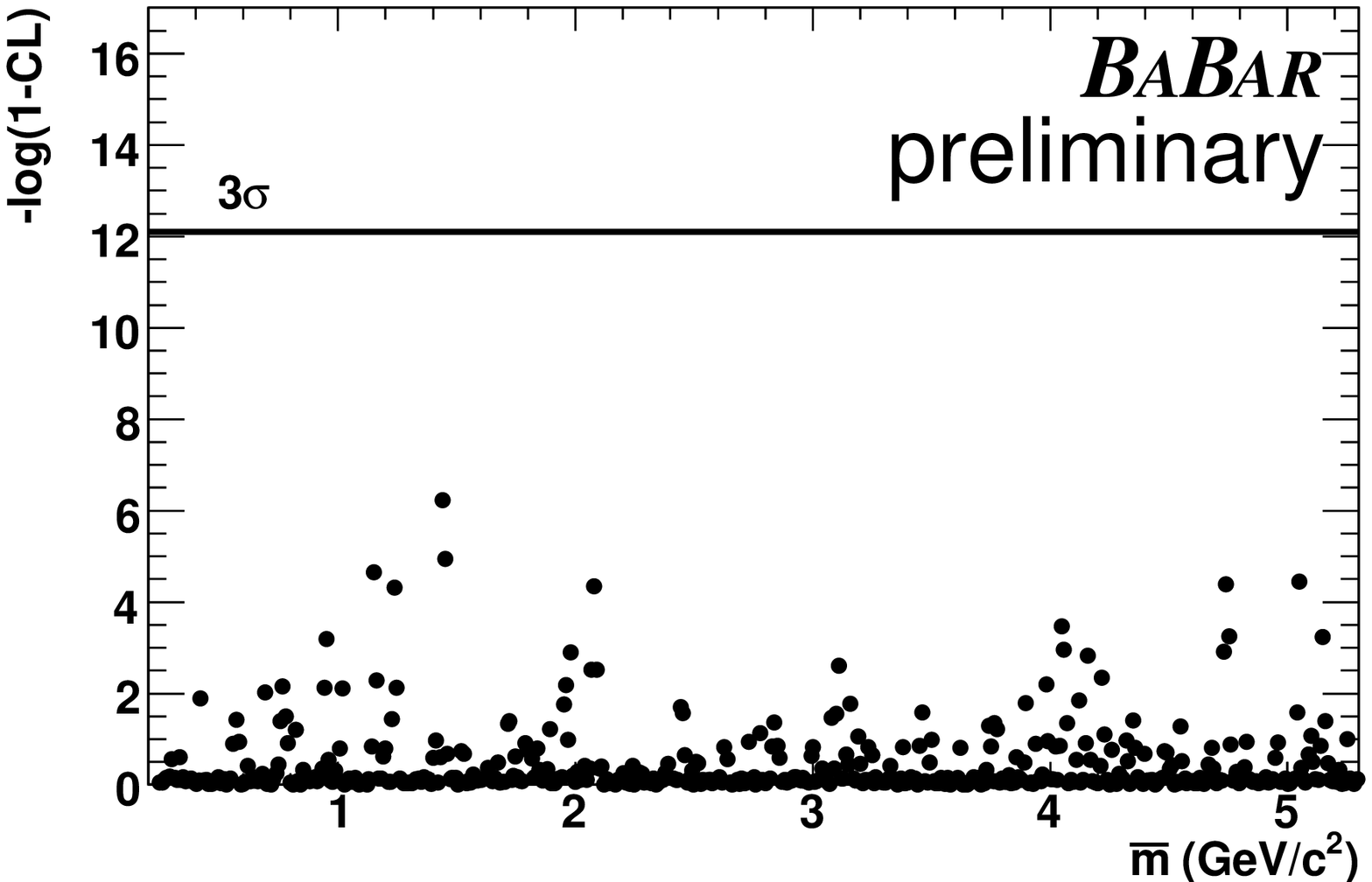}
              \epsfxsize6cm\epsffile{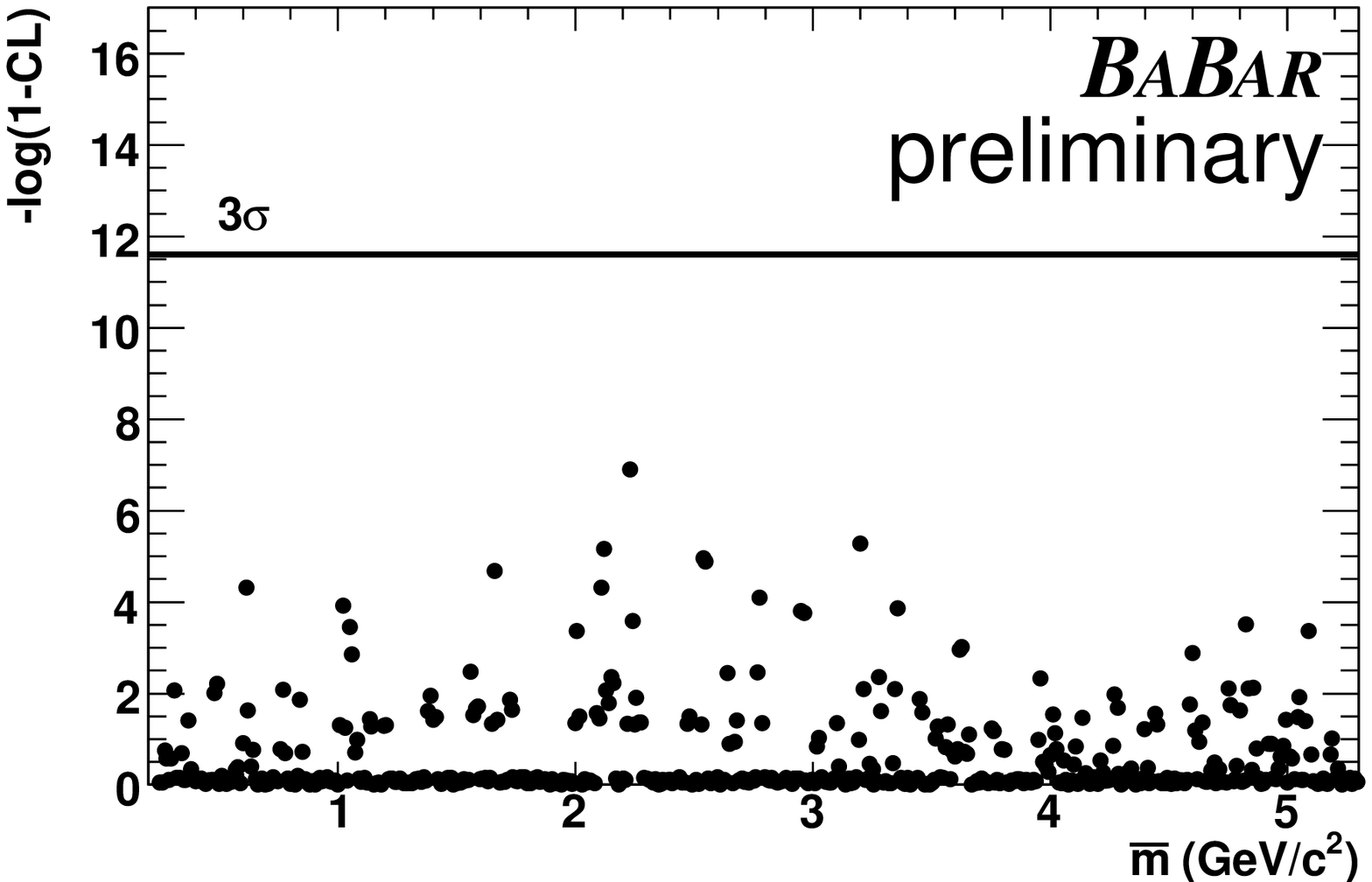}}
  \vspace{-0.1cm}
\caption{\label{fig:UnblindSig}
The value of $-ln(1-CL)$  versus $\mbar$ for  (left to right) $\epem\epem$, $\epem\mupmum$, and $\mupmum\mupmum$. 
}
\end{figure}

The spectra for the entire dataset (including the 10\% test sample) show no significant signal in any of $\foure$, $\twoetwomu$, $\fourmu$ final states, or the combination of the three.  The summary of results is shown in Table \ref{tab:resultsSumm}.  The distribution of observed signal events, after background subtraction, for all bins in $\mbar$ is shown in Figure \ref{fig:UnblindObs}. The values of $-ln(1-CL)$ versus $\mbar$ are shown in Figure \ref{fig:UnblindSig} and show no bins above the 3$\sigma$ value, shown on the plots.  The raw distribution of  $-ln(1-CL)$ compared to toy simulations with only background is shown in Figure \ref{fig:UnblindRawSig} and is in good agreement. The plots in Figure~\ref{fig:UnblindMaxSig} compare the values of the $-ln(1-CL_{max})$ observed in data with the distribution found in toy simulation.  
\begin{table}[thb]
  \begin{center}
    \caption{\label{tab:resultsSumm}
        Summary of the $-ln(1-CL{max})$ observed in data.  
}
%    \vspace{-0.8cm}
\begin{tabular}{lcc}
\hline
\hline  
&&\\[-0.2cm]
                 & $-ln(1-CL_{max})$  & $\mbar_{max}$ (GeV)   \\

\hline
&&\\[-0.2cm]
  $\epem\epem$   &5.88          &       5.27    \\
$\epem\mupmum$   &6.26          &       1.44    \\
$\mupmum\mupmum$ &6.94          &       2.23    \\
Combined         &7.15          &       1.66    \\
\hline
\hline
\end{tabular}
  \end{center}
\end{table}


\begin{figure}[tb]
  \centerline{\epsfxsize15cm\epsffile{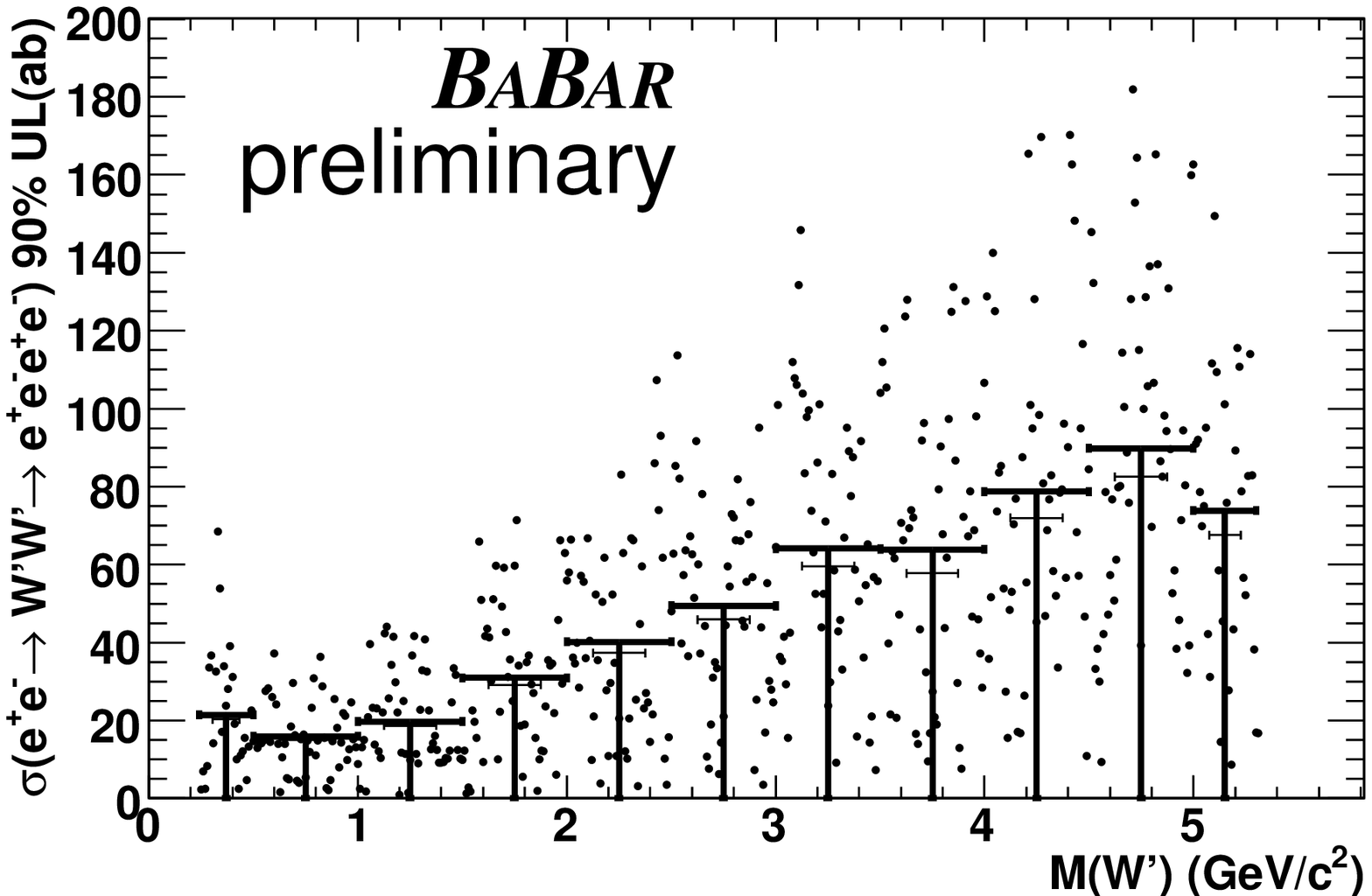}}
        \centerline{    \epsfxsize15cm\epsffile{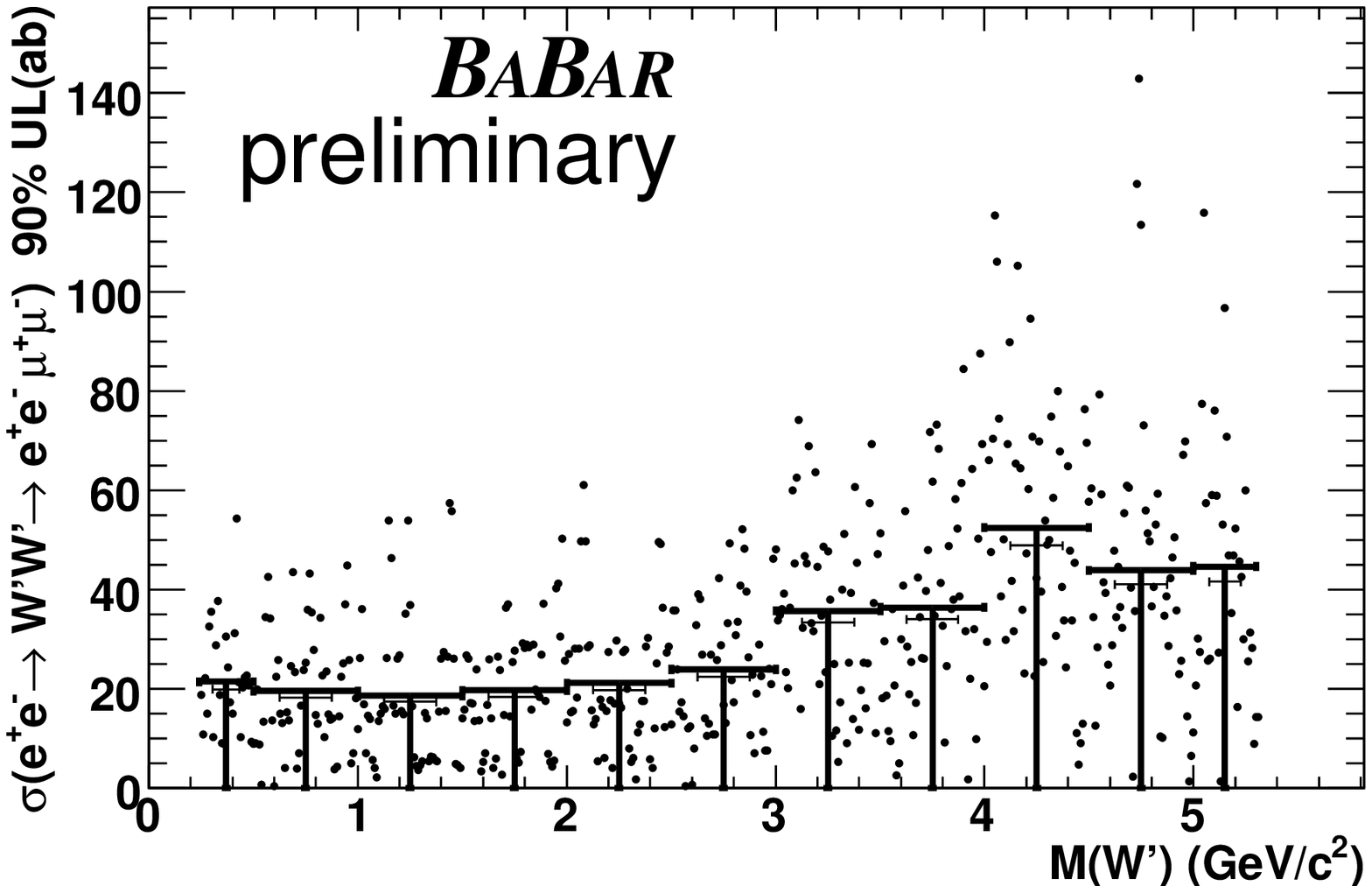}}
  \vspace{-0.1cm}
\caption{\label{fig:UnblindCSUL}
The cross section  90\% upper limit versus $\mbar$ for  (top to bottom) $\epem\to\Wp\Wp\to\epem\epem$ and  $\epem\to\Wp\Wp\to\epem\mupmum$.  The points are the upper limit for each $\mbar$ bin while the lines are the average of the limits over many bins.  
}
\end{figure}

\begin{figure}[tb]
     \centerline{          \epsfxsize15cm\epsffile{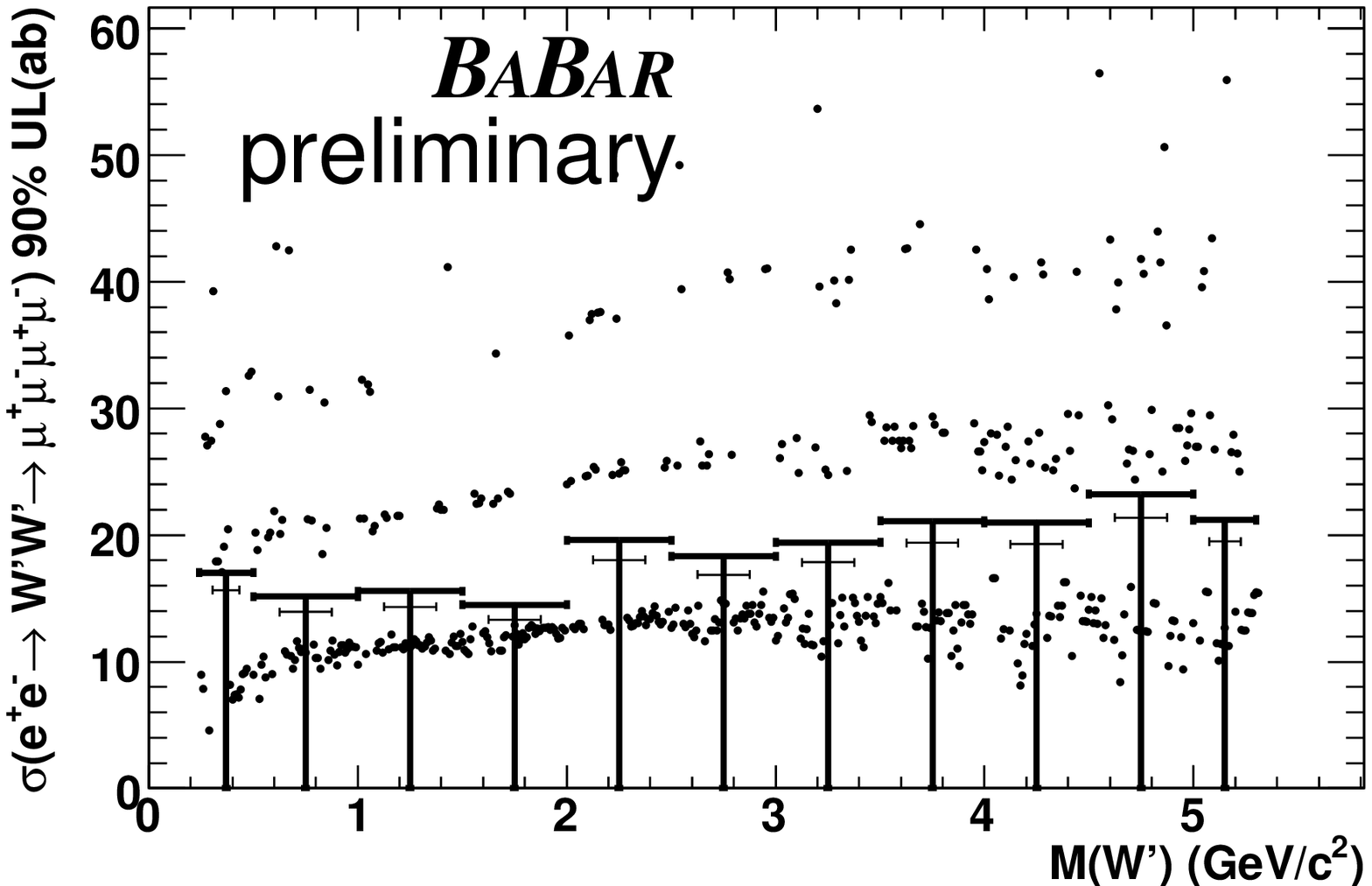}}
  \centerline{\epsfxsize15cm\epsffile{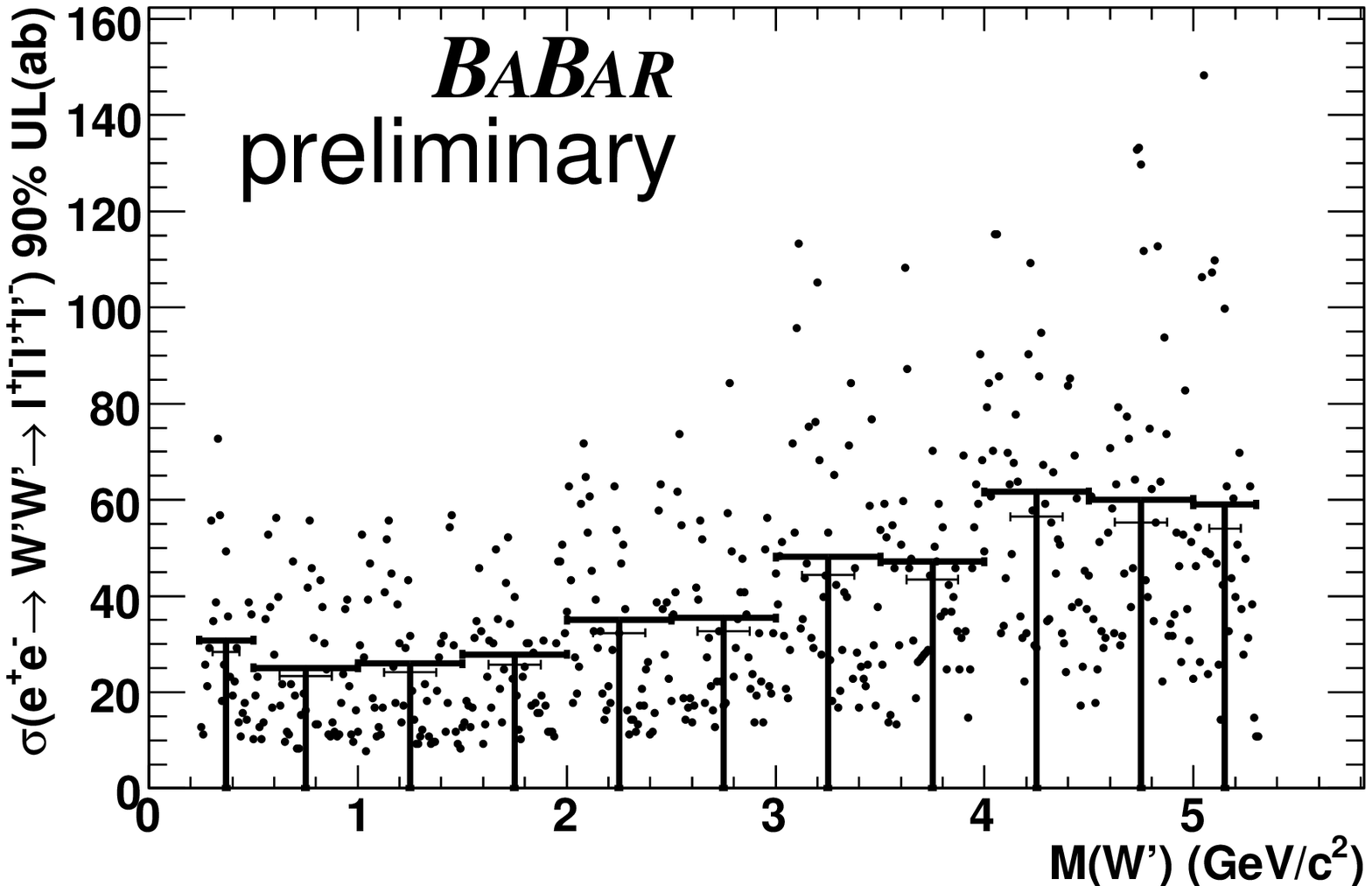}}
  \vspace{-0.1cm}
\caption{\label{fig:UnblindCSULCombined}
The cross section  90\% upper limit versus $\mbar$ for  (top to bottom)  $\epem\to\Wp\Wp\to\mupmum\mupmum$ and the combined $\epem\to\Wp\Wp\to\lplm l^{\prime +}l^{\prime -}$ assuming lepton universality. The points are the upper limit for each $\mbar$ bin while the lines are the average of the limits over many bins.  The band structure evident in the $\fourmu$ plot is due to the very low number of events in this mode.  
}
\end{figure}

% Figure \ref{fig:UnblindSigEvtsLimits} shows the 69\% interval (error bars) and 90\% upper limit (smooth curve) of the number of signal events for the three modes. 
Correcting for efficiency (Figure \ref{fig:effvsmass}, using linear interpolation between points and including the 90\% cut on $\dm$) and scaling by the luminosity, we obtain a 90\% upper limit for the cross section as shown in Figures \ref{fig:UnblindCSUL} and \ref{fig:UnblindCSULCombined}.  The points in these plots are the upper limit for each bin in $\mbar$ while the solid lines are the averages of the upper limits in the $\mbar$ region shown.  We set upper limits of $\sigma(\epem\to\Wp\Wp\to\foure)<(15-70)\ab$, $\sigma(\epem\to\Wp\Wp\to\twoetwomu)<(15-40)\ab$, and $\sigma(\epem\to\Wp\Wp\to\fourmu)<(11-17)\ab$ depending on $\Wp$ mass (taking the ranges from the averaged limits).  

Assuming lepton universality ($BR(\Wp\to\epem)=BR(\Wp\to\mupmum)$), we combine the three modes to obtain  upper limits for the 
reaction $\epem\to\Wp\Wp\to\lplm\lplmP$.  We obtain this limit by combining the individual profile likelihood functions for the 
three decay modes as a function of $\epem\to\Wp\Wp\to\lplm\lplmP$ cross section. 
The combined upper limit is shown in Figure \ref{fig:UnblindCSULCombined};  we set upper limits for  $\sigma(\epem\to\Wp\Wp\to\lplm\lplmP<(25-60)\ab$.   

%\begin{figure}[tb]
%  \centerline{\epsfxsize6cm\epsffile{figures/full-scan-4e.eps}
%              \epsfxsize6cm\epsffile{figures/full-scan-2e2mu.eps}
%              \epsfxsize6cm\epsffile{figures/full-scan-4mu.eps}}
%  \vspace{-0.1cm}
%\caption{\label{fig:UnblindSigEvtsLimits}
%The 69\% interval (error bars) and 90\% upper limit (smooth curve) of the number of signal events versus $\mbar$ for  (left to right) $\epem\epem$, $\epem\mupmum$, and $\mupmum\mupmum$. 
%}
%\end{figure}

\begin{figure}[tb]
     \centerline{          \epsfxsize15cm\epsffile{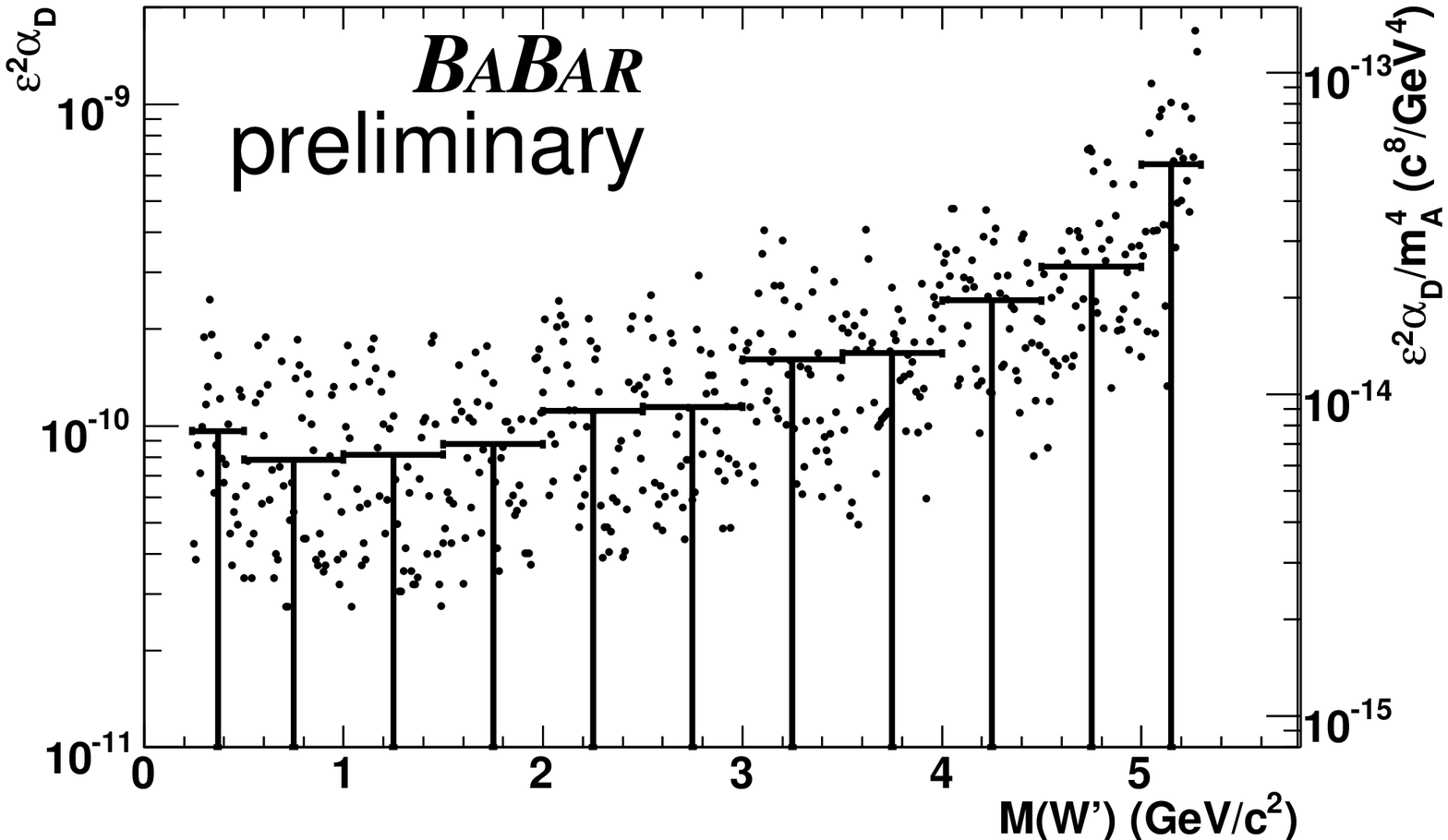}}
  \vspace{-0.1cm}
\caption{\label{fig:eps2alpha}
The  90\% upper limit on  $\varepsilon^2\alpha_D$ (left axis) or   $\frac{\epsilon^2\alpha_D}{m_{A^\prime}^4}$ (right axis) versus $m(\Wp)$.   The points are the upper limit for each $m(\Wp)$ bin while the lines are the average of the limits over many bins.  
}
\end{figure}

From the combined upper limit, we derive limits on the  possible couplings between the Standard Model and dark sectors.  The cross section for $\epem\to\Wp\Wp$ has been calculated by Essig \ea\cite{essig}.  For a dark photon $A^\prime$ mass less than the center of mass energy, $E_{cm}$,  the cross section is given by:
\beqn
\sigma(\epem\to\Wp\Wp)_{low}=N_c \frac{4\pi}{3}\frac{\varepsilon^2\alpha_D\alpha}{E_{cm}^2}\sqrt{1-\frac{4 m_{\Wp}^{2}}{E_{cm}^2}}\left(1+\frac{2 m_{\Wp}^2}{E_{cm}^2}\right)
\eeqn
while for an   $A^\prime$ mass larger than $E_{cm}$  the cross section is:
\beqn
\sigma(\epem\to\Wp\Wp)_{high}=N_c \frac{4\pi}{3}\frac{\varepsilon^2\alpha_D\alpha}{E_{cm}^2}\frac{E_{cm}^4}{m_{A^\prime}^4}\sqrt{1-\frac{4 m_{\Wp}^2}{E_{cm}^2}}\left(1+\frac{2 m_{\Wp}^2}{E_{cm}^2}\right)
\eeqn
where $N_c$ is the number of colors in the dark sector, $\varepsilon$ is the mixing parameter between the SM and the dark sector, and $\alpha_D$ is the dark sector coupling constant.  
Figure \ref{fig:eps2alpha} shows the upper limits we obtain on $\varepsilon^2\alpha_D$ assuming low $A^\prime$ mass, or on  $\frac{\epsilon^2\alpha_D}{m_{A^\prime}^4}$ assuming large $A^\prime$ mass.  For most of the mass range, we exclude values of $\varepsilon^2\alpha_D$ above $2\times 10^{-10}$ in the low  $A^\prime$ mass scenario or values of $\frac{\epsilon^2\alpha_D}{m_{A^\prime}^4}$ above $2\times 10^{-14}$ in the high $A^\prime$ mass scenario. In the model of Ref~\cite{essig}, these limits exclude the preferred parameter region for $A^\prime$ masses above $1.0\gevcc$.

 We would like to thank Rouven Essig, Philip Schuster, and Natalia 
Toro for useful discussions and for generating the signal Monte Carlo samples. 
We are grateful for the 
extraordinary contributions of our \pep2\ colleagues in
achieving the excellent luminosity and machine conditions
that have made this work possible.
The success of this project also relies critically on the 
expertise and dedication of the computing organizations that 
support \babar.
The collaborating institutions wish to thank 
SLAC for its support and the kind hospitality extended to them. 
This work is supported by the
US Department of Energy
and National Science Foundation, the
Natural Sciences and Engineering Research Council (Canada),
the Commissariat \`a l'Energie Atomique and
Institut National de Physique Nucl\'eaire et de Physique des Particules
(France), the
Bundesministerium f\"ur Bildung und Forschung and
Deutsche Forschungsgemeinschaft
(Germany), the
Istituto Nazionale di Fisica Nucleare (Italy),
the Foundation for Fundamental Research on Matter (The Netherlands),
the Research Council of Norway, the
Ministry of Education and Science of the Russian Federation, 
Ministerio de Educaci\'on y Ciencia (Spain), and the
Science and Technology Facilities Council (United Kingdom).
Individuals have received support from 
the Marie-Curie IEF program (European Union) and
the A. P. Sloan Foundation.

\end{document}